\documentclass[]{aastex631}

\usepackage{graphicx}	% Including figure files
\usepackage{amsmath}	% Advanced maths commands
\usepackage{amssymb}	% Extra maths symbols
\usepackage{hyperref}
\usepackage{subfigure}
\usepackage{romannum}

\newcommand{\diff}{{\text{d}}}
\allowdisplaybreaks
\usepackage{color}

\shorttitle{Magnetically Driven Winds in TDE disks}
\shortauthors{Mageshwaran et al.}
\graphicspath{{./}{figures/}}
\begin{document}

\title{Evolution of Tidal Disruption Event Disks with Magnetically Driven Winds}

\correspondingauthor{Mageshwaran Tamilan and Kimitake Hayasaki}
%\email{tmageshwaran2013@gmail.com,tmageshwaran@chungbuk.ac.kr}

\author[0000-0002-6187-4073]{Mageshwaran Tamilan}
\affiliation{Department of Space Science and Astronomy, Chungbuk National University, Cheongju 361-763, Korea}
\email{tmageshwaran2013@gmail.com, tmageshwaran@chungbuk.ac.kr, kimi@chungbuk.ac.kr}

\author[0000-0003-4799-1895]{Kimitake Hayasaki}
\affiliation{Department of Space Science and Astronomy, Chungbuk National University, Cheongju 361-763, Korea}
%\email{kimi@chungbuk.ac.kr}

\author[0000-0001-9734-9601]{Takeru K. Suzuki}
\affiliation{School of Arts and Sciences, The University of Tokyo, 3-8-1, Meguro, Tokyo 153-8902, Japan}

%% Note that the \and command from previous versions of AASTeX is now
%% depreciated in this version as it is no longer necessary. AASTeX 
%% automatically takes care of all commas and "and"s between authors names.

%% AASTeX 6.31 has the new \collaboration and \nocollaboration commands to
%% provide the collaboration status of a group of authors. These commands 
%% can be used either before or after the list of corresponding authors. The
%% argument for \collaboration is the collaboration identifier. Authors are
%% encouraged to surround collaboration identifiers with ()s. The 
%% \nocollaboration command takes no argument and exists to indicate that
%% the nearby authors are not part of surrounding collaborations.

%% Mark off the abstract in the ``abstract'' environment. 
\begin{abstract}
We present a time-dependent, one-dimensional, magnetically-driven disk wind model based on magnetohydrodynamic (MHD) equations, in the context of tidal disruption events (TDEs). We assume that the disk is geometrically thin and gas-pressure dominated, and explicitly accounts for magnetic braking and turbulent viscosity through an extended alpha-viscosity prescription. We find a particular wind solution for a set of basic equations that satisfies the necessary and sufficient conditions for vertically unbound MHD flows. The solution shows that the disk evolves with mass loss due to wind and accretion from the initial Gaussian density distribution. We confirm that the mass accretion rate follows the power law of time $t^{-19/16}$ at late times in the absence of wind, which matches the classical solution of \citet{1990ApJ...351...38C}. We find that the mass accretion rate is steeper than the $t^{-19/16}$ curve when the wind is present. Mass accretion is also induced by magnetic braking, known as the wind-driven accretion mechanism, which results in a faster decay with time of both the mass accretion and loss rates. In the disk emission, the ultraviolet (UV) luminosity is the highest among the optical, UV, and X-ray luminosities. While the optical and X-ray emission is observationally insignificant without magnetic braking, the X-ray emission is brighter at late times, especially in the presence of magnetic braking. This provides a possible explanation for observed delayed X-ray flares. Our model predicts that late-time bolometric light curves steeper than $t^{-19/16}$ in UV-bright TDEs are potentially compelling indicators of magnetically driven winds.
\end{abstract}

%% Keywords should appear after the \end{abstract} command. 
%% The AAS Journals now uses Unified Astronomy Thesaurus concepts:
%% https://astrothesaurus.org
%% You will be asked to selected these concepts during the submission process
%% but this old "keyword" functionality is maintained in case authors want
%% to include these concepts in their preprints.

%% From the front matter, we move on to the body of the paper.
%% Sections are demarcated by \section and \subsection, respectively.
%% Observe the use of the LaTeX \label
%% command after the \subsection to give a symbolic KEY to the
%% subsection for cross-referencing in a \ref command.
%% You can use LaTeX's \ref and \label commands to keep track of
%% cross-references to sections, equations, tables, and figures.
%% That way, if you change the order of any elements, LaTeX will
%% automatically renumber them.
%%
%% We recommend that authors also use the natbib \citep
%% and \citet commands to identify citations.  The citations are
%% tied to the reference list via symbolic KEYs. The KEY corresponds
%% to the KEY in the \bibitem in the reference list below. 

%%%%%%%%%%%%%%%%%%%%%%%%%%%%%%%%%%%%%%%%%%%%%%%%%%

%%%%%%%%%%%%%%%%% BODY OF PAPER %%%%%%%%%%%%%%%%%%

%
%%%%%%%%%%%
\section{Introduction} 
\label{sec:intro}
%%%%%%%%%%%
%

%A tidal disruption event (TDE) occurs when a star approaches a supermassive black hole (SMBH) closely enough to disrupt the star by the SMBH tidal force. A roughly half mass of the disrupted star is bound to the SMBH, whereas the other mass is unbound to escape away from the SMBH. The bound debris fallbacks to the SMBH at a rate of -5/3 power law of time \citep{1988Natur.333..523R}. 

A tidal disruption event (TDE) occurs when a star comes close enough to a supermassive black hole (SMBH) to be disrupted by the tidal force of the SMBH. Roughly half of the mass of the disrupted star is bound to the SMBH, while the other half is unbound and escapes from the SMBH. The bound debris falls back to the SMBH at a rate of -5/3 power law of time \citep{1988Natur.333..523R}. 
However, the mass fallback rate deviates from the $t^{-5/3}$ law at early times due to the stellar density \citep{2009MNRAS.392..332L}, stellar rotation \citep{2019ApJ...872..163G} and stellar orbital eccentricity \citep{2013MNRAS.434..909H,2018ApJ...855..129H,2020ApJ...900....3P,2022ApJ...924...34C,2023ApJ...959...19Z}. Relativistic apsidal precession causes the fallback debris to collide between the head and tail, resulting in the formation of an accretion disk due to its energy dissipation \citep{2013MNRAS.434..909H,2016MNRAS.461.3760H,2016MNRAS.455.2253B}.  Once the disk viscosity dominates the evolution of the formed disk at late times, the mass accretion rate is likely to deviate from the mass fallback rate, since the viscous timescale is typically longer than the debris orbital period.

%The relativistic apsidal precession makes the fallback debris collide between the head and tail parts, resulting in the formation of an accretion disk due to its energy dissipation \citep{2013MNRAS.434..909H,2016MNRAS.461.3760H,2016MNRAS.455.2253B}. Once the disk viscosity dominates the evolution of the formed disk at late times, the mass accretion rate is likely to deviate from the mass fallback rate as the viscous timescale is usually longer than the debris orbital period.

Self-similar solutions for a time-dependent, geometrically thin, and gas pressure-dominated TDE disk with $\alpha$-viscosity have been developed by \citet{1990ApJ...351...38C}. They found that the mass accretion rate is proportional to $t^{-19/16}$ at late times for zero inner stable circular orbit (ISCO) stress at the inner boundary of the disk, which is flatter than $t^{-5/3}$. Recently, \citet{2019MNRAS.489..132M} derived solutions for a general relativistic time-dependent, geometrically thin disk in Kerr spacetime in the context of TDEs. They demonstrated that the late-time bolometric luminosity follows a significantly flatter power law with time than $t^{-19/16}$ for the finite ISCO stress boundary.
 \citet{2017ApJ...838..149A} indicates that the late time light curves of most X-ray TDEs agree with this lower power law index than the classical $t^{-19/16}$ solution, while some other observations suggest a significant deviation from this agreement. For example, the bolometric luminosity of ASASSN-18pg is flatter at late times than the $t^{-5/3}$ law \citep{2020ApJ...898..161H}, and the bolometric luminosity of AT2019qiz shows a steep decrease with time, scaling as $t^{-2.54}$ \citep{2020MNRAS.499..482N}. Future observations may reveal the diversity of the power law index.

%Self-similar solutions for a time-dependent, geometrically thin, and gas-pressure dominant TDE disk with $\alpha-$viscosity were developed by \citet{1990ApJ...351...38C}. They found that the mass accretion rate is proportional to $t^{-19/16}$ at late times for zero inner stable circular orbit (ISCO) stress at the inner boundary of the disk, which is flatter than $t^{-5/3}$. Recently, \citet{2019MNRAS.489..132M} derived solutions for a general relativistic time-dependent, geometrically thin disk in Kerr spacetime in the context of TDEs. They demonstrated that the late time bolometric luminosity follows a significantly flatter power law of time than $t^{-19/16}$ for the finite ISCO stress boundary. \citet{2017ApJ...838..149A} indicates that the late time light curves of most X-ray TDEs agree with this lower power-law index than the classical $t^{-19/16}$ solution, while some other observations suggest a significant deviation from this agreement. For example, the bolometric luminosity of ASASSN-18pg is shallower at late times than the $t^{-5/3}$ law \citep{2020ApJ...898..161H} and the bolometric luminosity of AT2019qiz exhibits a precipitous diminution over time, scaling as $t^{-2.54}$ \citep{2020MNRAS.499..482N}. These observations suggest that future observations can exhibit the diversity of the power law index.

The geometrically thin disk approximation breaks down when the mass accretion rate exceeds the Eddington rate, where the radiation pressure is dominant. A geometrically thin and radiation pressure dominant disk causes the Lightman-Eardely thermal instability \citep{1974ApJ...187L...1L}. Therefore, the advection cooling term in the energy equation is needed to construct a thermally stable slim disk model \citep{1988ApJ...332..646A}. 
The strong radiation pressure in the super-Eddington disk causes an outflow from the disk \citep{2009MNRAS.400.2070S,2007ApJ...660..541G,2015MNRAS.448.3514C,2019ApJ...885...93F}. The disk wind obscures the disk emission if the wind is optically thick, reprocessing the X-rays to lower wavelengths. This is thought to make the optical and UV emissions observationally dominant \citep{2020SSRv..216..114R}. Even if the mass accretion rate is at sub-Eddington, the magnetic field in the disk can trigger the vertical outflow from the disk.

%The strong radiation pressure in the super-Eddington disk causes an outflow from the disk \citep{2009MNRAS.400.2070S,2007ApJ...660..541G,2015MNRAS.448.3514C,2019ApJ...885...93F}. The disk wind obscures the disk emission if the wind is optically thick, reprocessing the X-ray radiation to lower wavelengths. This is thought to make optical/UV emissions dominant observationally \citep{2020SSRv..216..114R}. Even if the mass accretion rate is at sub-Eddington, the magnetic field in the disk can trigger the vertical outflow from the disk. 

Stars possess magnetic fields that are thought to arise from a dynamo within the stars. Using Zeeman Doppler imaging, \citet{2016MNRAS.457..580F} found that the magnetic fields of 15 young solar-type stars in the mass range 0.7-1.2 $M_{\odot}$ and ages 20-250 Myr have complex large-scale geometries with mean field strengths of 14-140 Gs. Magnetic fields are also ubiquitous in massive stars; $\sim 10^{3}~{\rm Gs}$ in O stars \citep{2002MNRAS.333...55D,2008MNRAS.387L..23P} and 
$\sim 3 \times 10^4~{\rm Gs}$ in A and B stars \citep{1999A&A...343..865B}. A white dwarf, which is a possible object subject to tidal disruption by intermediate-mass black holes (IMBHs), typically has $\sim 10^9~{\rm Gs}$ \citep{2003ApJ...595.1101S}.

%Stars possess magnetic fields that are thought to arise from a dynamo within the stars. \citet{2016MNRAS.457..580F} using Zeeman Doppler imaging found that the magnetic fields of 15 young solar-type stars in the mass range 0.7-1.2 $M_{\odot}$ with ages 20 to 250 Myr have complex large-scale geometries with an average field strength of 14 to 140 Gs. Magnetic fields are also ubiquitously detected in massive stars; $\sim 10^{3}~{\rm Gs}$ in O stars \cite{2002MNRAS.333...55D,2008MNRAS.387L..23P} and $\sim 3 \times 10^4~{\rm Gs}$ in A and B stars \cite{1999A&A...343..865B}. A white dwarf, which is a possible object subject to a tidal disruption by intermediate-mass black holes (IMBHs), has typically $\sim 10^9~{\rm Gs}$ \cite{2003ApJ...595.1101S}.

The magnetic field strength and configuration within the star are altered by stellar tidal disruption. \citet{2017MNRAS.464.2816B} have shown that the magnetic field can promote the circularization process; in particular, both the circularization timescale and the circularization radius decrease with $v_{\rm A}/v_{\rm c}$, where $v_{\rm A}$ and $v_{\rm c}$ are the Alfv\'{e}n velocity and the circular orbital velocity, respectively. This is because the stellar debris loses angular momentum due to magnetic stresses during debris circularization. For a black hole mass of $M = 10^{6}M_{\odot}$, the circularization timescale is reduced by an order of magnitude for $v_{\rm A}/v_{\rm c} = 0.3$.

% The magnetic field strength and configuration within the star change by stellar tidal disruption. \citet{2017MNRAS.464.2816B} have shown that the magnetic field can promote the circularization process; specifically, both the circularization timescale and circularization radius decrease with $v_{\rm A}/v_{\rm c}$, where $v_{\rm A}$ and $v_{\rm c}$ are the Alfv\'{e}n velocity and circular orbital velocity, respectively. This is because the stellar debris loses angular momentum due to magnetic stresses during the debris circularization. For a black hole mass with $M = 10^{6}M_{\odot}$, the circularization timescale reduces by an order of magnitude for $v_{\rm A}/v_{\rm c} = 0.3$. 

\citet{2017MNRAS.469.4879B} found that the magnetic field evolution significantly depends on the pericenter of the stellar orbit. The magnetic field strength increases sharply near the pericenter when the star is close to the SMBH for a deep penetrating encounter. This is due to the strong compression of the star prior to the perturbation. They showed that the magnetic field never becomes dynamically critical for complete disruption of a star with an initial stellar magnetic field of 1 Gs. Instead, the full disruption of a star with a magnetic field of $10^6~{\rm Gs}$ produces debris streams where the magnetic pressure is comparable to the gas pressure a few tens of hours after the disruption and is crucial for the TDE dynamics. 

%\citet{2017MNRAS.469.4879B} found that the magnetic field evolution demonstrates a significant dependence on the pericenter of the stellar orbit.  The magnetic field strength increases sharply close to the pericenter when the pericenter is significantly close to the black hole for a deep penetrating encounter. This is due to the strong compression of the star before the disruption. They showed that the magnetic field never becomes dynamically crucial for the full disruption of a star with an initial stellar magnetic field of 1 Gs. Instead, the full disruption of a star with a magnetic field of $10^6~{\rm Gs}$ produces debris streams where the magnetic pressure is comparable to the gas pressure a few tens of hours after disruption and is crucial for TDE dynamics. 

\citet{2017ApJ...834L..19G} performed numerical magnetohydrodynamic (MHD) simulations of tidally disrupted stars. They showed that as the disrupted debris expands, the component of the magnetic fields perpendicular to the direction of debris stretching decreases, while the magnetic field strength of the component parallel to the stretching increases.The magnetic field configuration in any disk-like structure that forms from the debris is likely to be toroidal. 

%\citet{2017ApJ...834L..19G} performed the numerical magnetohydrodynamic (MHD) simulations of tidally disrupted stars. They showed that as the disrupted debris expands, the component of the magnetic fields perpendicular to the direction of debris stretching decreases, whereas the magnetic field strength of the component parallel to stretching increases. The magnetic field configuration in any disk-like structure that forms from the debris will likely be toroidal. 

The magnetic field in the accretion disk impacts the accretion dynamics and results in an astrophysical outflow or jets depending on the strength of the magnetic field. The strong magnetic field in the disk is required to have a jet. For example, the magnetic field required for the jetted TDE Swift J1644+57 is $B \approx 10^8 ~{\rm Gs}$ \citep{2014MNRAS.437.2744T}. Such strongly jetted TDEs could result from the complete disruption of a star with a high magnetic field strength, or after multiple encounters by partial disruption, where the magnetic field of the surviving core is amplified after each passage \citep{2017MNRAS.469.4879B}. 

%For example, the required magnetic field for jetted TDE Swift J1644+57 is $B \approx 10^8 ~{\rm Gs}$ \citep{2014MNRAS.437.2744T}. Such strong jetted TDEs could arise from the full disruption of a star with high magnetic field strength or after multiple encounters via partial disruption where the magnetic field of the surviving core is amplified after each passage \citep{2017MNRAS.469.4879B}. 

The differential rotation of electrically conducting fluids around a central object causes a magnetorotational instability. (MRI; \citealp{Velikhov1959,1961hhs..book.....C,1991ApJ...376..214B,1998RvMP...70....1B}), which induces Maxwell and Reynolds stresses due to the resultant MHD turbulence. The turbulent Maxwell and Reynolds stresses transport angular momentum outwards, driving mass accretion. In addition, vertical outflows are also initiated by the turbulent MHD pressure \citep{2009ApJ...691L..49S,2014ApJ...784..121S}. The mass loss from such disk winds reduces the accretion rate in the inner disk and the disk luminosity.

%The differential rotation of electrically conducting fluids around a central object causes a magnetorotational instability (MRI; \citealp{Velikhov1959,1961hhs..book.....C,1991ApJ...376..214B,1998RvMP...70....1B}), inducing Maxwell and Reynolds stresses due to the resultant MHD turbulence.  
%The turbulent Maxwell and Reynolds stress transport angular momentum outward, which drives mass accretion. In addition, vertical outflows are also launched by MHD turbulent pressure \citep{2009ApJ...691L..49S,2014ApJ...784..121S}. The mass loss by such disk winds reduces the accretion rate in the inner disk and the luminosity from the disk.

%\citet{2018ApJ...859L..20D} performed a general relativistic radiation magnetohydrodynamics (GRRMHD) simulation of a disk around a black hole of mass $M = 5 \times 10^6 M_{\odot}$ and spin $j = 0.8$. The initial weak poloidal field (plasma beta of $\sim 20 - 30$) amplifies via MRI, and a magnetically arrested disk (MAD) is formed up to the radius $r \sim 80 r_{\rm g}$, where 

\citet{2018ApJ...859L..20D} performed a general relativistic radiation magnetohydrodynamics (GRRMHD) simulation of a disk around a black hole of mass $M = 5 \times 10^6 M_{\odot}$ and spin $j = 0.8$. The initial weak poloidal field (plasma beta of $\sim 20 - 30$) is amplified by MRI, and a magnetically arrested disk (MAD) is formed up to radius $r \sim 80 r_{\rm g}$, where 
\begin{eqnarray}
r_{\rm g} = \frac{G M}{c^2}
\sim 1.5 \times 10^{11}\,{\rm cm}\,
\left(
\frac{M}{10^6M_\odot}
\right)
\label{eq:rg}
\end{eqnarray}
is the gravitational radius, $G$ is the gravitational constant, and $c$ is the speed of light. The accumulated magnetic flux is $\Phi \sim 10^{31}~{\rm Gs~cm^{-2}}$ and induces a jet due to the high spinning black hole. Along with this, a wide and fast wind is launched from the magnetized disk supported by the radiation pressure, and the outflowing wind reprocesses the radiation from the disk. 
The recent GRRMHD simulations by \citet{2019MNRAS.483..565C} of a super-Eddington disk that is formed by the disruption of a solar mass star by $10^6 M_{\odot}$ black hole mass found that a rapidly spinning black hole and MAD accretion are necessary to produce a jetted TDE, which agrees with the simulation by \citet{2018ApJ...859L..20D}.

The time evolution of the mass accretion rate and the mass outflow rate is the key to determining the radiation reprocessing dynamics of the disk \citep{2020ApJ...894....2P,2020SSRv..216..114R}. The numerical simulations have shown that an outflow with or without a jet arises from the disk \citep{2018ApJ...859L..20D, 2019MNRAS.483..565C, 2023MNRAS.518.3441C}, but little is known about the long-term evolution of the magnetically driven disk-wind system. Moreover, the classical solution demonstrated that the power-law index is $n=-19/16$ for the late-time variation of the TDE light curves \citep{1990ApJ...351...38C}. However, some observations indicate steeper slopes than $n=-19/16$. Another motivation for our study is to propose a natural theoretical model to explain this steeper decline at late times.

%The time evolution of the mass accretion rate and the mass outflow rate is key to determining the disk radiation reprocessing dynamics \citep{2020ApJ...894....2P,2020SSRv..216..114R}. The numerical simulations have shown that an outflow with or without a jet arises from the disk (\citet{2018ApJ...859L..20D, 2019MNRAS.483..565C, 2023MNRAS.518.3441C}), but there is little known about the long-term evolution of the magnetically driven disk-wind system. In addition, the classical solution demonstrated the power-law index is $n=-19/16$ for the late-time variation of the TDE light curves \citep{1990ApJ...351...38C}. However, some observations indicate the steeper slopes than $n=-19/16$. Another motivation for our study is to propose a natural theoretical model to explain this steeper decline at late times.

We construct numerical models of one-dimensional gas pressure dominated TDE disks with magnetically driven outflows based on 
\citet{2016A&A...596A..74S} to study their long-term evolution.  In section~\ref{sec:model}, we describe the basic equations of our model and the initial and boundary conditions. In section~\ref{sec:results} we present numerical solutions for the basic equations: radial profiles of the disk and wind quantities, as well as the evolution of mass accretion and loss rates, disk mass, and angular momentum. For comparison with observations, we also compute disk spectra and light curves. We discuss our results in section~\ref{sec:discussion}. Section~\ref{sec:conclusion} is devoted to our conclusions.

%We construct numerical models of one-dimensional, gas-pressure dominant TDE disks with magnetically driven outflows based on \citet{2016A&A...596A..74S} for studying their long-term evolution.  In Section~\ref{sec:model}, we describe the basic equations of our model and initial and boundary conditions. Section~\ref{sec:results} presents numerical solutions for the basic equations: radial profiles of disk and wind quantities as well as the evolution of mass accretion and loss rates, disk mass, and angular momentum. For comparison with observations, we also calculate disk spectra and light curves. We discuss our results in Section~\ref{sec:discussion}. Section~\ref{sec:conclusion} is devoted to our conclusions. 

%
%%%%%%%%%%%
\section{Model} 
\label{sec:model}
%%%%%%%%%%%
%
Following \citet{2016A&A...596A..74S}, we develop a one-dimensional, axisymmetric, time-dependent model with an outflow in cylindrical coordinates $\{r,~\phi,~z\}$, based on the standard accretion disk model (i.e., a vertically integrated, gas-pressure dominant, geometrically thin viscous accretion disk) with Keplerian rotation $\Omega=\sqrt{GM/r^3}$.

%Following \citet{2016A&A...596A..74S}, we develop a one-dimensional, axisymmetric time-dependent model with an outflow in cylindrical coordinates $\{r,~\phi,~z\}$ based on the standard accretion disk model (i.e., a vertically integrated, gas-pressure dominant, geometrically thin viscous accretion disk) with a Keplerian rotation $\Omega=\sqrt{GM/r^3}$.
%
%%%%%%%%%%%%%%%
\subsection{Basic equations}
\label{sec:basiceq}
%%%%%%%%%%%%%%%
%

The pressure of the disk is dominated by the gas pressure, $p_{\rm gas} = k_{\rm B} \rho T / \mu m_{\rm p}$, where $\rho$ is the density, $T$ is the mid-plane temperature of the disk, $k_{\rm B}$ is the Boltzmann constant, $m_{\rm p}$ is the proton mass, and $\mu$ is the mean molecular weight, assumed to be $0.65$. This is the mean molecular weight of the ionized gas for the Sun. We derive the vertical scale height of the disk to be $ H = c_{\rm s}/\Omega$ from the hydrostatic equilibrium for the geometrically thin disk, where the sound speed estimated at the disk midplane is given by $c_{\rm s} = \sqrt{p_{\rm gas}/\rho} = \sqrt{k_{\rm B} T / \mu m_{\rm p}}$ from the equation of state.

% The pressure of the disk is dominated by the gas pressure, $p = k_{\rm B} \rho T / \mu m_{\rm p}$, where $\rho$ is the density, $T$ is the mid-plane temperature of the disk, $k_{\rm B}$ is the Boltzmann constant, $m_{\rm p}$ is the proton mass, and $\mu$ is the mean molecular weight taken to be ionized solar mean molecular weight of $0.65$. We derive the vertical scale height of the disk to be $ H = c_{\rm s}/\Omega$ from the hydrostatic equilibrium for the geometrically thin disk, where the sound speed estimated at the disk midplane is given by $c_{\rm s} = \sqrt{p/\rho} = \sqrt{k_{\rm B} T / \mu m_{\rm p}}$ from the equation of state.

%By utilizing the MHD mass and momentum conservation equations given in \citet{1998RvMP...70....1B} with some approximations (see  Appendix \ref{app_sdeq} for the detail), we derive the evolution of the disk's surface density, $\Sigma =  \int_{-H}^{H} \rho \, \diff z \simeq 2 H \rho$,as \citep{2016A&A...596A..74S}

Using the MHD mass and momentum conservation equations given in \citet{1998RvMP...70....1B} with some approximations (see Appendix \ref{app_sdeq} for details), we derive the evolution of the disk surface density, $\Sigma = \int_{-H}^{H} \rho \, \diff z \simeq 2 H \rho$,
as \citep{2016A&A...596A..74S}
\begin{equation}
\frac{\partial \Sigma}{\partial t} - \frac{2}{r}\frac{\partial}{\partial r}\left[\frac{1}{r\Omega}\left\{\frac{\partial}{\partial r} \left(\bar{\alpha}_{r\phi} r^2 \Sigma  c_{\rm s}^2 \right) + \bar{\alpha}_{z\phi} r^2 \rho c_{\rm s}^2 \right\}\right] + \dot\Sigma_{\rm w}=0
\label{eq:sdevo}
\end{equation}
where $\dot\Sigma_{\rm w}\equiv2\rho v_{z,H}$ is the vertical mass flux, $v_{z,H}\equiv v_z(r,H)$ is the vertical velocity evaluated at the disk scale height $H$, and $\bar{\alpha}_{r\phi}$ and $\bar{\alpha}_{z\phi}$ are introduced as constant parameters due to the MHD turbulence and disk winds, respectively. According to equations~(\ref{eq:alrphi0}) and (\ref{eq:alzphi0}), these {$\alpha$} parameters are given by
\begin{eqnarray}
\bar\alpha_{r\phi}
&\equiv&
\frac{1}{c_{\rm s}^2}
\int_{-H}^{H} \rho \left[v_r \delta v_{\phi} - \frac{B_r B_{\phi}}{4\pi \rho}\right] 
\, dz 
\biggr/
\int_{-H}^{H} \rho\, dz,
\label{eq:alrphi}
\\
\bar\alpha_{z\phi}
&\equiv&
\frac{1}{c_{\rm s}^2}
\biggr[
v_z \delta v_{\phi} - \frac{B_z B_{\phi}}{4\pi \rho}
\biggr]_{z=H} \label{eq:alzphi} \\
&=&
\frac{4}{(3\tau)^{1/4}}\frac{1}{c_{{\rm s},H}^2}
\biggr[
v_z \delta v_{\phi} - \frac{B_z B_{\phi}}{4\pi \rho}
\biggr]_{z=H} \nonumber \\
&\approx&
\frac{1}{c_{{\rm s},H}^2}
\biggr[
v_z \delta v_{\phi} - \frac{B_z B_{\phi}}{4\pi \rho}
\biggr]_{z=H} \nonumber
\end{eqnarray}
where $\delta v_\phi$ is an azimuthal velocity perturbation, $(B_r, B_\phi, B_z)$ is each component of the $B$ field in cylindrical coordinates, $c_{{\rm s}, H}$ is the sound velocity evaluated at $H$, the relation between $c_{\rm s}$ and $c_{{\rm s}, H}$ is given by $c_{{\rm s},H}^2/c_{\rm s}^2=2/(3\tau)^{1/4}\sim\mathcal{O}(1)$ for $\tau\sim 10 - 100$ with the optical depth $\tau$, and the quantities in the brackets of the equation~(\ref{eq:alzphi}) are evaluated at $H$. In fact, $\bar{\alpha}_{r\phi}$ and $\bar{\alpha}_{z\phi}$ control how much angular momentum flux is removed radially by accretion and vertically by the winds from the top and bottom of the disk, respectively. The latter parameter corresponds to the magnetic braking; MHD disk winds carry away a significant fraction of the angular momentum from the accretion dis, which triggers mass accretion in the disk. The magnetic braking is quantitatively controlled by the magnetic field strength and the disk-wind flux \citep{1982MNRAS.199..883B,2023ASPC..534..567P}; in our model setup, the effect of magnetic braking is prescribed in the single parameter $\bar{\alpha}_{z\phi}$.

For the sake of simplicity, we assume that the second term of equation~(\ref{eq:sdevo}) is proportional to the thermal mass flux at the disk mid-plane:
\begin{equation}
\dot\Sigma_{\rm w}= C_{\rm w}\,\rho c_{\rm s},
\label{vzeqn}
\end{equation}
%where $C_{\rm w}$ is introduced by \citet{2010ApJ...718.1289S} as a non-dimensional proportionality coefficient, which is a function of radius and time. The MHD energy equation for an optically thick, Keplerian disk is given by (see Appendix~\ref{app_eneq})
where $C_{\rm w}$ is introduced by \citet{2010ApJ...718.1289S} as a dimensionless proportionality coefficient that is a function of radius and time. The MHD energy equation for an optically thick, Keplerian disk is given by (see Appendix~\ref{app_eneq})
\begin{equation}
C_{\rm w} \rho c_{\rm s} \left[E_{\rm w} + \frac{r^2 \Omega^2}{2} \right] + Q_{\rm rad} 
= 
\frac{3}{2} \bar{\alpha}_{r\phi} \Omega \Sigma c_{\rm s}^2 + \bar{\alpha}_{z\phi} r \Omega \rho c_{\rm s}^2,
\label{eq:ene0}
\end{equation}
where $Q_{\rm rad}$ is the radiative cooling rate:
\begin{eqnarray}
Q_{\rm rad} = \frac{64 \sigma T^4 }{3 \kappa_{\rm es} \Sigma}
\label{eq:qrad}
\end{eqnarray}
with the Stefan-Boltzmann constant $\sigma$ and the Thomson scattering opacity $\kappa_{\rm es}=0.34~{\rm cm^{-2}g^{-1}}$, and $E_{\rm w}$ is given by equation~(\ref{eq:ew0}) as
\begin{equation}
E_{\rm w} = \frac{1}{2} v^2 + \Phi + \frac{\gamma c_{\rm s}^2}{\gamma -1} + \frac{B_{\phi}^2 + B_r^2}{4\pi \rho} - \frac{B_z}{4\pi\rho v_z} \left(v_{\phi}B_{\phi} + v_r B_r\right),
\label{eq:ew}
\end{equation}
where $\Phi$ is the gravitational potential and $\gamma$ is the specific heat ratio.

%The necessary condition to make a disk wind blow is $E_{\rm w} \geq 0$, ensuring that the wind material can reach infinity with a positive velocity in the vertical direction. To obtain a particular wind solution, we impose $E_{\rm w} = 0$ on equation~(\ref{eq:ene0}) so that we get  
The necessary condition for a disk wind to blow is $E_{\rm w} \geq 0$, which ensures that the wind material can reach infinity with a positive velocity in the vertical direction. To obtain a particular wind solution, we impose $E_{\rm w} = 0$ on equation~(\ref{eq:ene0}) so that we get 
\begin{equation}
C_{\rm w,0} \frac{\Sigma r^2 \Omega^3}{4} + Q_{\rm rad} = \frac{3}{2} \bar{\alpha}_{r\phi}  \Omega \Sigma c_{\rm s}^2 + \frac{1}{2}\bar{\alpha}_{z\phi} r \Sigma \Omega^2 c_{\rm s},
\label{eq:ene1}
\end{equation}
where $C_{\rm w,0}=C_{\rm w}(E_{\rm w}=0)$ is the normalized mass flux at $E_{\rm w}=0$ and we adopt the thin disk approximation:

\begin{eqnarray} 
c_{\rm s} \approx \Omega{H}.
\label{eq:thindisk}
\end{eqnarray}

%By performing local MHD shearing box simulations for a protoplanetary disk, \cite{2009ApJ...691L..49S,2010ApJ...718.1289S} elucidated the mass flux of disk winds at a wind onset region where magnetic energy stands roughly commensurate with thermal energy, and subsequently they deduced $C_{\rm w, sim}$ as $10^{-5}\sim10^{-4}$ \citep{2016A&A...596A..74S}. Nonetheless, the estimated value of $C_{\rm w, sim}$ may exhibit overestimation due to the limitation of the vertical simulation domain \citet{2010ApJ...718.1289S} indicated that the vertical mass flux diminishes by a factor of $2-3$ by performing simulations with an augmented vertical box size. Consequently, we designate $C_{\rm w}$ such that
By performing local MHD shearing box simulations for a protoplanetary disk, \cite{2009ApJ...691L..49S,2010ApJ...718.1289S} elucidated the mass flux of disk winds in a wind onset region where the magnetic energy is roughly equal to the thermal energy, and subsequently they deduced $C_{\rm w, sim}$ as $10^{-5}\sim10^{-4}$ \citep{2016A&A...596A..74S}. Nevertheless, the estimated value of $C_{\rm w, sim}$ may be overestimated due to the limitation of the vertical simulation domain. \citet{2010ApJ...718.1289S} indicated that the vertical mass flux diminishes by a factor of $2-3$ when simulations are performed with an augmented vertical box size. Consequently, we denote $C_{\rm w}$ such that
\begin{equation}
C_{\rm w} = {\rm Min}[C_{\rm w, sim},\,C_{\rm w,0}]
\label{eq:cwtot}
\end{equation}
conservatively.

%It is found from equations~(\ref{eq:sdevo}), (\ref{eq:qrad}), (\ref{eq:ene1}), and (\ref{eq:cwtot}) that there are two basic equations with respect to three variables ($\Sigma, T, C_{\rm w,0}$). Therefore, we introduce an additional equation to make a closure of those equations as
From equations~(\ref{eq:sdevo}), (\ref{eq:qrad}), (\ref{eq:ene1}), and (\ref{eq:cwtot}), we see that there are two basic equations with respect to three variables ($\Sigma, T, C_{\rm w,0}$). Therefore, we introduce an additional equation to make a closure of these equations as
\begin{equation}
Q_{\rm rad} 
= 
\epsilon_{\rm rad} 
\left[
\frac{3}{2}
\bar{\alpha}_{r\phi}
\Omega \Sigma c_{\rm s}^2 
+ 
\frac{1}{2}
\bar{\alpha}_{z\phi} r \Sigma \Omega^2 c_{\rm s}
\right],
\label{eq:qradf}
\end{equation}
where $\epsilon_{\rm rad}$ represents a parameter that quantifies the fraction of accretion energy flux converted into radiative cooling flux. Here, the accretion energy is defined as the sum of the viscous heating and the gravitational energy liberated by the inward mass flows, which are driven by the outward transport of angular momentum due to MHD turbulence (characterized by $\bar{\alpha}_{r\phi}$) and the removal of angular momentum via magnetic braking (characterized by $\bar{\alpha}_{\phi z}$). The parameter $\epsilon_{\rm rad}$ lies within the range $0<\epsilon_{\rm rad}\le1$, where $\epsilon_{\rm rad} = 1$ results in $C_{\rm w,0} = 0$ and $v_z = 0$, corresponding to a standard disk solution without wind \citep{1990ApJ...351...38C,2002apa..book.....F,kato_black-hole_2008}. Note that the wind carries away a fraction $(1-\epsilon_{\rm rad})$ of the accretion energy.
%The $\epsilon_{\rm rad}$ parameter ranges $0\le\epsilon_{\rm rad}\le1$, where $\epsilon_{\rm rad} = 1$ results in $C_{\rm w,0} = 0$ and $v_z=0$, corresponding to a standard disk solution without wind \citep{1990ApJ...351...38C,2002apa..book.....F}.

From equations~(\ref{eq:sdevo}), (\ref{eq:qrad}), (\ref{eq:ene1}), and (\ref{eq:cwtot}), the basic equations governing the magnetically-driven disk wind are summarized as follows:
\begin{eqnarray}
&&
\frac{\partial \Sigma}{\partial t} 
= 
\frac{2}{r}\frac{\partial}{\partial r}\left[\frac{1}{r\Omega}\frac{\partial}{\partial r} 
\left(
\bar{\alpha}_{r\phi}
r^2 \Sigma c_{\rm s}^2 
\right)
\right] 
+
\frac{1}{r} \frac{\partial}{\partial r} \left( \bar{\alpha}_{z\phi} r\Sigma c_{\rm s}\right) -\frac{1}{2}C_{\rm w} \Sigma \Omega, 
\label{eq:sigma} \\
&&
\frac{1}{4}C_{\rm w,0} \Sigma r^2 \Omega^3 
+ 
\frac{64 \sigma T^4 }{3 \kappa_{\rm es} \Sigma} 
= 
\frac{3}{2}
\bar{\alpha}_{r\phi}
\Omega \Sigma c_{\rm s}^2 + \frac{1}{2} \bar{\alpha}_{z\phi} r \Omega^2 \Sigma c_{\rm s},
\label{eq:ene2} \\
&&
C_{\rm w,0} = \left(1-\epsilon_{\rm rad}\right)\left[6 \bar{\alpha}_{r\phi} \frac{c_{\rm s}^2}{r^2\Omega^2} + 2 \bar{\alpha}_{z\phi} \frac{c_{\rm s}}{r\Omega}\right],
\label{eq:cw} \\
&&
c_{\rm s} =  \sqrt{
\frac{k_{\rm B} T}{\mu m_{\rm p}}
}
\label{eq:cs}
\end{eqnarray}
where equation (\ref{eq:sigma}) describes the temporal evolution of the surface density, equation (\ref{eq:ene2}) gives the radial profile of the mid-plane temperature of the disk, and equation (\ref{eq:cw}) determines $C_{\rm w,0}$, for which we employ equations (\ref{eq:ene1}) and (\ref{eq:qradf}) in its derivation. Equation~(\ref{eq:cs}), which originates from the ideal gas equation of state, gives the relation between $c_{\rm s}$ and $T$. We solve equations (\ref{eq:sigma})-(\ref{eq:cs}) simultaneously to get the time evolution of the disk with wind.
%where equation (\ref{eq:sigma}) delineates the temporal evolution of the surface density, equation (\ref{eq:ene2}) gives the radial profile of the disk's mid-plane temperature, and equation (\ref{eq:cw}) determines $C_{\rm w,0}$, for which we employ equations (\ref{eq:ene1}) and (\ref{eq:qradf}) in its derivation. Equation~(\ref{eq:cs}), which originates from the equation of state for the ideal gas, gives the relation between $c_{\rm s}$ and $T$. We solve equations (\ref{eq:sigma})-(\ref{eq:cs}) simultaneously to get the time evolution of disk with wind.

%
%%%%%%%%%%%%%%%%%%%%%%%%%%%%%%%%%
\subsection{Initial and boundary conditions with model parameters}
%\subsection{Model setup and parameters}
\label{sec:inicon}
%%%%%%%%%%%%%%%%%%%%%%%%%%%%%%%%%
%

%
% 1st paragraph
%
In our models, we postulate an SMBH mass of $M=10^6\,M_\odot$ with the solar-type star. The tidal disruption radius is then given by 
\begin{eqnarray}
r_{\rm t}
&& 
\simeq 
\left(\frac{M}{M_{\star}}\right)^{1/3}R_{\star}
\nonumber \\
&&
\sim 
7.0\times10^{12}\,{\rm cm}
\left(\frac{M}{10^6\,M_\odot}\right)^{1/3}
\left(\frac{M_\star}{M_\odot}\right)^{-1/3}
\left(\frac{R_\star}{R_\odot}\right)
\nonumber 
\label{eq:rt}
\end{eqnarray}
\citep{1975Natur.254..295H}, where $M_{\star}$ and $R_{\star}$ are the stellar mass and radius, respectively.

%
% 2nd paragraph
%
The conservation of the angular momentum of stellar debris provides the circularization radius of stellar debris to be  \citep{2009MNRAS.400.2070S,2016MNRAS.461.3760H}
\begin{eqnarray}
r_{\rm c} 
&&
= \frac{(1+e_*) r_{\rm t}}{\beta}=2r_{\rm t}
\nonumber \\
&&
\sim 
1.4\times10^{13}\,{\rm cm}
\left(\frac{M}{10^6\,M_\odot}\right)^{1/3}
\left(\frac{M_\star}{M_\odot}\right)^{-1/3}
\left(\frac{R_\star}{R_\odot}\right)
\nonumber 
%&&
%\sim 94~r_{\rm g},
\label{eq:rc}
\end{eqnarray}
where $e_*$ and $\beta$ are, respectively, the stellar orbital eccentricity and the penetration factor, which is the ratio of the tidal disruption radius to the pericenter radius. We adopt $e_*=1.0$ and $\beta=1.0$ as standard TDE cases. Note that the circularization radius is also expressed as $r_{\rm c}\simeq 94\,r_{\rm g}(M/10^6\,M_\odot)^{-2/3}(M_\star/M_\odot)^{-1/3}(R_\star/R_\odot)$ in units of the gravitational radius (see equation~\ref{eq:rg}). Magnetic stress may reduce the circularization radius \citep{2017MNRAS.464.2816B} during the circularization process, but we do not take it into account because doing so would require detailed knowledge of the magnetic field strength and configuration in the debris. Additionally, we neglect any mass and angular momentum loss due to debris stream-stream collisions for simplicity, assuming that a ring-structured accretion disk is formed with half of the disrupted debris bound to the SMBH, $M_*/2$. The surface density of the disk initially has a Gaussian distribution with a peak at $r_{\rm c}$.
%where $e_*$ and $\beta$ are, respectively, a stellar orbital eccentricity and the penetration factor, which is the ratio of the tidal disruption to pericenter radii, and we adopt $e_*=1.0$ and $\beta=1.0$ as standard TDE cases. Note that the circularization radius is also expressed as {$r_{\rm c}\simeq 94\,r_{\rm g}(M/10^6\,M_\odot)^{-2/3}(M_\star/M_\odot)^{-1/3}(R_\star/R_\odot)$} in units of the gravitational radius (see equation~\ref{eq:rg}). The magnetic stress can reduce the circularization radius \citep{2017MNRAS.464.2816B} during the circularization process, but we take no account of it because the details will require the magnetic field strength and configuration in the debris. Also, we consider no mass and angular momentum loss due to the debris stream-stream collision for simplicity so that a ring-structured accretion disk is formed with half of the disrupted debris bound to the SMBH, $M_*/2$. The disk's surface density initially has a Gaussian distribution with a peak at $r_{\rm c}$:
%
\begin{equation}
\Sigma = \Sigma_0 \exp
\left[
-\frac{(r - r_{\rm c})^2}{\varpi^2}
\right]
\label{sigini}
\end{equation}
\citep{1990ApJ...351...38C}, where $\varpi$ is the width of the Gaussian disk, which is taken to be $3\,r_{\rm g}$\footnote{There is no remarkable difference between $\bar\omega=3r_{\rm g}$ and $\bar\omega=0.1r_{\rm c}\sim9.4r_{\rm g}$ once the disk evolves after $t/\tau_0=0.01$, although their initial distributions are distinguishable}. 
In addition, $\Sigma_0$ shows the surface density of the initial disk at $r = r_{\rm c}$: 
\begin{equation}
\Sigma_0 = 
M_{i}
\biggr/
\left(
2\pi 
\int_{r_{\rm in}}^{r_{\rm out}} r~ \exp
\left[
-\frac{(r - r_{\rm c})^2}{\varpi^2}
\right] 
\, \diff r
\right),
\label{eq:sig0}
\end{equation}
where $M_i$ is the initial disk mass, which is assumed be $0.5M_\odot$ as a fiducial value\footnote{The effect of $M_i$ on the results is discussed in section~\ref{sec:inidiskmass}}, $r_{\rm in}=6\,r_{\rm g}$ is the inner boundary of the calculation domain, corresponding to the inner stable circular orbit (ISCO) radius for a non-spinning black hole, and $r_{\rm out}=10^4\,r_{\rm g}$ is the outer radius of the disk, corresponding to the outer boundary of the calculation domain. As an additional boundary condition, we set the surface density at the inner and outer radii of the disk to zero \citep{1990ApJ...351...38C}. The matter at the inner radius is accreted onto the black hole, while the matter at the outer radius is expelled from the system. Since $r_{\rm out} \gg r_{\rm in}$, the viscous timescale for the matter to reach the outer radius is much longer than the evolution time of our calculations, ensuring that the outer boundary has no influence on the structure and evolution of the system. Furthermore, zero surface density at $r_{\rm in}$ implies vanishing viscous stress at the inner radius. These two conditions ensure that the total angular momentum of the disk is conserved in the absence of disk wind.
%where \red{$M_i$ is an initial disk mass, which is taken to be $0.5M_\odot$ as a fiducial value\footnote{The effect of $M_i$ on the results will be discussed in section~\ref{sec:inidiskmass}}}, $r_{\rm in}=6\,r_{\rm g}$ is the inner boundary of the calculation domain taken to be the inner stable circular orbit (ISCO) radius for a non-spinning black hole, and $r_{\rm out}=10^4\,r_{\rm g}$ is the disk's outer radius to corresponds to the outer boundary of the calculation domain. As an additional boundary condition, we impose the surface density to be zero at the disk's inner and outer radius \citep{1990ApJ...351...38C}. The matter at the inner radius is accreted onto the black hole and the matter at the outer radius is expelled from the system. Because $r_{\rm out} \gg r_{\rm in}$, the viscous timescale for the matter to reach the outer radius is much longer than the evolutionary time of our calculations so that the outer boundary gives no influence of the structure and evolution of the system. Moreover, the zero surface density at $r_{\rm in}$ implies vanishing viscous stress at the inner radius. These two conditions ensure the total angular momentum of the disk is conserved if the disk wind is absent.

\begin{table}
\caption{
Four models are employed to evaluate the evolution of magnetically-driven accretion disk-winds. Model I represents a geometrically thin disk with no disk wind, while Models II-IV depict disks with outflows, each characterized by different parameters.
}
\label{table:models}
\begin{tabular}{c c c c c c}
\hline
&&&&&\\
Model & $\epsilon_{\rm rad}$  & $\bar{\alpha}_{r\phi}$ & $\bar{\alpha}_{z\phi}$ & Wind \\
&& {Turbulent } & Magnetic \\
&& {viscosity} & braking  \\
&&&&&\\
\hline 
&&&&&\\
I & 1.0 & 0.1 & 0 & Off \\
&&&&&\\
II & 0.1 & 0.1 & 0 & On \\
&&&&&\\
III & 0.5 & 0.1 & 0 & On \\
&&&&&\\
IV & 0.5 & 0.1 & 0.001 & On \\
&&&&&\\
\hline
\end{tabular}
\end{table}

\begin{figure}
\centering
\gridline{\subfigure[]{\includegraphics[scale = 0.65]{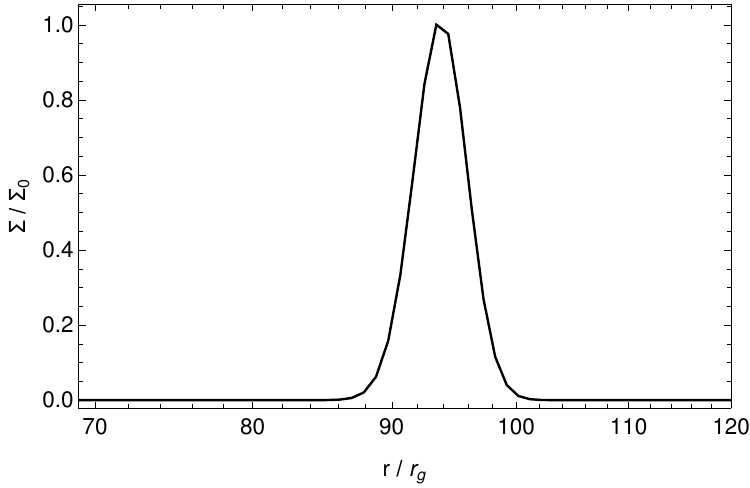}}
\subfigure[]{\includegraphics[scale = 0.65]{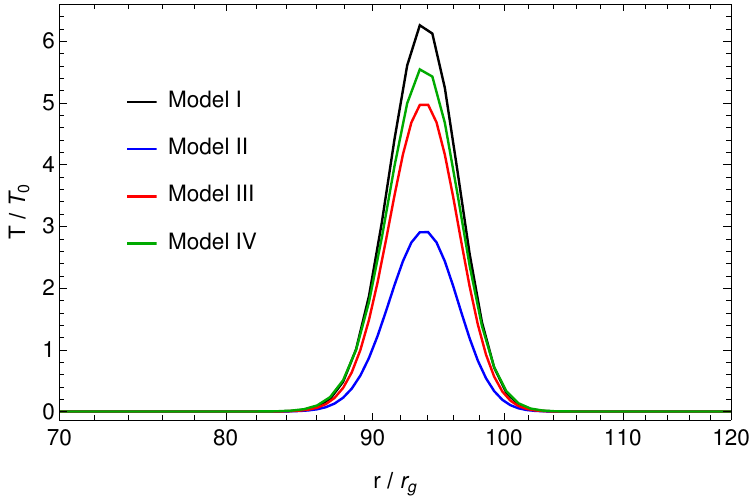}}}
\caption{
Initial radial distribution of the surface density (see equation~\ref{sigini}) and corresponding radial profiles of the disk mid-plane temperature for the four models. These profiles are displayed in panels (a) and (b), respectively. In panel (a), the surface density is normalized to $\Sigma_0 = 1.45 \times 10^{7}{\rm g\,cm^{-2}}$, and in panel (b), the disk temperature is normalized to $T_0 = 1.0 \times 10^6{\rm K}$. Note that the radial surface density profile is identical for all models.
%Initial radial distribution of the surface density (see equation~\ref{sigini}) and the corresponding radial profiles of the disk mid-plane temperature for the four models. These two profiles are shown in (a) and (b). In panel (a), the normalization of the surface density is represented as $\Sigma_0 = 1.45 \times 10^{7}~{\rm g~cm^{-2}}$, while the normalization of the disk temperature is denoted by $T_0=1.0\times10^6\,{\rm K}$ in panel (b). Note that the surface density radial profile is the same for all the models.
}
\label{inidist} 
\end{figure}

Table~\ref{table:models} shows the four models we calculated. Model I represents the standard disk with no wind, while the other models simulate the evolution of magnetically-driven accretion disk winds with different sets of parameters for ($\epsilon_{\rm rad}$, $\bar{\alpha}_{z\phi}$). For all models, we adopt $\bar{\alpha}_{r\phi} = 0.1$ and $C_{\rm w,sim} = 2 \times 10^{-5}$ \citep{2016A&A...596A..74S}. Figure~\ref{inidist} depicts the initial surface density distribution given by equation~(\ref{sigini}) and the corresponding mid-plane temperature for all models. According to equation (\ref{eq:qradf}), a decrease in $\epsilon_{\rm rad}$ reduces the radiative flux, resulting in a lower temperature. In contrast, an increase in $\bar{\alpha}_{z\phi}$ leads to an increase in the radiative flux, thereby raising the disk temperature.

%A decrease in $\epsilon_{\rm rad}$ reduces the radiative flux, leading to a drop in temperature, while an increase in $\bar{\alpha}_{z\phi}$ enhances the viscous heating flux, resulting in a higher disk temperature.

%Table~\ref{table:models} shows the four models that we calculate. Model I shows the standard disk with no wind, whereas the others model the evolution of magnetically-driven accretion disk winds with different sets of parameters of ($\epsilon_{\rm rad}$, $\bar{\alpha}_{z\phi}$). For all the models, we adopt $\bar{\alpha}_{r\phi} = 0.1$ and $C_{\rm w,sim} = 2 \times 10^{-5}$ \citep{2016A&A...596A..74S}. Figure~\ref{inidist} depicts the initial surface density distribution given by equation~(\ref{sigini}) and the corresponding mid-plane temperature for all the models. The decrease in $\epsilon_{\rm rad}$ reduces the radiative flux, causing a drop in the temperature, while $\bar{\alpha}_{z\phi}$ increases the viscous heating flux so that the disk temperature is higher. 

%
%%%%%%%
%  Figure 1 
%%%%%%%
%
%%%%%%%%
%

%The $\alpha$ viscosity prescription $\nu=\alpha{c_{\rm s}}H$ and the thin disk approximation (see equation~\ref{eq:thindisk}) gives the time normalization at $r_{\rm c}$ as 
The $\alpha$ viscosity prescription $\nu=\alpha{c_{\rm s}}H$ and the thin disk approximation (see equation~\ref{eq:thindisk}) provide the time normalization at $r_{\rm c}$ as
\begin{eqnarray}
\tau_0 
&&
\equiv 
\frac{r_{\rm c}^2}{\nu}
\approx
\frac{1}{\bar\alpha_{r\phi}\Omega}
\left(\frac{H}{r_{\rm c}}\right)^{-2}
\nonumber \\
&&
\sim
14.3\,{\rm yr}
\left(
\frac{\bar\alpha_{r\phi}}{0.1}
\right)^{-1}
\left(
\frac{H/r_{\rm c}}{0.01}
\right)^{-2}
\label{eq:t0}
\end{eqnarray}
throughout the paper, where we adopt the disk aspect ratio as a constant, i.e., $H/r_{\rm c}=0.01$, for the purpose of comparing the results among the different models.

%
%
%%%%%%%%%%%
\section{Results}
\label{sec:results}
%%%%%%%%%%%
%
In this section, we show the solutions for {equations~(\ref{eq:sigma})-(\ref{eq:cs})} with the initial and boundary conditions described in Section~\ref{sec:inicon}.
%
%%%%%%%%%%%%%%%%%%%%%%%%%%%%%%%%%
\subsection{Evolution of radial profiles of disk and wind quantities}
\label{eq:radprofiles}
%%%%%%%%%%%%%%%%%%%%%%%%%%%%%%%%%
%

\begin{figure}
\centering
\gridline{\subfigure[]{\includegraphics[scale = 0.67]{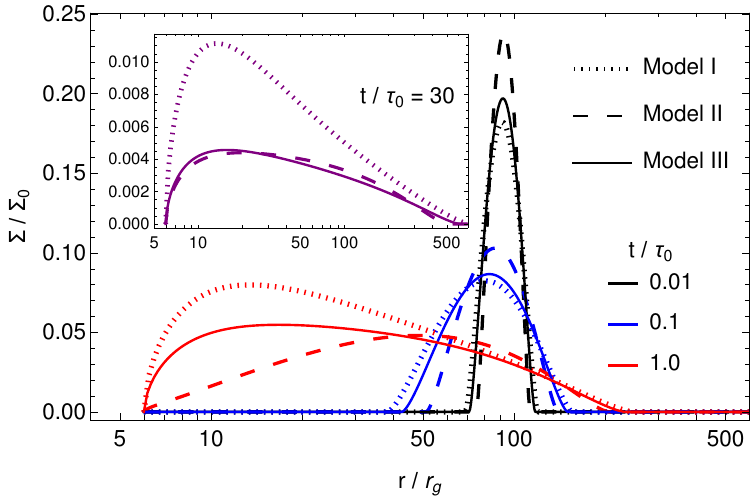}}
\subfigure[]{\includegraphics[scale = 0.65]{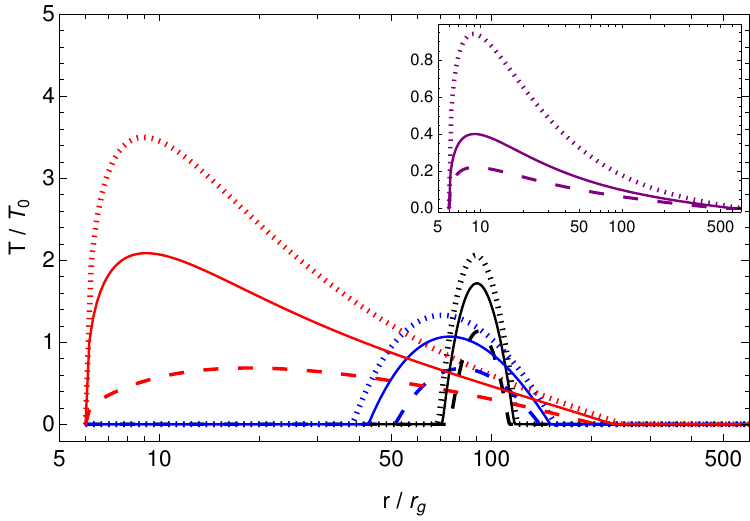}}}
\gridline{\subfigure[]{\includegraphics[scale = 0.67]{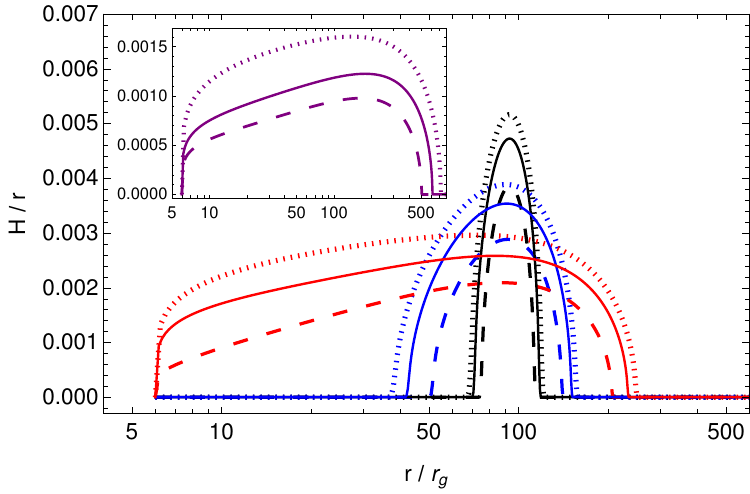}}
\subfigure[]{\includegraphics[scale = 0.65]{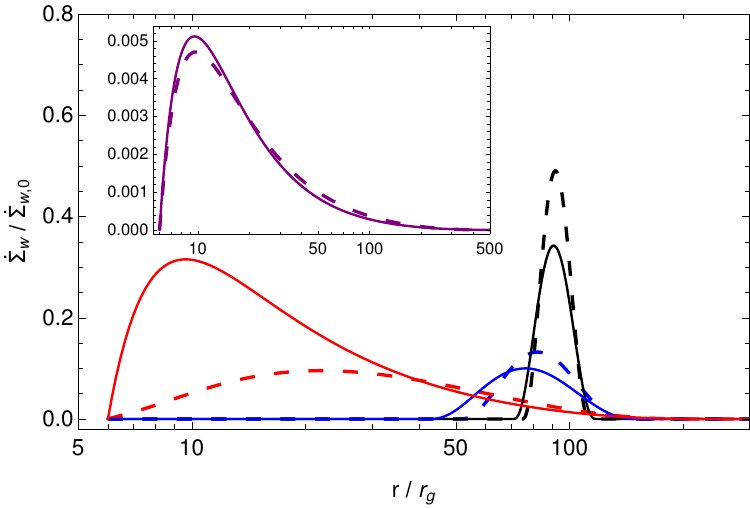}}}
\caption{
Radial profiles of the surface density, disk mid-plane temperature, disk aspect ratio, and vertical mass flux for Models I through III. For all three models, $\bar\alpha_{z\phi} = 0$, with different values of $\epsilon_{\rm rad}$: 1.0 (Model I), 0.1 (Model II), and 0.5 (Model III). Panel (a) shows the profiles with $\tau_0 = 14.3~{\rm yr}$ and $\Sigma_0 = 1.45 \times 10^{7}{\rm gcm^{-2}}$, panel (b) shows the profiles with $T_0 = 1.0 \times 10^6{\rm K}$, panel (c) shows the disk aspect ratio, and panel (d) shows the profiles with $\dot\Sigma_{\rm w,0} = 6.4 \times 10^{-3}{\rm gs^{-1}~cm^{-2}}$. The inset in each panel displays the corresponding physical quantity at a very late time, $t/\tau_0 = 30$.
}
\label{diskprop_erad}
\end{figure}

Figure~\ref{diskprop_erad} shows the radial profiles of the surface density, disk mid-plane temperature, disk aspect ratio, and wind mass flux for Models I through III at three early time epochs. Each panel includes an inset displaying the radial profile of each quantity at a later time epoch, $t/\tau_0=30$. These models differ in the values of $\epsilon_{\rm rad}$, while $\bar{\alpha}_{z\phi}$ is kept constant at zero. The initial Gaussian disk spreads with time both radially inward and outward due to disk viscosity. Mass accretion increases the surface density and temperature near the inner radius at later times. The disk aspect ratio remains well below unity, indicating that the disk is geometrically thin at all time epochs.

%Figure~\ref{diskprop_erad} presents the radial profiles of the surface density, disk mid-plane temperature, scale-height-to-radius ratio, and the wind mass flux for Models I through III at the three early time epochs with the inset per panel displaying the radial profile of each quantity at a very late time epoch, $t/\tau_0=30$. These models are distinguished by varying values of $\epsilon_{\rm rad}$ while holding $\bar{\alpha}_{z\phi}=0$. The initial Gaussian disk spreads with time radially inward and outward due to the disk viscosity. The mass accretion increases the surface density and temperature near the inner radius at late times. The scale-height-to-radius ratio is much smaller than unity and implies that the disk is geometrically thin at all time epochs.

Panels (a) to (c) show that disk temperature and $H/r$ increase in the order of Models I, III, and II, indicating that these quantities are higher for larger values of $\epsilon_{\rm rad}$. Since the viscous timescale is estimated as $r^2/\nu \propto \bar\alpha_{r\phi}^{-1} T^{-1}$, it correspondingly decreases in the same order. This suggests that the disk evolves more rapidly as $\epsilon_{\rm rad}$ increases, because a larger fraction of the liberated energy is transferred to the thermal component, resulting in a higher temperature.

%It is noted from panels (a) to (c) that the disk temperature and $H/r$ are higher in the order of Models 1, 3, and 2, indicating that these quantities are higher as $\epsilon_{\rm rad}$ is larger. Since the viscous timescale is estimated to be $r^2/\nu\propto\bar\alpha_{r\phi}^{-1}T^{-1}$, it is shorter in the same order as above. This suggests that the disk evolves more rapidly as $\epsilon_{\rm rad}$ is larger, because a larger fraction of the liberated energy is transferred to the thermal component for larger $\epsilon_{\rm rad}$ so that the temperature is higher. 

Now, focusing on the radial profile of the surface density for each model, as shown in panel (a), equation~(\ref{eq:cw}) indicates that the energy available for mass loss increases as $\epsilon_{\rm rad}$ decreases, leading to a higher vertical mass flux. This is consistent with the result that the vertical mass flux of Model II is larger than that of Model III at $t/\tau_0=0.01$, as shown in panel (d). If mass loss proceeds efficiently, the surface density should decrease more rapidly. However, the surface density of Model II is higher than that of Model III, which appears to contradict this expectation

%Now, we focus on the radial profile of the surface density of each model, as seen in panel (a). From equation~(\ref{eq:cw}), the energy available for mass loss is larger as $\epsilon_{\rm rad}$ is smaller, indicating that the vertical mass flux is larger. This is consistent with the fact that the vertical mass flux of Model II is larger than that of Model III at $t/\tau_0=0.01$, as seen in panel (d). If mass loss proceeds efficiently, the surface density should decrease more rapidly. However, the surface density of Model II is higher than that of Model III, which is opposite to this consideration.

How can we interpret the seemingly contradictory results at $t/\tau_0=0.01$? These results can be understood by considering two factors. First, although $(1-\epsilon_{\rm rad})$ determines the efficiency of the mass-loss rate, the actual mass-loss rate also depends on the surface density. This implies that a higher efficiency does not necessarily lead to a higher mass-loss rate. Second, there is a competition between viscous accretion and wind mass loss in the evolution of the disk's surface density. The viscous effect dominates the early evolution of Models II and III. Specifically, since the viscous timescale is shorter in Model III than in Model II, the surface density of Model III evolves more rapidly, resulting in the opposite trend at early times compared to later times.

\begin{figure*}
\centering
\gridline{\subfigure[]{\includegraphics[scale = 0.71]{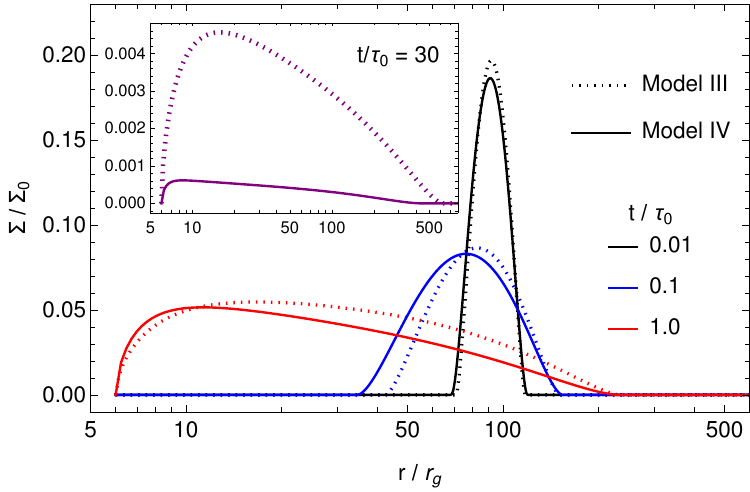}}
\subfigure[]{\includegraphics[scale = 0.69]{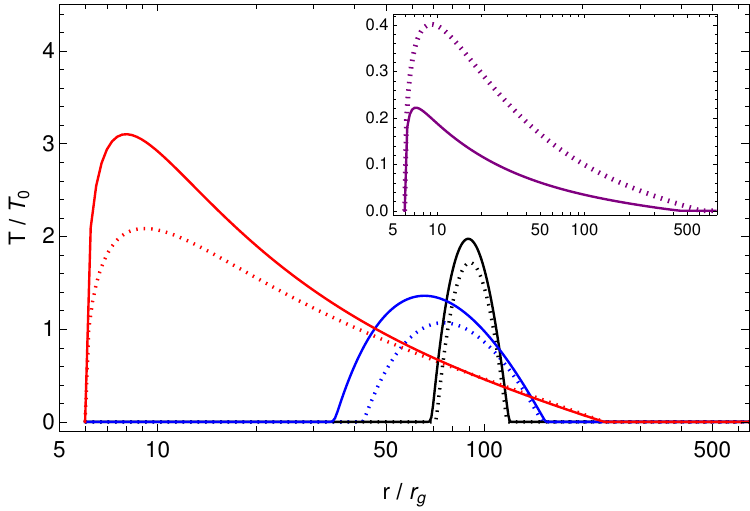}}}
\gridline{\subfigure[]{\includegraphics[scale = 0.71]{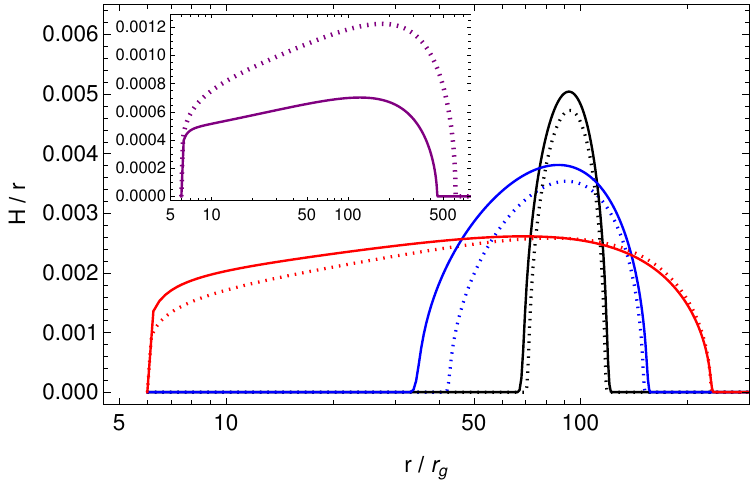}}
\subfigure[]{\includegraphics[scale = 0.69]{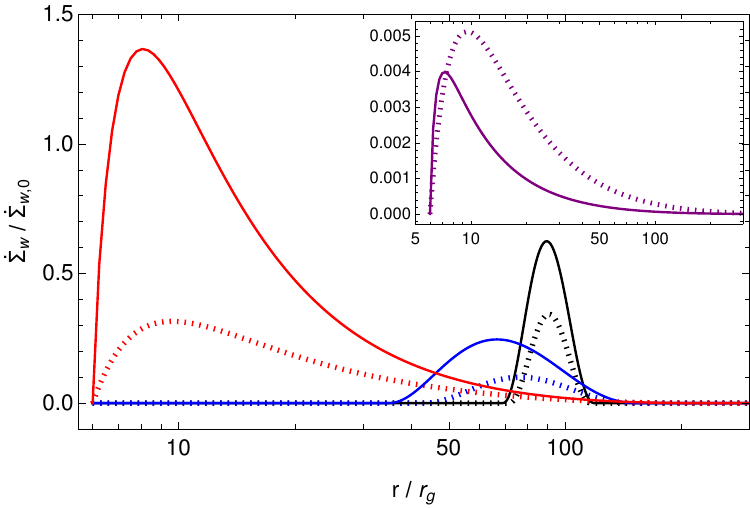}}}
\caption{\label{diskprop_azphi} 
The same format as Figure~\ref{diskprop_erad} but for Models III and IV.
}
\end{figure*}

Figure~\ref{diskprop_azphi} presents the same format as Figure~\ref{diskprop_erad}, but for the scenario with a non-zero value of $\bar{\alpha}_{z\phi}$, referred to as Model IV. For comparative purposes, the characteristics of Model III are also shown in the same figure. According to panels (a) and (d) of the figure at early times, there is little difference between Models III and IV in terms of surface density, while the vertical mass flux shows a significant difference. This indicates that $\bar{\alpha}_{z\phi}$ has minimal impact on surface density evolution but significantly affects the vertical mass flux, which will be discussed in more detail later in this subsection. From panels (b) and (c), we observe that $\bar{\alpha}_{z\phi}$ has a mild influence on the disk mid-plane temperature and the disk aspect ratio evolution. In contrast to the behavior at $t/\tau_0=1.0$, the surface density, disk mid-plane temperature, $H/r$, and $\dot\Sigma{\rm w}$ of Model IV are overall lower than those of Model III at $t/\tau_0=30$. This is due to the significantly rapid decay of the mass accretion and loss rates caused by magnetic braking from early to late times (see also Figures~\ref{Mdots} and \ref{Mwinds}).

%Figure~\ref{diskprop_azphi} delineates the same format as Figure~\ref{diskprop_erad} but for the scenario with a non-zero value of $\bar{\alpha}_{z\phi}$, denoted as Model IV. For comparative elucidation, the characteristics of Model III are concurrently exhibited within the same figure. According to panels (a) and (d) of the figure at early times, there is little difference between Models III and IV in the surface density, while the vertical mass flux demonstrates a remarkable difference. This means $\bar{\alpha}_{z\phi}$ affects little the surface density evolution but impacts the vertical mass flux significantly, which will be explained in more detail later in this subsection. From panels (b) and (c), we note that $\bar{\alpha}_{z\phi}$ has a mild influence on the disk mid-plane temperature and scale-height-to-radius ratio evolution. Contrary to the behavior at $t/\tau_0=1.0$, the surface density, disk mid-plane temperature, $H/r$, and $\dot\Sigma_{\rm w}$ of Model IV are overall lower than those of Model III at $t/\tau_0=30$. This is because of the remarkably rapid decay of the mass accretion and loss rate due to magnetic braking from early to late times (see also Figures~\ref{Mdots} and \ref{Mwinds}).

The mass accretion rate is obtained using equation (\ref{rsvreqn}) as 
\begin{equation}
\dot{M} = -2 \pi r \Sigma v_r = \frac{4 \pi}{r \Omega} \left[\frac{\partial}{\partial r}
\left(
\bar{\alpha}_{r\phi} r^2\Sigma c_s^2
\right) 
+ 
\bar{\alpha}_{z\phi} r^2 \rho c_s^2
\right].
\label{mdotrate}
\end{equation}
The mass accretion rate is positive for $v_r < 0$, indicating a radial inflow of the gas to the black hole, whereas the mass accretion rate is negative for $v_r > 0$, indicating a radial outflow.

As a result of wind blowing vertically, mass is lost at a rate of
\begin{equation}
\dot{M}_{\rm w} = \int \dot{\Sigma}_{\rm w} 2 \pi r \, \diff r = \pi \sqrt{G M} \int_{r}^{r_{\rm out}} \frac{C_{\rm w} \Sigma}{\sqrt{r}} \, \diff r,
\label{mwindrate}
\end{equation}
where we used equation (\ref{vzeqn}) with equation~(\ref{eq:thindisk}) for the derivation.

%
%%%%%%%
%%%%%%
%

%
%%%%%%%
%%%%%%
%

\begin{figure*}
\centering
\gridline{\subfigure[]{\includegraphics[scale = 0.6]{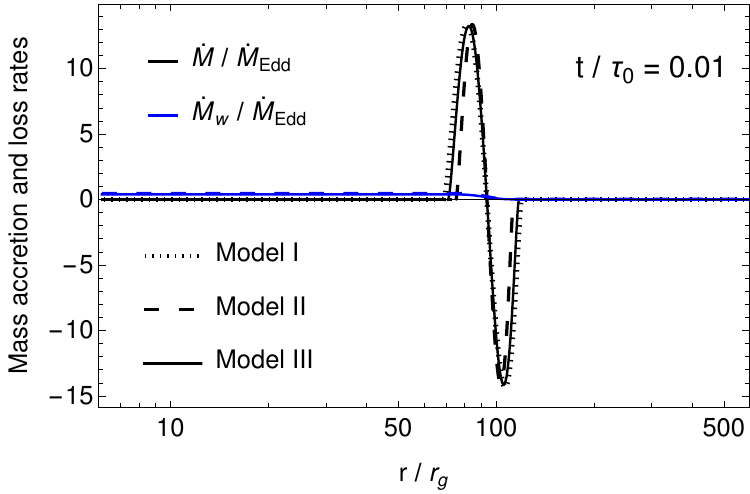}}
\subfigure[]{\includegraphics[scale = 0.6]{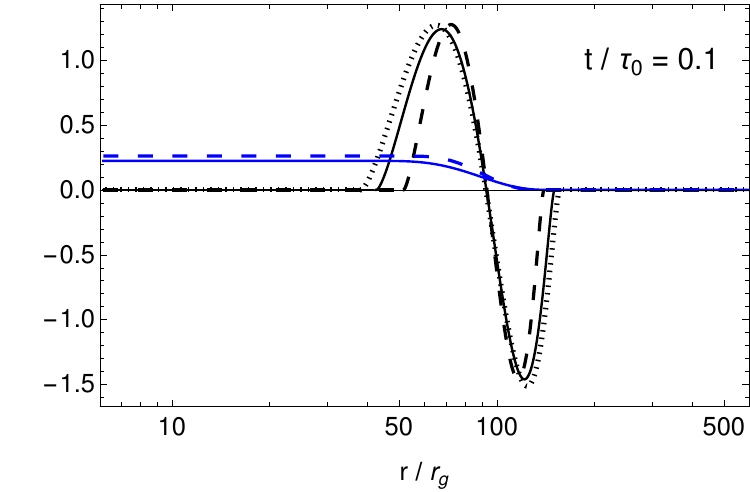}}}
\gridline{\subfigure[]{\includegraphics[scale = 0.6]{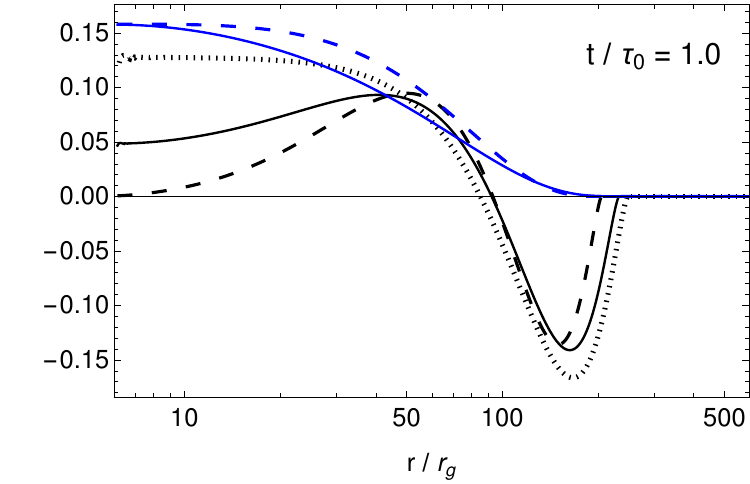}}
\subfigure[]{\includegraphics[scale = 0.6]{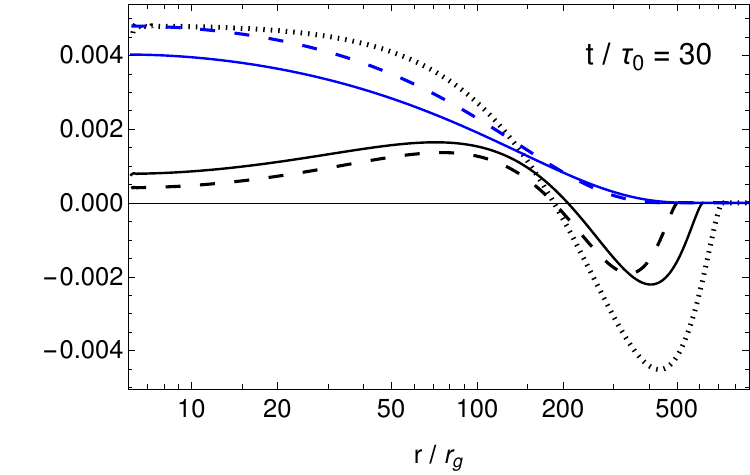}}}
\caption{\label{Mdot_rad_erad}
Radial dependence of the mass accretion rate (solid lines) and the mass wind rate (dashed lines) for Models I through III, as obtained using equations (\ref{mdotrate}) and (\ref{mwindrate}), respectively. Panels (a) to (d) show the results at $t/\tau_0 = 0.01$, $t/\tau_0 = 0.1$, $t/\tau_0 = 1.0$, and $t/\tau_0 = 30$, respectively. The mass accretion rate is represented by black solid lines and the mass wind rate by blue dashed lines, with both rates normalized by the Eddington accretion rate. Different line styles indicate different models. Note that there is no blue line for Model I, as there is no wind present for $\epsilon_{\rm rad} = 1.0$.
}
\end{figure*}

\begin{figure*}
\centering
\gridline{\subfigure[]{\includegraphics[scale = 0.6]{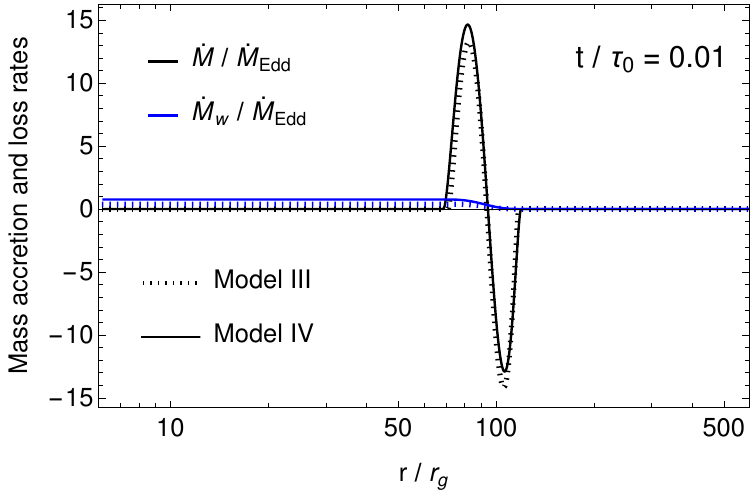}}
\subfigure[]{\includegraphics[scale = 0.6]{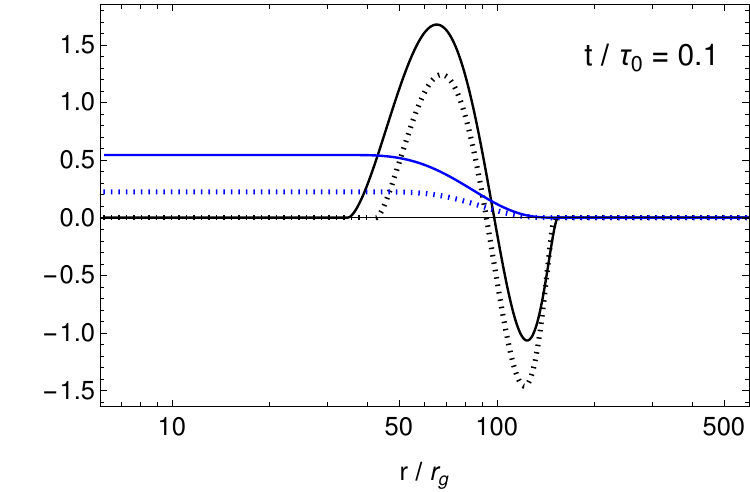}}}
\gridline{\subfigure[]{\includegraphics[scale = 0.6]{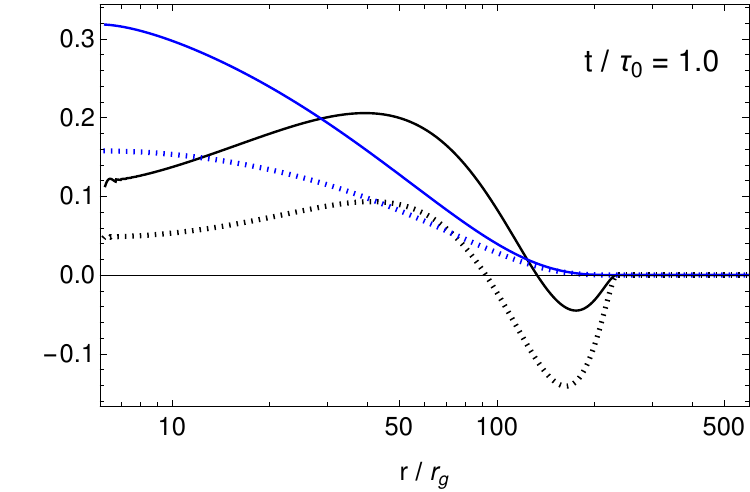}}
\subfigure[]{\includegraphics[scale = 0.62]{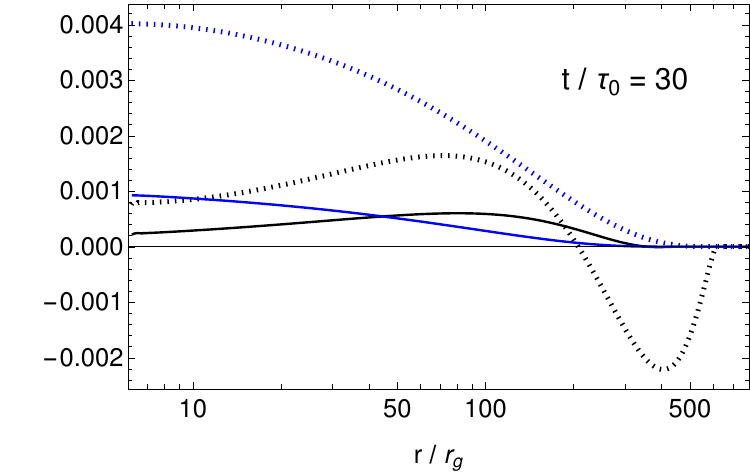}}}
\caption{
The same format as Figure~\ref{Mdot_rad_erad} but for Models III and IV.
\label{Mdot_rad_az} 
}
\end{figure*}

Figure~\ref{Mdot_rad_erad} shows the radial profiles of the mass accretion rate ($\dot{M}$) and the mass loss rate ($\dot{M}_{\rm w}$) at different time epochs for Models I through III. Both rates are normalized by the Eddington accretion rate:
\begin{eqnarray}
\dot{M}_{\rm Edd}
&&
= \frac{L_{\rm Edd}}{\eta c^2} \sim 
2.6 \times 10^{-2}
\,
M_\odot/{\rm yr}
\,
\left(\frac{\eta}{0.1}\right)^{-1} 
\left(
\frac{M}{10^6\,M_\odot}
\right),
\label{eq:edd}
\end{eqnarray}
where
\begin{eqnarray}
L_{\rm Edd}=\frac{4\pi GM c }{\kappa_{\rm es}}
\approx1.5\times10^{44} \,{\rm erg~s^{-1}}
\left(\frac{M}{10^6M_{\odot}}\right)
\label{eq:ledd}
\end{eqnarray}
is the Eddington luminosity, and $\eta$ is the radiative efficiency, with a fiducial value of $0.1$ for the Eddington accretion rate. A positive value of $\dot{M}$ indicates an inward accretion flow, while a negative value of $\dot{M}$ denotes an outward decretion flow. The mass loss rate represents the outflowing wind from the disk's upper and lower surfaces and is always positive.
%is the Eddington luminosity and $\eta$ is the radiative efficiency with the fiducial value $0.1$ for the Eddington accretion rate. The positive value of $\dot{M}$ indicates an accretion flow inwardly, whereas the negative value of $\dot{M}$ designates a decreation flow outwardly. The mass loss rate represents the outflowing wind from the disk's upper and lower surfaces, which is always positive. 

%%the referee's suggestion%%It is noted from the figure that the mass accretion rate of Model II is lower than that of Model III at $t/\tau_0=1.0$, while the mass loss rates of both Models II and III are nearly equal at the inner radii. However, the mass loss rate of Model III decays more rapidly than that of Model II as the disk radius gets larger up to $100\,r_{\rm g}$. At a very late time ($t/\tau_0=30$), the mass accretion rate of Model II is lower than that of Model III over the whole disk. In contrast, the mass loss rate of Model II is higher than that of Model III at $r/r_{\rm g}\lesssim200$.

Figure~\ref{Mdot_rad_az} depicts the radial profiles of the mass accretion and loss rates for Models III and IV at $t/\tau_0=0.01$, $t/\tau_0=0.1$, $t/\tau_0=1.0$, and $t/\tau_0=30$, respectively. The figure shows that, at early times, both the mass accretion and loss rates of Model IV are higher across the disk's entire region up to $100\,r_{\rm g}$ compared to those of Model III. This is due to the magnetocentrifugal effect resulting from the non-zero value of $\bar{\alpha}_{z\phi}$ \citep{1982MNRAS.199..883B}, which directly removes angular momentum from the disk and facilitates mass accretion onto the central object. This wind-driven accretion leads to additional liberation of gravitational energy, increasing the kinetic energy of the disk winds. Consequently, the mass loss rate also increases due to the presence of $\bar{\alpha}_{z\phi}$. The enhanced mass accretion and loss rates of Model IV compared to Model III at early times rapidly deplete the disk's mass in Model IV, leading to a lower surface density at late times (see the inset of Figure~\ref{diskprop_azphi}a for $t/\tau_0 = 30$). Consequently, this depletion causes the mass accretion and loss rates of Model IV to become lower than those of Model III at $t/\tau_0 = 30$.

%
%%%%%%%%%%%
%%      Figure 6
%%%%%%%%%%%
%
\begin{figure*}
\centering
\gridline{\subfigure[]{\includegraphics[scale = 0.75]{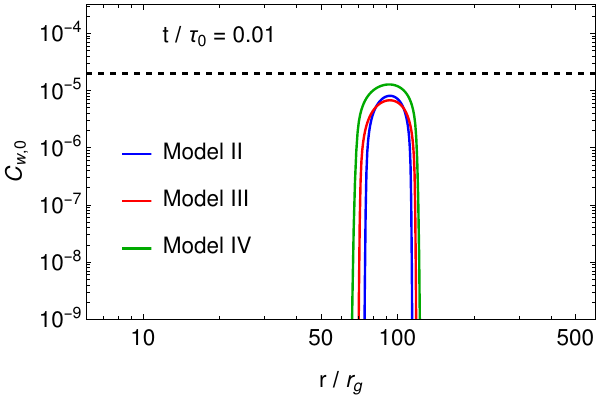}}
\subfigure[]{\includegraphics[scale = 0.75]{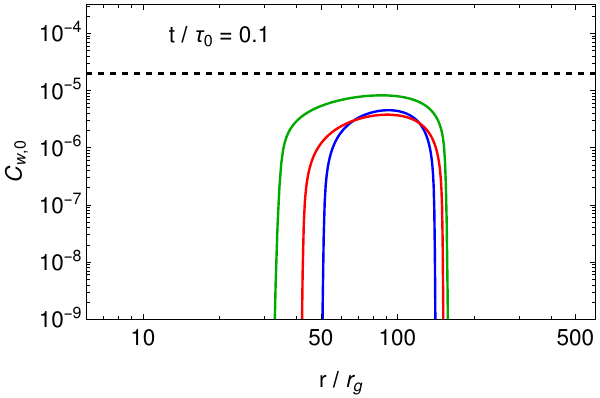}}}
\gridline{\subfigure[]{\includegraphics[scale = 0.75]{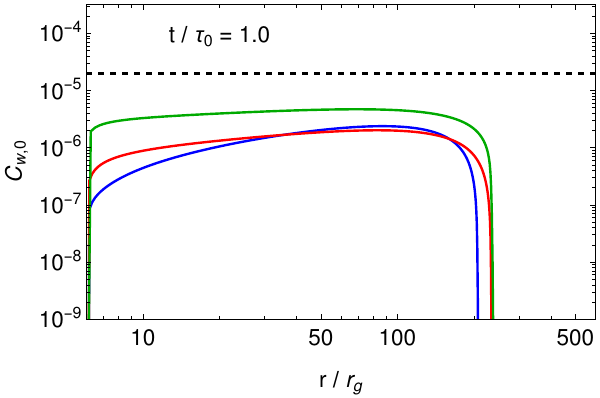}}
\subfigure[]{\includegraphics[scale = 0.75]{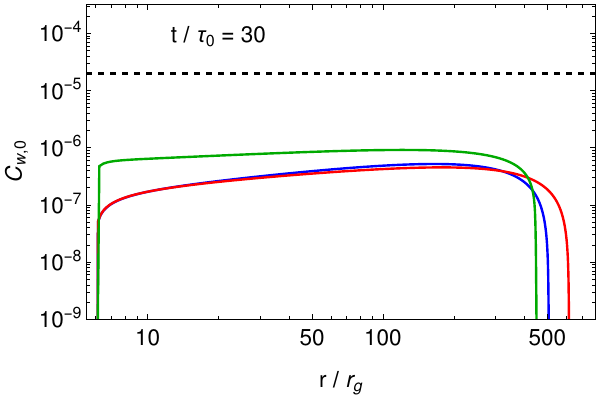}}}
\caption{
Radial profiles of $C_{\rm w,0}$, obtained using equation (\ref{eq:cw}), for Models II through IV at different time epochs. Panels (a) to (d) show profiles at $t/\tau_0 = 0.01$, $t/\tau_0=0.1$, $t/\tau_0=1.0$, and $t/\tau_0=30$, respectively, displayed chronologically. In each panel, different colors represent different models. The black dashed line indicates the $C_{\rm w,sim} = 2\times 10^{-5}$ line, obtained from simulations \citep{2016A&A...596A..74S}, which serves as the fiducial value.
%Radial profiles of $C_{\rm w,0}$, which are obtained by equation (\ref{eq:cw}), for Model II through IV at different time epochs. Panels (a) to (d) are displayed chronologically with $t/\tau_0 = 0.01$, $t/\tau_0=0.1$, $t/\tau_0=1.0$, and $t/\tau_0=30$, respectively. For each panel, the different colors are the different models. The black dashed line represents the $C_{\rm w,sim} = 2\times 10^{-5}$ line obtained from the simulations \citep{2016A&A...596A..74S} as the fiducial value.
}
\label{cwe_erad}
\end{figure*}

%Let us consider the radial distribution of $C_{\rm w,0}$, which is approximated by using equation~(\ref{eq:cw}) with equation (\ref{eq:thindisk}) as
Let us consider the radial distribution of $C_{\rm w,0}$, which is approximated by using equation~(\ref{eq:cw}) along with equation (\ref{eq:thindisk}) as follows:
\begin{eqnarray}
C_{\rm w,0}
&&
\approx
2(1-\epsilon_{\rm rad})
\biggr[
3
\bar\alpha_{r\phi}
\left(
\frac{H}{r}
\right)^2
+
\bar\alpha_{z\phi}
\left(
\frac{H}{r}
\right)
\biggr]
\label{eq:cw02}
\end{eqnarray}
This expression allows us to analytically obtain $C_{\rm w,0}<C_{\rm w,sim}$ when $\bar\alpha_{r\phi}=0.1$, $\bar\alpha_{z\phi}=0$ or $\bar\alpha_{z\phi}=1.0\times10^{-3}$, and ${H/r}\lesssim5\times10^{-3}$. Figure~\ref{cwe_erad} displays the radial dependence of $C_{\rm w, 0}$. As shown in the figure, for all models and time epochs, $C_{\rm w, 0}$ is slightly smaller than the fiducial value, which is consistent with the analytical estimation. Additionally, $C_{\rm w,0}$ decreases with time over the entire disk. Assuming $\bar\alpha_{r\phi}=0$, from equations~\ref{eq:cwtot} and \ref{eq:cw02}, we find that the vertical mass flux is simply proportional to $\bar\alpha_{z\phi}(H/r)$ through $C_{\rm w,0}$. This is the primary reason why the vertical mass flux is higher at early times than in the case of $\bar\alpha_{z\phi}=0$, as shown in Figure~\ref{diskprop_azphi}.

%This allows us to evaluate $C_{\rm w,0}<C_{\rm w,sim}$ for $\bar\alpha_{r\phi}=0.1$, $\bar\alpha_{z\phi}=0$ or $\bar\alpha_{z\phi}=1.0\times10^{-3}$, and ${H/r}\lesssim5\times10^{-3}$. Figure~\ref{cwe_erad} displays the radial dependence of $C_{\rm w, 0}$. From the figure, we note for all the models and time epochs, $C_{\rm w, 0}$ is slightly smaller than the fiducial value. This is consistent with the analytical estimation. Also, we find that $C_{\rm w,0}$ decreases with time over the whole of the disk. Assuming that $\bar\alpha_{r\phi}=0$, from equations~\ref{eq:cwtot} and \ref{eq:cw02}, we see that the vertical mass flux is simply proportional to $\bar\alpha_{z\phi}(H/r)$ through $C_{\rm w,0}$. This is the main reason why the vertical mass flux is higher than the $\bar\alpha_{z\phi}=0$ case at early times as shown in Figure~\ref{diskprop_azphi}.

%
%%%%%%%%%%%%%%%%%%%%%%%%%%%
\subsection{Evolution of mass accretion and mass-loss rates}
\label{sec:timeevo}
%%%%%%%%%%%%%%%%%%%%%%%%%%%
%

\begin{figure}
\centering
\gridline{\subfigure[]{\includegraphics[scale = 0.64]{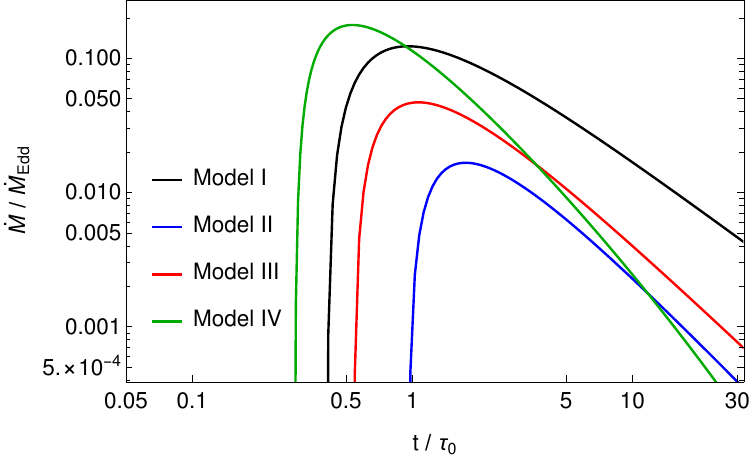}}
\subfigure[]{\includegraphics[scale = 0.61]{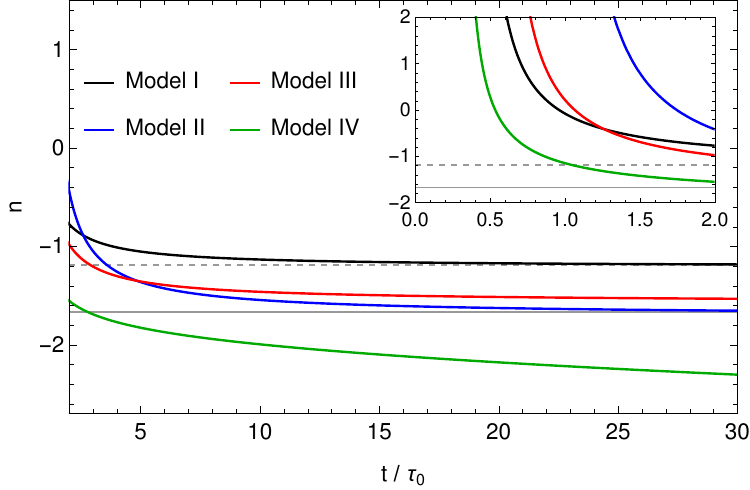}}}
\caption{\label{Mdots} 
Time evolution of the mass accretion rates estimated at the disk's inner radius for all models, along with the corresponding slopes. In panel (a), the mass accretion rate is normalized by the Eddington accretion rate. Different line colors represent different models. In panel (b), the slope is represented by the power-law index of time. The gray solid and dashed lines correspond to $-5/3$ (mass fallback rate case; \citealp{2009MNRAS.392..332L}) and $-19/16$ (classical solution case; \citealp{1990ApJ...351...38C}), respectively. The inset plot displays the slope at early times.
}
\end{figure}

Panel (a) of Figure~\ref{Mdots} shows the time evolution of the mass accretion rate estimated at $r_{\rm in}$ for Models I through IV. The peak mass accretion rate increases with $\epsilon_{\rm rad}$ or $\bar{\alpha}_{z\phi}$. It is important to note that $\bar{\alpha}_{z\phi}$ governs magnetic braking in the disk, while $\epsilon_{\rm rad}$ controls the fraction of the total heating flux in the disk that is converted into radiative cooling flux.
Higher values of these two parameters result in more efficient angular momentum loss, thereby increasing the mass accretion rate. When $\bar{\alpha}_{z\phi} = 0$, the mass accretion rate overall increases as $\epsilon_{\rm rad}$ approaches 1.

%Panel (a) of Figure~\ref{Mdots} represents the time evolution of the mass accretion rate estimated at $r_{\rm in}$ for Models I through IV. The peak mass accretion rate increases with $\epsilon_{\rm rad}$ or $\bar{\alpha}_{z\phi}$. Recalling $\bar{\alpha}_{z\phi}$ governs magnetic braking in the disk, while $\epsilon_{\rm rad}$ controls how much fraction of the viscous heating flux is converted to radiative cooling flux. The higher values of these two parameters make the loss of angular momentum more efficient, increasing the mass accretion rate as a result. In case of $\bar{\alpha}_{z\phi} = 0$, the mass accretion rate overall increases as $\epsilon_{\rm rad}$ is close to 1. 

%The presence of wind ($0<\epsilon_{\rm rad}<1$) causes the slopes of the $\dot{M}$ decline curves to steepen with time, as evident from the panel. Assuming $\dot{M}\propto{t^n}$, the slope is characterized by the power-law index of time in the mass accretion curve, which is given by

The presence of a wind ($0<\epsilon_{\rm rad}<1$) causes the slopes of the $\dot{M}$ decay curves to steepen with time, as evident from the panel. Assuming $\dot{M}\propto{t^n}$, the slope is expressed by the power-law index $n$ of the mass accretion rate curve: 
\begin{eqnarray}
n=\frac{d\ln(\dot{M})}{d\ln{t}}
\label{eq:pindex}
\end{eqnarray}
Panel (b) of Figure~\ref{Mdots} shows the time dependence of $n$ for $0\le{t/\tau_0}\le30$. The small inset illustrates the variation of $n$ at very early times. In all models, $n$ initially rises sharply and then rapidly decreases, demonstrating the quick transition from an increasing to a decreasing $\dot{M}$, as shown in panel (a). In Models I to III, $n$ asymptotes to specific values at late times. In the absence of wind ($\epsilon_{\rm rad}=1$), $n$ asymptotes to $-19/16$ at late times, which corresponds to the solution by \citet{1990ApJ...351...38C}. When wind is present, the power-law index asymptotes to $-1.62$ in Model II, while it asymptotes to $-1.52$ in Model III. This suggests that the slope becomes slightly steeper as $\epsilon_{\rm rad}$ decreases. Comparing Model III and IV, the slope is significantly steeper without saturation at the non-zero value of $\bar{\alpha}_{z\phi}$.
%Panel (b) of Figure~\ref{Mdots} shows the time dependence of $n$ for $0\le{t/\tau_0}\le30$. The small inset illustrates the time variation of $n$ at very early times. For all models, $n$ initially rises sharply and then rapidly decreases, demonstrating the rapid transition from an increase to a decrease in $\dot{M}$, as seen in panel (a). In Models I to III, $n$ asymptotes to specific values at late times.
%In the absence of the wind ($\epsilon_{\rm rad}=1$), $n$ asymptotes to $-19/16$ at late times, which corresponds to a solution of \citet{1990ApJ...351...38C}. In the presence of the wind, the power-law index asymptotes to $-1.62$ in Model II, while it asymptotes to $-1.52$ in Model III. This suggests that the slope is slightly steeper as the $\epsilon_{\rm rad}$ decreases. According to the comparison between Model III and IV,  the slope is significantly steeper without saturation at the non-zero value of $\bar{\alpha}_{z\phi}$. 

\begin{figure}
\centering
\gridline{\subfigure[]{\includegraphics[scale = 0.64]{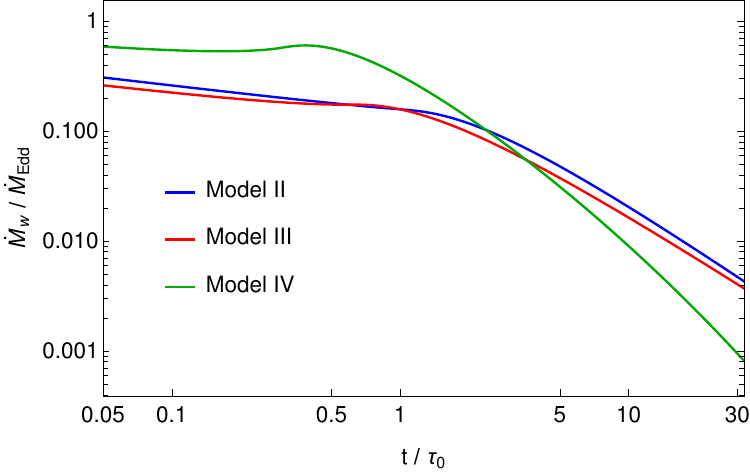}}
\subfigure[]{\includegraphics[scale = 0.61]{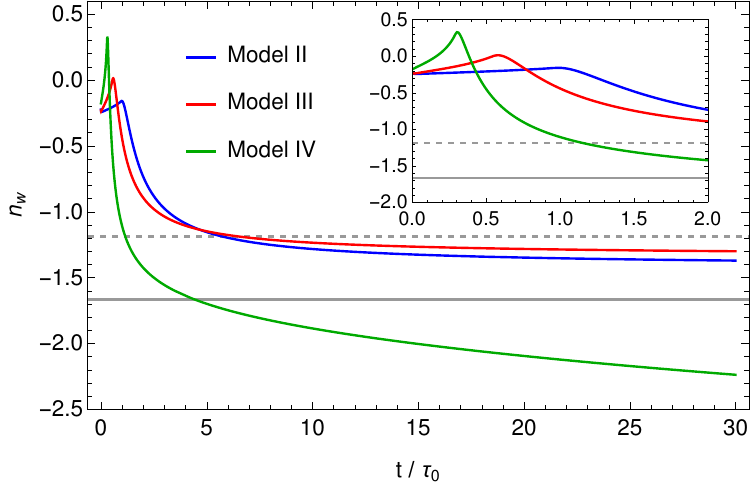}}}
\caption{
Same format as Figure~\ref{Mdots}, but for the mass loss rates for Models II through IV. Note that there is no mass loss in Model I.
}
\label{Mwinds} 
\end{figure}

Panel (a) of Figure~\ref{Mwinds} shows the mass loss rate for Models II through IV at different time epochs. From the figure, it can be observed that the mass loss rates of Models II and III slowly decay at early times, followed by a rapid decrease. In contrast, the mass loss rate of Model IV exhibits a small peak around $t/\tau_0 \approx 0.4 $, followed by a rapid decay after the peak. Magnetic braking, characterized by $\bar{\alpha}_{z\phi}$, significantly affects the magnitude and slope of the mass loss rate, while the effect of $\epsilon_{\rm rad}$ is less pronounced.

%Panel (a) of Figure~\ref{Mwinds} represents the mass loss rate for Models II through IV at the different time epochs. It is noted from the figure that the mass loss rates of Model II and III slowly decay at early times and subsequently rapidly decrease, while the mass loss rate of Model IV has a small peak around $t/\tau_0 \approx 0.4 $ and rapidly decays after the peak. Magnetic braking, which is characterized by $\bar{\alpha}_{z\phi}$, has a significant impact on the magnitude and slope of the mass loss rate, whereas the effect of $\epsilon_{\rm rad}$ is not so remarkable. 

The mass loss rates of Models II and III follow a power-law evolution at late times. The power law index of time of the mass loss rate, similar to equation~(\ref{eq:pindex}), is given by
\begin{eqnarray}
n_{\rm w}=\frac{d\ln(\dot{M}_{\rm w})}{d\ln{t}}
\label{eq:pindex-w}
\end{eqnarray}
Panel (b) of Figure~\ref{Mwinds} shows the time dependence of $n_{\rm w}$ for $0 \leq t/\tau_0 \leq 30$. The small inset shows the time variation of $n_{\rm w}$ at very early times. In Models II and III, $n_{\rm w}$ asymptotes to lower values than $n$, while $n_{\rm w}$ in Model IV shows no saturation within our calculation time.

%Panel (b) of Figure~\ref{Mwinds} shows the time dependence of $n_{\rm w}$ for $0\le{t/\tau_0}\le30$. The small inset shows the time variation of $n_{\rm w}$ very early times. In Models II and III, $n_{\rm w}$ asymptotes the lower values than that of $n$, while $n_{\rm w}$ of Model IV exhibits no saturation within our calculation time.

%%referee's suggestion%%In comparison with Figure~\ref{Mdots}, the magnitudes of mass loss rates for Models II to IV at early times are higher than those of the peak mass accretion rates for corresponding models. Considering the differences in the magnitude and power law index of time between the mass accretion and loss rates, we find that {the mass loss rate varies with time differently from the mass accretion rate.}

%
%%%%%%%%
%% Figures
%%%%%%%%
%
\begin{figure}
\centering
\gridline{\subfigure[]{\includegraphics[scale = 0.62]{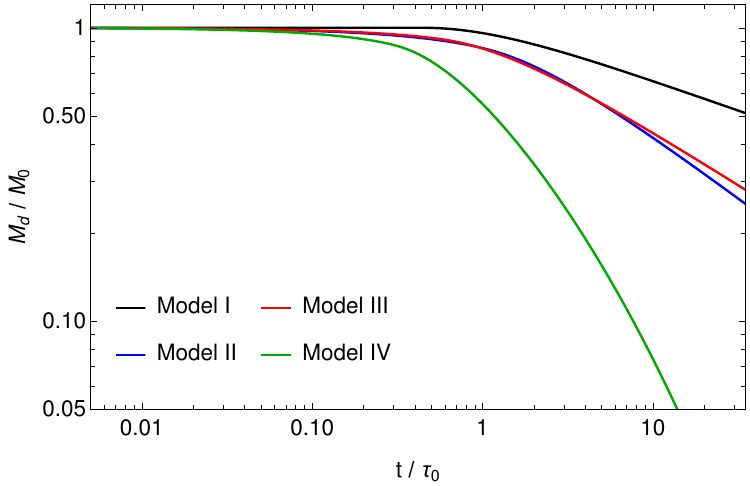}}
\subfigure[]{\includegraphics[scale = 0.62]{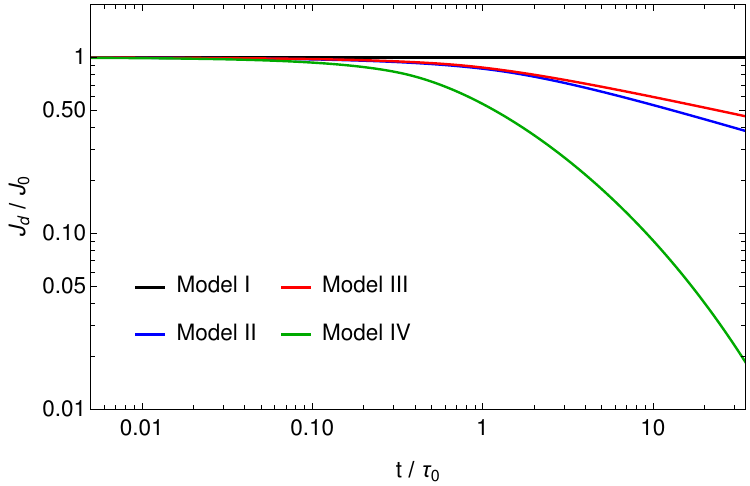}}}
\caption{
Time evolution of the disk mass and angular momentum for Models I through IV. The different colors represent the different models. The disk mass is normalized by the mass of the bound debris, given by $M_0 = M_\star/2 \sim 1.0 \times 10^{33}\,{\rm g}\,(M_\star/M_\odot)$. The normalization of the disk angular momentum is determined by the angular momentum of the bound debris, $J_0 = M_0\sqrt{2 G M r_t} \sim 4.3 \times 10^{55}\,{\rm g\,cm^2\,s^{-1}}$.
%Time evolution of the disk mass and angular momentum in Models I through IV. The different colors represent the different models. The mass of the bound debris gives the normalization of the disk mass as $M_0=M_\star/2\sim1.0\times10^{33}\,{\rm g}\,(M_\star/M_\odot)$. The normalization of the disk angular momentum is given by the angular momentum of the bound debris as $J_{0} = M_0\sqrt{2 G M r_t} = 4.3 \times 10^{48}~{\rm kg~m^2~s^{-1}}$.
}
\label{MdJd} 
\end{figure} 

The disk mass and angular momentum are given by 
\begin{eqnarray}
M_{\rm d} &=& 2\pi \int_{r_{\rm in}}^{r_{\rm out}} r\Sigma  \, \diff r, 
\nonumber 
\\
J_{\rm d} &=& 2\pi \int_{r_{\rm in}}^{r_{\rm out}} r^3 \Omega \Sigma  \, \diff r,
\nonumber
\end{eqnarray}
respectively.

Panel (a) of Figure~\ref{MdJd} shows the time evolution of the total disk mass. The initial disk mass corresponds to half of the stellar mass, and this value is used for mass normalization. As the disk evolves, its mass decreases with time. In Model I, the decrease is attributed to mass loss via accretion, whereas in Models II to IV, the disk mass decreases more rapidly due to both accretion and wind. In particular, a comparison between Models II and III indicates that $M_{\rm d}$ shows only a weak dependence on $\epsilon_{\rm rad}$. Furthermore, comparing Models III and IV demonstrates that wind-driven accretion with a non-zero value of $\bar{\alpha}_{z\phi}$ significantly reduces the disk mass.
%Panel (a) of Figure~\ref{MdJd} shows the time evolution of the total disk mass. The initial disk mass corresponds to half of the stellar mass, so we use it as the mass normalization. As the disk evolves, the disk mass decreases at late times due to mass loss via accretion in Model I, while it more rapidly decreases due to both accretion and wind for Models II to IV. It is noted from the comparison between Models II and III that $M_{\rm d}$ weakly depends on $\epsilon_{\rm rad}$. In addition, the comparison between Models III and IV clearly demonstrates that wind-driven accretion with non-zero $\bar\alpha_{z\phi}$ substantially reduces the disk mass.

Panel (b) of Figure~\ref{MdJd} shows the time evolution of the total angular momentum of the disk. In the absence of the wind ($\epsilon_{\rm rad} = 1$), the total angular momentum of the disk is conserved due to the zero viscous torque at the inner boundary of the disk, even though mass is lost, as indicated in panel (a). This confirms that our numerical calculations are accurate. When the wind is present, it carries away angular momentum from the disk, resulting in a decrease in $J_{\rm d}$ for $0 < \epsilon_{\rm rad} < 1$. The dependence of $J_{\rm d}$ on $\epsilon_{\rm rad}$ and $\bar\alpha_{z\phi}$ exhibits the same trend as the $M_{\rm d}$ evolution for Models II-IV.

%Panel (b) of Figure~\ref{MdJd} displays the time evolution of the total angular momentum of the disk.  In the case of the absence of the wind ($\epsilon_{\rm rad} = 1$), the total disk angular momentum is conserved due to the zero viscous torque at the disk's inner boundary, although the mass is lost as seen in Panel (a). This ensures that our numerical calculations have been done correctly. The wind carries off the angular momentum from the disk, causing a decline in $J_{\rm d}$ for $0<\epsilon_{\rm rad}<1$. The dependence of $J_{\rm d}$ on $\epsilon_{\rm rad}$ and $\bar\alpha_{z\phi}$ demonstrates the same tendency as the $M_{\rm d}$ evolution for Models II-IV.

%
%%%%%%%%%%%%%%%%%%%%%%
\subsection{Disk spectra and light curves}
\label{sec:spectra}
%%%%%%%%%%%%%%%%%%%%%%
%

%In this subsection, we calculate the disk spectra for all the models and describe the spectral properties of the disk and wind system. The effective temperature is given by $T_{\rm eff} = (Q_{\rm rad} / 2 \sigma)^{1/4}$ using the radiative flux $Q_{\rm rad}$. Equation (\ref{eq:qrad}) gives the relation between $T_{\rm eff}$ and the mid-plane temperature $T$, which is numerically calculated. Because the disk is highly optically thick, the observed flux is given with Planck's Blackbody distribution by 

%In this subsection, we calculate the disk spectra for all models and describe the spectral properties of the disk and wind system. The effective temperature is given by $T_{\rm eff} = (Q_{\rm rad} / 2 \sigma)^{1/4}$. Equation (\ref{eq:qrad}) relates $T_{\rm eff}$ to the mid-plane temperature $T$, which is calculated numerically. Since the disk is highly optically thick, the observed flux is given using Planck’s blackbody distribution by

In this subsection, we compute the disk spectra for all models and describe the spectral properties of the disk in the combined disk and wind system. Since the effective temperature is given by $T_{\rm eff} = (Q_{\rm rad} / 2 \sigma)^{1/4}$ and equation (\ref{eq:qrad}) relates $T_{\rm eff}$ to the disk mid-plane temperature $T$, $T_{\rm eff}$ can be determined numerically. Because the disk is highly optically thick, the observed flux is given using Planck’s blackbody distribution by
\begin{equation}
F_{\rm \nu, obs}
=
2\pi
\frac{\cos \theta_{\rm los}}{D_{\rm L}^2} 
\int_{r_{\rm in}}^{r_{\rm out}} B_{\nu}(T_{\rm eff},~\nu) r \, \diff r,
\label{eq:fluxdensity}
\end{equation}
where $\theta_{\rm los}$ is the angle between the observer's line of sight and the disk normal vector, $D_{\rm L}$ is the luminosity distance from the source to the observer. The disk luminosity of a certain frequency range of $\nu_{\rm l}\le\nu\le\nu_{\rm u}$ is expressed by 
%where $\theta_{\rm los}$ is the angle between the observer line of sight and disk normal vector, $D_{\rm L}$ is the luminosity distance of the source to the observer. The disk luminosity of a certain frequency domain of $\nu_{\rm l}\le\nu\le\nu_{\rm u}$ is given by 
\begin{eqnarray}
L 
&=&
\int_{\nu_{\rm l}}^{\nu_{\rm u}}
\,
L_{\nu} 
\, 
\diff \nu,
\label{eq:dlumi}
\end{eqnarray} 
where $L$ goes to the bolometric luminosity, $L_{\rm b}$, at $\nu_{\rm l}\rightarrow0$ and $\nu_{\rm u}\rightarrow\infty$, and $L_\nu$ is the spectral luminosity:
\begin{eqnarray}
L_{\nu} 
&=&
4 \pi D_{\rm L}^2 F_{\rm \nu, obs}
=
8\pi^2\cos \theta_{\rm los} 
\int_{r_{\rm in}}^{r_{\rm out}} B_{\nu}(T_{\rm eff},~\nu) r \, \diff r,
\label{eq:lnu}
\end{eqnarray}
where the right-hand side is derived using equation~(\ref{eq:fluxdensity}). In our calculations, we adopt $\theta_{\rm los} = 0^{\circ}$, corresponding to an observer viewing the entire disk from a pole-on perspective.
%In our calculations, we adopt $\theta_{\rm los} = 0^{\circ}$, such that an observer is viewing the full disk on the pole-on view.

\begin{figure}
\centering
\gridline{\subfigure[]{\includegraphics[scale = 0.77]{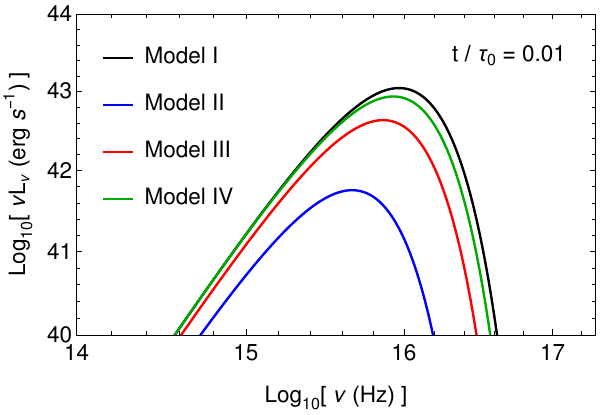}}
\subfigure[]{\includegraphics[scale = 0.77]{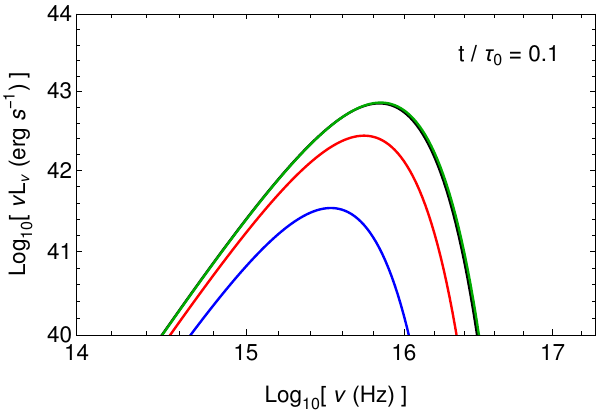}}}
\subfigure[]{\includegraphics[scale = 0.77]{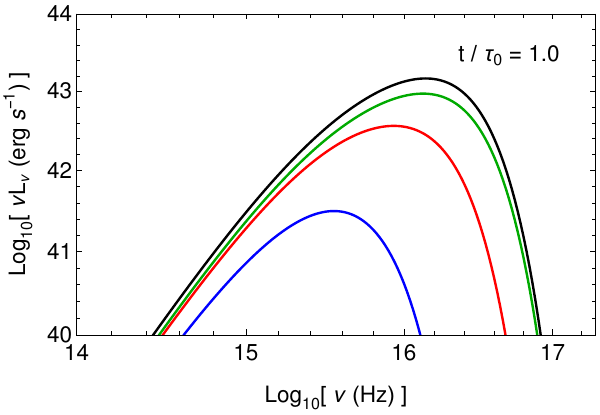}}
\caption{
Frequency distributions of the spectral luminosities (see equation~\ref{eq:lnu}) for Models I through IV. Different models are represented in different colors. Panels (a), (b), and (c) display the $t/\tau_0=0.01$ case, the $t/\tau_0=0.1$ case, and the $t/\tau_0=1.0$ case, respectively.
}
\label{speclum} 
\end{figure}

%Figure \ref{speclum} designates the disk spectra of Models I through IV at different time epochs. 
Figure \ref{speclum} illustrates the disk spectra of Models I to IV at different times. As can be seen in the figure, the disk spectra exhibit peaks ranging from the far ultraviolet (UV) to the mid-UV at early times, with the peaks gradually shifting to the near-UV band over time. The peak frequency decreases and the spectral magnitude diminishes as $\epsilon_{\rm rad}$ decreases. Model IV shows a higher peak at a higher frequency compared to Model III due to the non-zero positive value of $\bar{\alpha}_{z\phi}$, which increases the radiative flux as shown in equation~(\ref{eq:qradf}), resulting in a brighter spectral luminosity for the disk.

\begin{figure}
\centering
\gridline{\subfigure[]{\includegraphics[scale = 0.65]{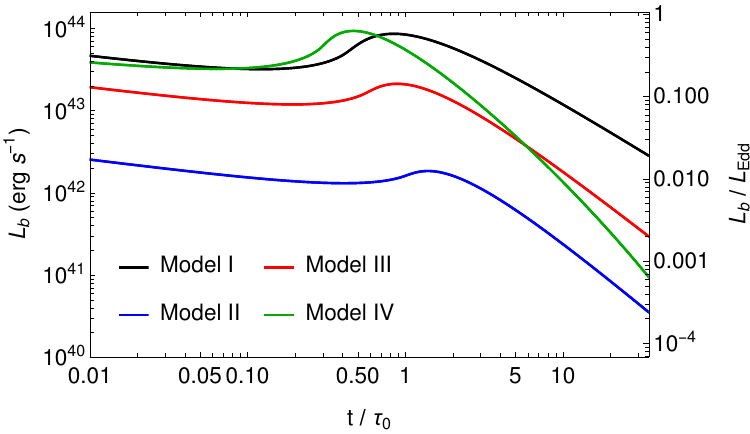}}
\subfigure[]{\includegraphics[scale = 0.65]{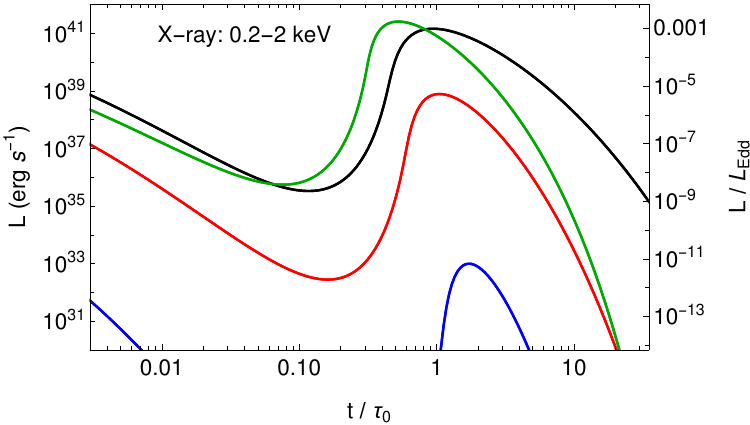}}}
\gridline{\subfigure[]{\includegraphics[scale = 0.67]{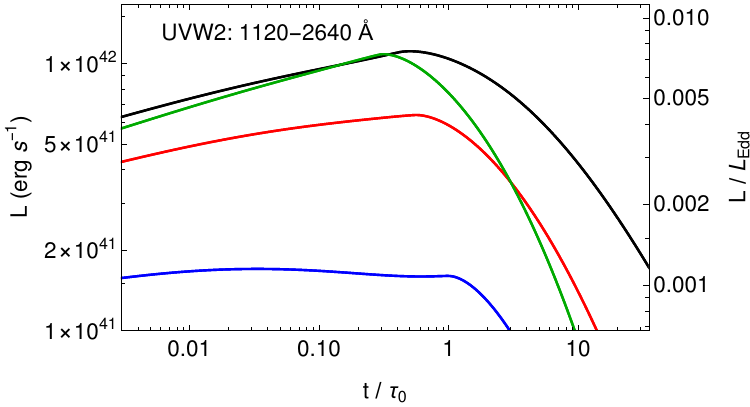}}
\subfigure[]{\includegraphics[scale = 0.69]{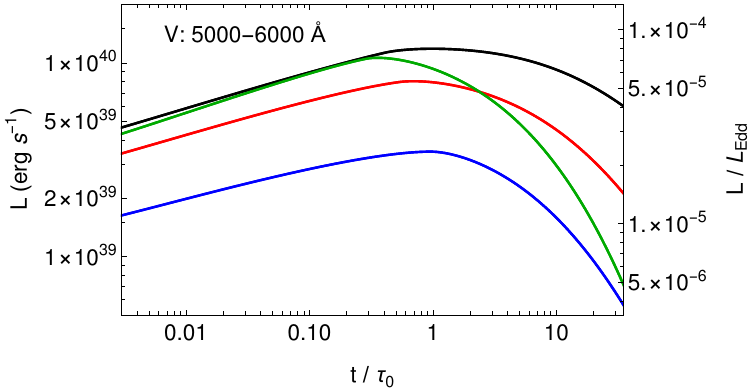}}}
\caption{
Time evolution of the disk luminosities of the three spectral bands and the bolometric luminosities for Models I through IV. Panels (a), (b), (c), and (d) show bolometric and three luminosities (X-ray, UV, and optical). Different colors indicate different models in all the panels.
}
\label{lum_band} 
\end{figure}

Figure~\ref{lum_band} shows the bolometric luminosities and the disk luminosities of the X-ray, UV, and optical bands calculated by equation~(\ref{eq:dlumi}). As shown in panel (a), the bolometric light curves, $L_{\rm b}$, are smaller than the Eddington luminosity (see equation~\ref{eq:ledd}) for all models over the long-term evolution. From the remaining panels, we see that for all models, the UV band luminosity is the brightest among the three bands. The X-ray luminosity varies significantly with time for all models, as shown in panel (b), while the optical and UV luminosities increase with time before $t/\tau_0=1.0$ and then decrease rapidly at later times due to the wind-driven disk mass depletion, as shown in panels (c) and (d). The bolometric light curves show a more gradual but similar temporal variability to the X-ray light curves.

As shown in panel (b) of Figure~\ref{lum_band}, the X-ray luminosity decreases with time at early stages and increases rapidly around $t/\tau_0\sim1.0$. This re-brightening occurs as the mass in the disk starts to accrete onto the black hole due to the viscous spreading of the initial Gaussian ring. In fact, the time corresponding to the peak of the X-ray re-brightening coincides with the peak time of the mass accretion rate, as seen in Figure~\ref{Mdots}. After the peak, the surface density near the inner radius decreases due to accretion and mass loss, resulting in a rapid decrease in luminosity. In contrast, the optical and UV luminosities increase with time at early stages. This is because the disk spectrum, as the initial Gaussian ring viscously spreads, peaks around the optical to UV range. After reaching the peaks, these luminosities decrease with time.

The comparison between Model IV and the other three models shows that the spectral luminosity for a non-zero $\bar{\alpha}_{z\phi}$ increases significantly at early times but decreases rapidly at late times compared to the cases of $\bar{\alpha}_{z\phi}=0$. This is due to the high rate of mass accretion and loss for $\bar{\alpha}_{z\phi} = 0.001$, which promotes efficient depletion of the disk mass, quickly leading to a low disk surface density and temperature at late times compared to other cases.

The X-ray luminosity is negligibly small for Model II, i.e., the case of the low value of $\epsilon_{\rm rad}$ with $\bar{\alpha}_{z\phi} = 0$. Comparing Models III and IV, we find that the non-zero value of $\bar{\alpha}_{z\phi}$ increases the X-ray luminosity by more than two orders of magnitude. This suggests that magnetic braking plays an important role in mass accretion from the outer to the inner radius.

%The comparison between Model IV and the other three models demonstrates that the spectral luminosity for a non-zero $\bar{\alpha}_{z\phi}$ increases significantly at early times but declines rapidly at late times compared to the {cases} of $\bar\alpha_{z\phi}=0$.  This is because of the high mass accretion and loss rate for $\bar{\alpha}_{z\phi} = 0.001$, promoting an efficient depletion of the disk mass. This rapidly results in a low disk surface density and temperature at late times compared to other cases.

%The X-ray luminosity is negligibly small for Model II, i.e., the case of the low value of  $\epsilon_{\rm rad}$ with $\bar{\alpha}_{z\phi} = 0$. According to the comparison between Models III and IV, we note the non-zero value of $\bar{\alpha}_{z\phi}$ makes the X-ray luminosity higher by more than two orders of magnitude. This suggests that magnetic braking plays an important role in mass accretion from the outer to inner radii.

%
%%%%%%%%%%%%%%%%%%%%%%%%%%%%%%%%%%%%%%%%%%%
\subsection{$\epsilon_{\rm rad}$-dependence of mass loss rates and power-law indices}
%%%%%%%%%%%%%%%%%%%%%%%%%%%%%%%%%%%%%%%%%%%
%

\begin{figure}
\centering
\includegraphics[scale = 0.65]{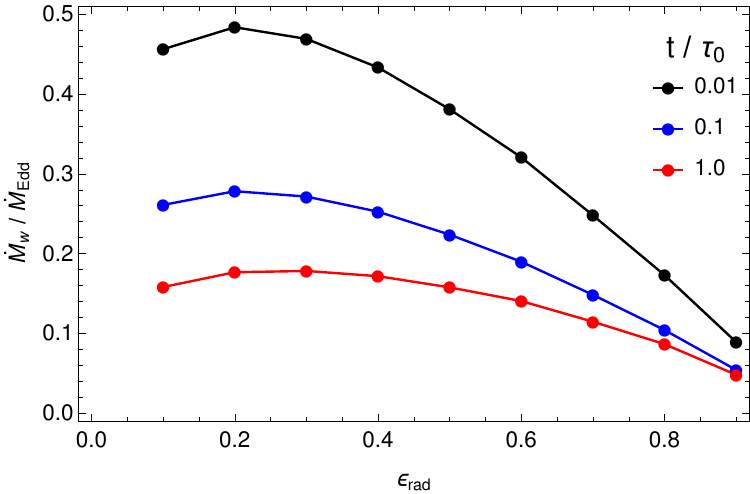}
\caption{
$\epsilon_{\rm rad}$-dependent mass loss rates at the three different times for $\bar{\alpha}_{z\phi} = 0$. Different colors indicate those different times. The mass loss rates are normalized by the Eddington accretion rate, $\dot{M}_{\rm Edd}$. The filled circle represents each data point in every $\epsilon_{\rm rad}=0.1$ step.
}
\label{MWerad} 
\end{figure}

Figure~\ref{MWerad} shows the dependence of the mass loss rates on $\epsilon_{\rm rad}$ at three different times for $\bar{\alpha}_{z\phi} = 0$. The mass loss rate decreases with time, as expected. The mass loss rates peak around $\epsilon_{\rm rad}$=0.2 at three different times. According to equation~(\ref{eq:cw}), $C_{\rm w,0}$ and the energy available for mass loss are higher for smaller $\epsilon_{\rm rad}$, suggesting that the mass loss rate is highest at the zero cooling limit, i.e., at $\epsilon_{\rm rad}=0$. However, one can see a local maximum in the mass loss rate at $\epsilon_{\rm rad} \approx 0.2$. This is because the surface density and temperature of smaller $\epsilon_{\rm rad}$ cases are lower than those of larger $\epsilon_{\rm rad}$ cases, particularly in the inner region. One may find this tendency in panels (a) and (b) of Figure~\ref{diskprop_erad}; the surface density of Model II ($\epsilon_{\rm rad} = 0.1$) is considerably lower than that of Model III ($\epsilon_{\rm rad}=0.5$) in $r/r_{\rm g} \lesssim 50$ due to the more efficient mass removal by the disk wind. Consequently, the temperature is also lower because of the reduced accretion heating. The lower surface density and temperature lead to the smaller mass flux of the disk wind (see equation~\ref{vzeqn}). Therefore, the total mass loss rate decreases for $\epsilon_{\rm rad} \rightarrow 0$ in the small $\epsilon_{\rm rad}$ ($\lesssim 0.2$) regime.

\begin{figure}
\centering
\includegraphics[scale = 0.65]{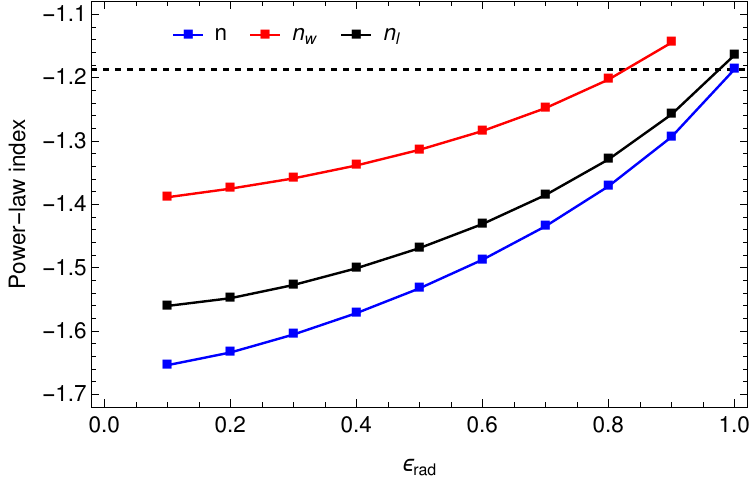}
\caption{
$\epsilon_{\rm rad}$-dependence of the slopes of mass accretion and loss rates for $\bar\alpha_{z\phi} = 0$. For all the cases, $n$, $n_{\rm w}$, and $n_{\rm l}$ are evaluated at $t/\tau_0=30$ and are denoted by blue, red, and black solid lines, respectively. Each data point in every $\epsilon_{\rm rad}=0.1$ step is indicated by the filled square. The horizontal black dashed line indicates the no wind case ($n=-19/16$) as a fiducial value \citep{1990ApJ...351...38C}.
}
\label{slope_n} 
\end{figure}

It is not trivial that the bolometric luminosity is proportional to the mass accretion rate if the mass loss by the disk wind is present. 
The power-law index of the bolometric lightcurve is defined by 
\begin{eqnarray}
n_{\rm l}=\frac{d\ln(L_{\rm b})}{d\ln{t}},
\nonumber
\label{eq:nl}
\end{eqnarray}
where $L_{\rm b}$ is given by equation~(\ref{eq:dlumi}) with $\nu_{\rm l}=0$ and $\nu_{\rm u}=\infty$. For comparison purposes, we show $\epsilon_{\rm rad}$-dependence of power law indices: $n$, $n_{\rm w}$, and $n_{\rm l}$ at a sufficiently late time, $t/\tau_0=30$, in Figure~\ref{slope_n}. It is noted from the figure that $|n|$ is higher than $|n_{\rm w}|$ and $|n_{\rm l}|$ takes a value between $|n|$ and $|n_{\rm w}|$ in the reasonable range of $\epsilon_{\rm rad}$. The steeper slope of the mass accretion rate indicates that it declines more rapidly than the mass loss rate by the disk wind and that the former is dominated by the latter at the late phase. 
This is because the disk wind removes the gas from early times so that the surface density in the inner region is lower for smaller $\epsilon_{\rm rad}$ as shown in Figure~\ref{diskprop_erad}~(a). Since the mass accretion rate is proportional to the surface density at the inner disk edge, it drops more rapidly with time for smaller $\epsilon_{\rm rad}$, which gives larger $|n|$ as shown in Figure~\ref{slope_n}. In contrast, the decrease in the mass loss rate is slower because disk winds still emanate from the outer region where sufficient mass remains. It results in moderately smaller $| n_{\rm w} |$. Panel (d) of Figure~\ref{Mdot_rad_erad} supports this interpretation. Considering the bolometric luminosity is affected through $Q_{\rm rad}$ (see equation~\ref{eq:qrad}) by both mass accretion and mass loss, it is natural that the resultant power-law index of the bolometric luminosity has some value between $|n|$ and $|n_{\rm w}|$ for the given range of $\epsilon_{\rm rad}$. As $\epsilon_{\rm rad}$ is larger, $n_{\rm l}$ asymptotes $n$. A slight deviation yet exists between $n$ and $n_{\rm l}$ even if the disk wind is absent ($\epsilon_{\rm rad}=1$). This is because of the time-dependent nature of the accretion disk. In fact, we confirm that $n_{\rm l}$ corresponds to $n$ in the steady-state limit ($t/\tau_0\simeq100$).

\pagebreak
%
%%%%%%%%%%%%%%%%%%%%%%
\subsection{Impact of the initial disk mass}
\label{sec:inidiskmass}
%%%%%%%%%%%%%%%%%%%%%%
%

The peak of the initial radial distribution of the surface density, $\Sigma_0$, increases with the initial disk mass $M_{\rm i}$, as seen in equation~(\ref{eq:sig0}), which increases the mass accretion and loss rates. Figures \ref{Mdot_rad_erad} and \ref{Mdot_rad_az} show that the mass accretion rate initially exceeds the Eddington accretion rate, indicating that the initial disk with a ring-like structure is a radiation-pressure dominated, super-Eddington accretion flow. Subsequently, the disk viscously spreads with time to begin accreting onto an SMBH at a sub-Eddington accretion rate after $t/\tau_0>0.5$, indicating that our formulation is adequate after that time.
However, our formulation is insufficient to describe the structure and evolution of the super-Eddington flow at a very early time. Therefore, we explore the effect of the initial disk mass, $M_{\rm i}$, on the mass accretion and loss rates at late times by comparing the disk model with a much lower initial mass $M_{\rm i}=0.005M_{\odot}$.

\begin{figure*}
\centering
\subfigure[]{\includegraphics[scale = 0.6]{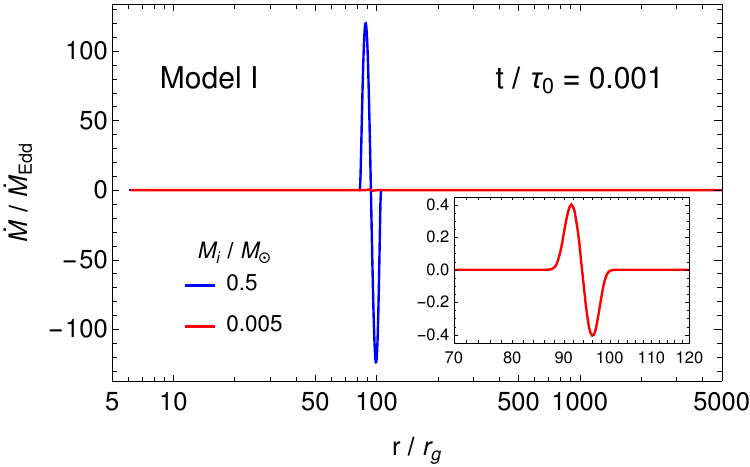}}
\subfigure[]{\includegraphics[scale = 0.6]{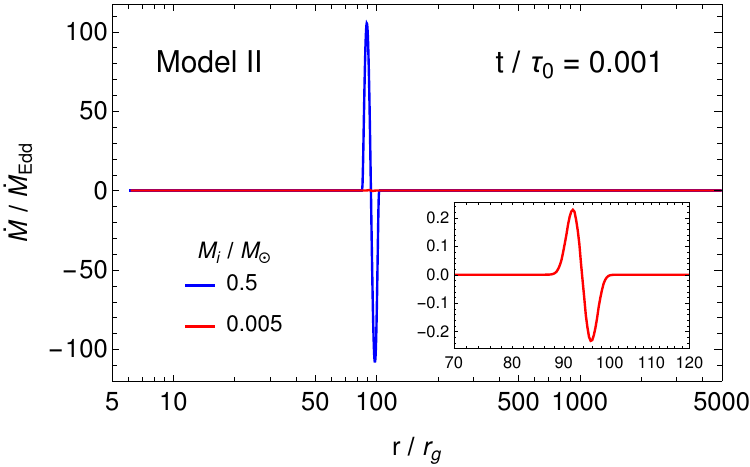}}
\subfigure[]{\includegraphics[scale = 0.6]{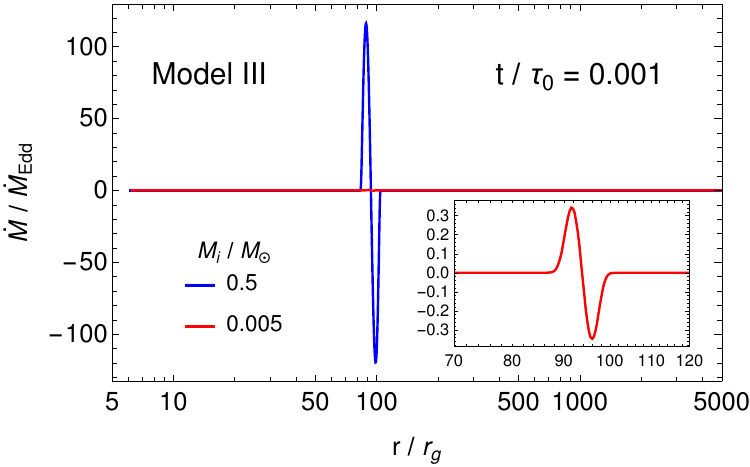}}
\subfigure[]{\includegraphics[scale = 0.6]{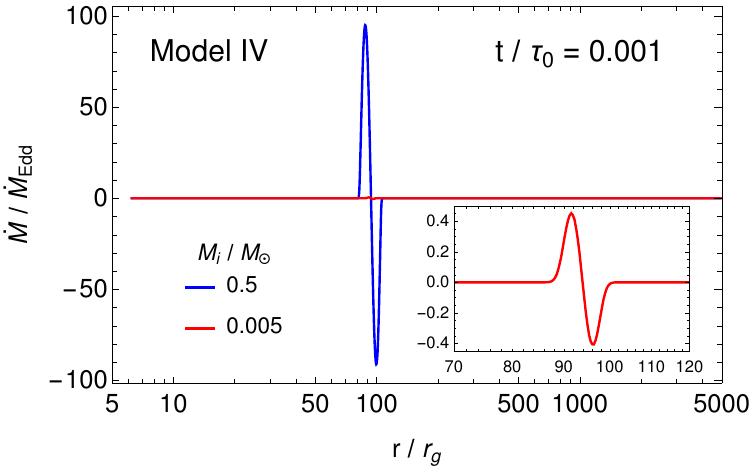}}
\caption{
Radial profiles of the mass accretion rates, which are obtained by equation~(\ref{mdotrate}), at a near initial time $t/\tau_0=0.001$. Panels (a) to (d) represent those of Model I to IV. For all panels, the blue and red solid lines correspond to an initial disk mass of $M_{\rm i} = 0.5M_{\odot}$ and $0.005M_{\odot}$, respectively. The mass accretion and loss rates are normalized by the Eddington accretion rate. 
}
\label{fig:mdot-r} 
\end{figure*}

\begin{figure*}
\centering
\subfigure[]{\includegraphics[scale = 0.6]{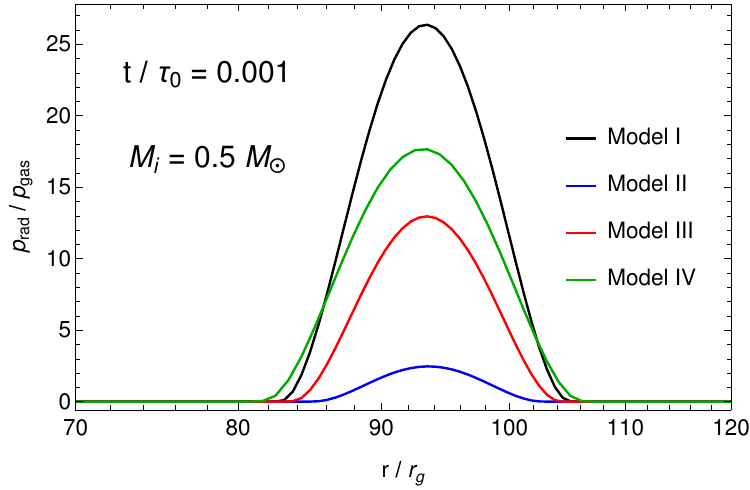}}
\subfigure[]{\includegraphics[scale = 0.62]{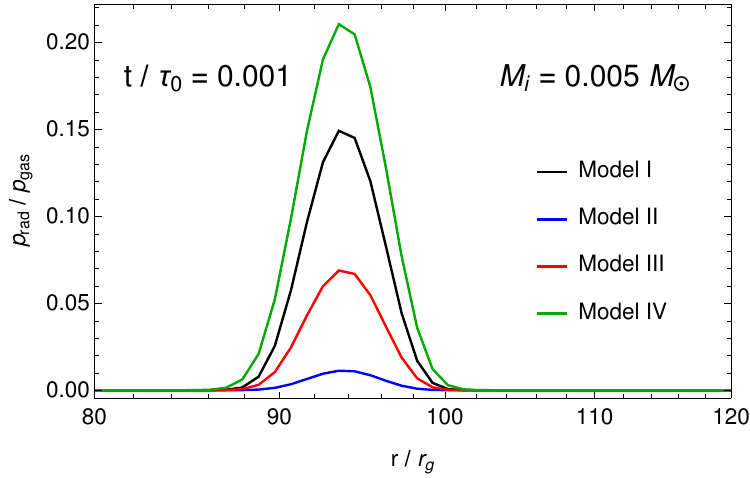}}
\caption{
Radial profile of the radiation to gas pressure ratio for Models I to IV at a near initial time $t/\tau_0=0.001$. Panels (a) and (b) correspond to an initial disk mass of $M_{\rm i} = 0.5M_{\odot}$ and $0.005M_{\odot}$, respectively. In both panels, the different colored lines indicate different models.
}
\label{fig:prpg} 
\end{figure*}

Figure~\ref{fig:mdot-r} compares the mass accretion rate of $M_{\rm i} = 0.5M_{\odot}$ with that of $0.005M_{\odot}$ at $t/\tau_0 = 0.001$. It is noted from the figure that the mass accretion rate for the $M_{\rm i}=0.005$ case is at a sub-Eddington rate from the beginning. Figure \ref{fig:prpg} shows the radiation-to-gas pressure ratio in the disk for Models I to IV. The radiation pressure is dominant for the $M_{\rm i} = 0.5M_{\odot}$ case, while the gas pressure is dominant for the $M_{\rm i} = 0.005M_{\odot}$ case. From equations (\ref{eq:qrad}), (\ref{eq:qradf}), and (\ref{eq:cs}), the ratio of radiation pressure to gas pressure is given as $p_{\rm rad}/p_{\rm gas}\propto\epsilon_{\rm rad}[(3/2)\bar{\alpha}_{r\phi}(H/r)+\bar{\alpha}_{z\phi}/2]r\Omega\Sigma$, where $p_{\rm rad}\propto\,T^4$. Note that the surface density decreases as the initial disk mass decreases. When the initial disk mass is extremely low, the contribution of the $\bar{\alpha}_{z\phi}$ term to $p_{\rm rad}/p_{\rm gas}$ becomes larger than that of the $\bar{\alpha}_{r\phi}$ term for moderate values of $\epsilon_{\rm rad}$. As a result, $p_{\rm rad}/p_{\rm gas}$ is larger in magnitude in Model IV than in Model I, as shown in panel (b) of Figure~\ref{fig:prpg}. Figures~\ref{fig:mdot-r} and \ref{fig:prpg} demonstrate that the disk with $M_{\rm i} = 0.005M_{\odot}$ is a geometrically thin, gas-pressure dominant.

\begin{figure}
\centering
\subfigure[]{\includegraphics[scale = 0.67]{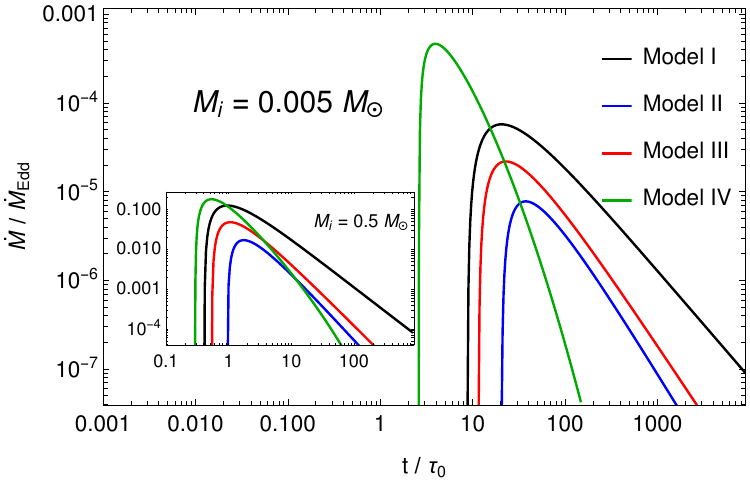}}\hspace{0.5cm}
\subfigure[]{\includegraphics[scale = 0.67]{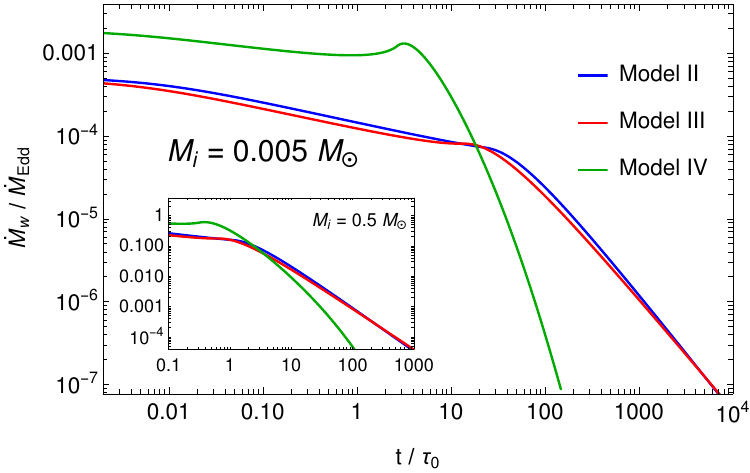}}
\caption{
Time evolution of the mass accretion rate estimated at the ISCO radius and the mass loss rate for the case of $M_{\rm i} = 0.005 M_{\odot}$.
Panels (a) and (b) represent the mass accretion rate curve and the mass loss rate curve for all four models, respectively. The different colored lines indicate different models. The Eddington accretion rate normalizes both the mass accretion and loss rates. The insets of panels (a) and (b) display the mass accretion and loss rates for the $M_{\rm i} = 0.5 M_{\odot}$ case, respectively.
}
\label{fig:mdot-mdotw} 
\end{figure}

The viscous timescale of a geometrically thin, gas-pressure dominated disk is proportional to $\Sigma^{-2/3}$, and the initial surface density is proportional to $M_{\rm i}$. Therefore, the viscous timescale is longer for disks with lower initial mass. As a result, the disk with $M_{\rm i} = 0.005 M_{\odot}$ evolves much more slowly than the disk with $M_{\rm i} = 0.5 M_{\odot}$ over the course of the calculation. Panels (a) and (b) of Figure~\ref{fig:mdot-mdotw} show the time evolution of the mass accretion rates at the ISCO radius for all four models, along with the corresponding mass loss rates. As predicted, in the case of $M_{\rm i} = 0.005 M_{\odot}$, the peak mass accretion rate is much lower, and the time at which this peak occurs is significantly delayed compared to the case of $M_{\rm i} = 0.5 M_{\odot}$. Similarly, the mass loss rate for the $M_{\rm i} = 0.005 M_{\odot}$ case is significantly lower. This lower outflow rate and longer viscous timescale extend the duration of the early, flatter phase of the mass loss rate compared to the $M_{\rm i} = 0.5 M_{\odot}$ case.
%The disk with $M_{\rm i} = 0.005M_{\odot}$ evolves much more slowly than the disk with $M_{\rm i} = 0.5M_{\odot}$ during the calculation time. This is because the viscous timescale of the sub-Eddington accretion flow is much longer than that of the super-Eddington accretion flow. Panels (a) and (b) of Figure~\ref{fig:mdot-mdotw} represent the time evolution of the mass accretion rates estimated at the ISCO radius for all four models and the corresponding mass loss rates. As predicted, for the case of $M_{\rm i} = 0.005 M_{\odot}$, the peak of the mass accretion rate is much lower, and the time at the peak is significantly delayed compared to the case of $M_{\rm i} = 0.5 M_{\odot}$. Similarly, the mass loss rate of the $M_{\rm i} = 0.005 M_{\odot}$ case is much lower. This lower outflow rate and longer viscous timescale makes the early flatter phase of the mass loss rate longer than that of the $M_{\rm i} = 0.5 M_{\odot}$ case.

\begin{figure}
\centering
\subfigure[]{\includegraphics[scale = 0.64]{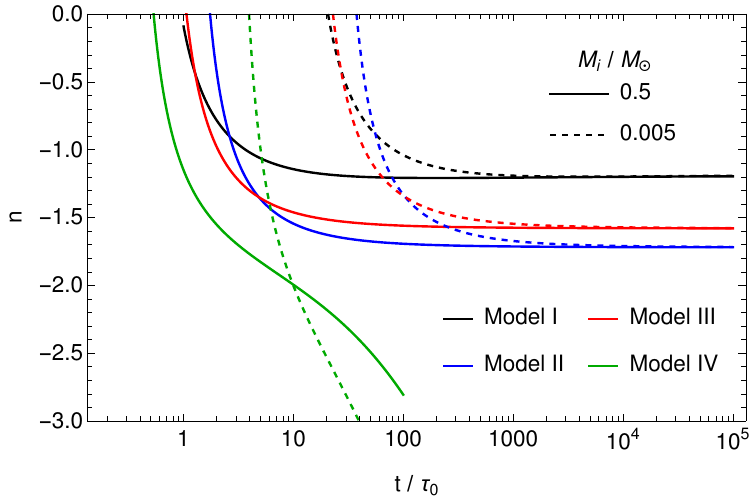}}\hspace{1.0cm}
\subfigure[]{\includegraphics[scale = 0.62]{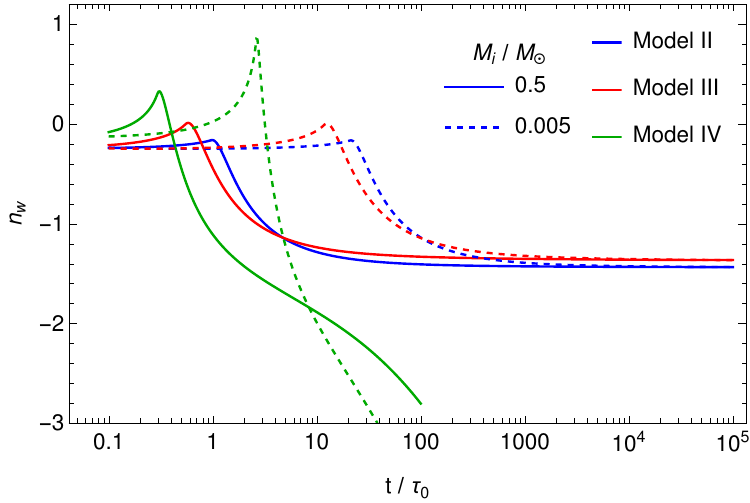}}
\caption{
Time evolution of the power-law indices of the mass accretion rate and mass loss rate curves. The indices $n$ and $n_{\rm w}$ are given by equations (25) and (26), respectively. The solid and dashed lines correspond to the cases $M_{\rm i} = 0.5 M_{\odot}$ and $0.005 M_{\odot}$, respectively. The different colored lines indicate different models.
\label{fig:nnw}
}
\end{figure}

Figure~\ref{fig:nnw} shows the time evolution of the power law indices of time of the mass accretion and loss rates, which are given by equations (\ref{eq:pindex}) and (\ref{eq:pindex-w}), respectively. The $n$ and $n_{\rm w}$ reach the same saturation value at the late times for $M_{\rm i} = 0.5M_{\odot}$ and $0.005M_{\odot}$ for Models I, II, and III. In contrast, the $n$ and $n_{\rm w}$ of Model IV corresponding to the non-zero value of $\bar{\alpha}_{z\phi}$, decrease more rapidly at late times for $M_{\rm i} = 0.005M_{\odot}$ than for $M_{\rm i} = 0.5M_{\odot}$. These results indicate that no initial disk mass impacts the late-time evolution of the disk without magnetic braking, while the initial disk mass affects the late-time evolution of the disk wind with magnetic braking.

%
% Fig 14
%

%
% Fig 15
%

%
% Fig 16
%

\pagebreak
%%%%%%%%%%%%%
\section{discussion}
\label{sec:discussion}
%%%%%%%%%%%%%

%We have derived the basic equations of a time-dependent, one-dimensional magnetically driven disk-wind model based on magnetohydrodynamical (MHD) equations in the context of TDEs and present a particular solution for them. We have started our calculations by making the initial condition: the surface density is distributed around the circularization radius in the Gaussian form. There are four key parameters to characterize the subsequent evolution of the disk and wind: ($\bar{\alpha}_{r\phi}$, $\bar{\alpha}_{z\phi}$, $\epsilon_{\rm rad}$, $C_{\rm w,0}$). The first two parameters have been introduced by extending the $\alpha$ parameter prescription, while the remaining two parameters $\epsilon_{\rm rad}$ and $C_{\rm w,0}$ are introduced to make the basic equations closed and to control the vertical mass flux, respectively. 

We have derived the basic equations for a time-dependent, one-dimensional, magnetically driven disk-wind model based on magnetohydrodynamic (MHD) equations in the context of TDEs and present a particular solution for these equations. We initiate our calculations with the initial condition that the surface density is Gaussian-distributed around the circularization radius. Four key parameters characterize the subsequent evolution of the disk and wind: ($\bar{\alpha}_{r\phi}$, $\bar{\alpha}_{z\phi}$, $\epsilon_{\rm rad}$, $C_{\rm w,0}$). The first two parameters are introduced by extending the $\alpha$ parameter prescription, while the remaining two parameters, $\epsilon_{\rm rad}$ and $C_{\rm w,0}$, are introduced to close the basic equations and to control the vertical mass flux, respectively.

%The viscosity parameter $\bar\alpha_{r\phi}$ corresponds to the Shakura-Sunuyaev viscosity parameter $\alpha=2\bar{\alpha}_{r\phi}/3$ for purely hydrodynamics. In fact, the first term of the right-hand side of equation~(\ref{eq:qradf}) reduces to the viscous heating rate of the standard disk model for $\bar\alpha_{z\phi}=0$, $C_{\rm w,0}=0$, and $\epsilon_{\rm rad}=1$. In our model, we have adopted $\bar\alpha_{r\phi}=0.1$, indicating $\alpha\simeq0.067$. \citet{2004MNRAS.347...67S} estimated $0.01 \leq \alpha \leq 0.03$ from the optical variabilities on time scales of months to years for 41 quasars. This range of $\alpha$ lies within the range of $\sim 0.005$ to $\sim 0.6$ predicted by various local shearing box MHD turbulent disk simulations \citep{1995ApJ...440..742H,1995ApJ...445..767M,1995ApJ...446..741B,2018ApJ...866..134G}. Our inferred value of $\alpha$ is reasonable because it is also within the range obtained by the MHD simulations.

The viscosity parameter $\bar\alpha_{r\phi}$ corresponds to the Shakura-Sunyaev viscosity parameter $\alpha=2\bar{\alpha}_{r\phi}/3$ for pure hydrodynamics. In fact, the first term on the right-hand side of equation~(\ref{eq:qradf}) reduces to the viscous heating rate of the standard disk model for $\bar\alpha_{z\phi}=0$, $C_{\rm w,0}=0$, and $\epsilon_{\rm rad}=1$. In our model, we used $\bar\alpha_{r\phi}=0.1$, which corresponds to $\alpha\approx0.067$. \citet{2004MNRAS.347...67S} estimated $0.01 \leq \alpha \leq 0.03$ from the optical variabilities on timescales of months to years for 41 quasars. This range of $\alpha$ is within the range of $\sim 0.005$ to $\sim 0.6$ predicted by various local shearing box MHD turbulent disk simulations \citep{1995ApJ...440..742H,1995ApJ...445..767M,1995ApJ...446..741B,2018ApJ...866..134G}. 
%Our inferred value of $\alpha$ is reasonable because it is also within the range obtained by the MHD simulations.

%The mass loss rates increase with $C_{\rm w}$ because the vertical mass flux is proportional to $C_{\rm w}$. This was originally introduced alternatively to solving the vertical component of the momentum equation, which gives the vertical component of the velocity. Therefore, $C_{\rm w}$ is not completely a free parameter, and the upper limit is obtained from the local MHD simulations. From Figure~\ref{cwe_erad}, we note that $C_{\rm w,0} < C_{\rm w, sim}$ for the given parameter ranges. However, $C_{\rm w,0}$ increases with $\bar{\alpha}_{z\phi}$ so that $C_{\rm w,0}$ can be higher than $C_{\rm w,sim}$ for high $\bar{\alpha}_{z\phi}$. This implies there can be an upper limit of $\bar{\alpha}_{z\phi}$. The magnetic braking is characterized by $\bar\alpha_{z\phi}$. We have adopted the two cases: $\bar\alpha_{z\phi}=0$ and $\bar\alpha_{z\phi}=0.001$ for our model. Magnetic braking enhances mass accretion by liberating gravitational energy as a result of the angular momentum removal. In the forthcoming paper, we will make the relation of $C_{\rm w}$ and $\bar\alpha_{z\phi}$ clearer, by which we will examine the effect of $\bar\alpha_{z\phi}$ on mass accretion and loss rates.

In our model, a part of disk heating due to the viscosity and magnetic braking goes to radiative cooling. The assigned energy to the cooling is controlled by an unknown parameter, $\epsilon_{\rm rad}$ (see equation~\ref{eq:qradf}). For given $C_{\rm w}$, we can evaluate $\epsilon_{\rm rad}$ by equations~(\ref{eq:cw}) and~(\ref{eq:cw02}) as
\begin{eqnarray}
\epsilon_{\rm rad}=1-\frac{C_{\rm w,0}}{2}
\biggr[
3
\bar\alpha_{r\phi}
\left(
\frac{H}{r}
\right)^2
+
\bar\alpha_{z\phi}
\left(
\frac{H}{r}
\right)
\biggr]^{-1}.
\nonumber 
\end{eqnarray}
For $C_{\rm w,0}=1.0\times10^{-5}$, $\bar\alpha_{r\phi}=0.1$, and $H/r=5\times10^{-3}$, we estimate $\epsilon_{\rm rad}\approx1/3$ for $\bar{\alpha}_{z\phi} = 0$ and $\epsilon_{\rm rad} \approx 0.6$ for $\bar{\alpha}_{z\phi} = 0.001$. By obtaining $C_{\rm w}$ from the local MHD simulations, we can solve a set of basic equations without $\epsilon_{\rm rad}$. To obtain the vertical velocity directly without introducing $C_{\rm w}$, we need to add the vertical component of the momentum conservation equation in the current basic equations and solve it simultaneously. This is our future work.

Adopting $\bar\alpha_{r\phi}=0.1$, $\bar\alpha_{z\phi}=0$, and $H/r=5\times10^{-3}$, we obtain $C_{\rm w,0}\sim1.5\times10^{-5} (1-\epsilon_{\rm rad})$ from equation~(\ref{eq:cw02}), which corresponds to a plasma beta of $\mathcal{O}(10)$ \citep{2010ApJ...718.1289S} when $\epsilon_{\rm rad} \ll 1$, indicating a strong magnetic field. According to \citet{2009ApJ...691L..49S,2010ApJ...718.1289S}, $C_{\rm w,0}$ becomes larger as the plasma beta decreases. 
%Local shearing box simulations show that stronger magnetic fields predict a larger mass flux, $C_{\rm w,0}$ \citep{2009ApJ...691L..49S,2010ApJ...718.1289S}. 
%Since the effect of the disk wind is negligible under the condition of weak magnetic fields, $\epsilon_{\rm rad} \sim 1$ such that $C_{\rm w}\ll1.5\times10^{-5}$, we expect the power-law index to have asymptotes $n=-19/16$. The power-law index is distributed over $-5/3\lesssim{n}<-19/16$ when the magnetic field is mildly strong, corresponding to $C_{\rm w,0}\lesssim1.5\times10^{-5}$ and $\epsilon_{\rm rad} < 1$. 
Since the effect of the disk wind is negligible if the magnetic field is such weak that $C_{\rm w}\ll1.5\times10^{-5}$, i.e., $\epsilon_{\rm rad} \sim 1$, we expect the power-law index to have asymptotes $n=-19/16$. Next, the power-law index is distributed over $-5/3\lesssim{n}<-19/16$ when the magnetic field is mildly strong, corresponding to $C_{\rm w,0}\lesssim1.5\times10^{-5}$, i.e., $\epsilon_{\rm rad} < 1$. This case is consistent with ASASSN-18pg, whose bolometric luminosity at late times is flatter than $t^{-5/3}$ \citep{2020ApJ...898..161H}. Note that $\epsilon_{\rm rad}=0$ gives an upper limit of $C_{\rm w,0}$, meaning that $C_{\rm w,0}$ is not greater than $1.5\times10^{-5}$ with the current parameter sets. However, $C_{\rm w,0}$ can be greater than $1.5\times10^{-5}$ if $\bar\alpha_{z\phi}$ takes a non-zero value, indicating a magnetic field strong enough for magnetic braking to work efficiently. This scenario can explain the late-time variation of the following TDEs: AT2019qiz, whose light curve decays rapidly with time, scaling as $t^{-2.54}$ \citep{2020MNRAS.499..482N}, and possibly ASASSN-14ae and ASASSN-15oi, whose light curves exhibit an exponential decline at late times \citep{2014MNRAS.445.3263H, 2016MNRAS.463.3813H}. We have discovered a new regime of power-law indices at late times: $-5/3 \lesssim n < -19/16$ for $\bar\alpha_{z\phi}=0$ or $n < -2$ for $\bar\alpha_{z\phi} > 0$. Our MHD disk-wind model naturally explains TDEs that deviate from the late-time light curves of X-ray TDEs, whose power-law index follows $n \lesssim -1.2$ as described in \citet{2017ApJ...838..149A}.

%As explained in section~\ref{sec:inidiskmass}, if the initial disk mass is large, the accretion rate will exceed the Eddington accretion rate in the early stages before accretion onto the black hole. In the absence of magnetic braking, this has no significant effect on the time evolution of the subsequent accretion rate or mass loss rate, but once magnetic braking is present, it affects their later evolution. 
As described in section~\ref{sec:inidiskmass}, if the initial disk mass is large, the accretion rate can exceed the Eddington limit during the early stages, prior to accretion onto the black hole. In the absence of magnetic braking, this excess does not significantly impact the late-time evolution of the accretion and mass loss rates. However, when magnetic braking is considered, it influences their subsequent evolution. In the super-Eddington accretion flow, a radiation pressure is dominant and the advection cooling term is significant in the energy equation, which is different from the geometrically thin disk case that we assume throughout this paper \citep{1988ApJ...332..646A,2002apa..book.....F,kato_black-hole_2008,2011ApJS..195....7X}. It is interesting how advection cooling and radiation pressure change the physical properties of the magentically driven disk wind. Therefore, in the future we will reformulate the basic equations for the current time-dependent model accordingly and specifically investigate the case of non-zero value of $\alpha_{z\phi}$.

We ideally need to perform multi-dimensional MHD simulations of the evolution of the MHD disk-wind system for deciding all the parameters, including the remaining $\bar{\alpha}_{r\phi}$ and $\bar{\alpha}_{z\phi}$. Especially, the magnetic field geometry of the accretion disk is so complicated that deciding the reasonable magnetic field geometry over the long term is challenging  \citep{2018ApJ...854....2J}. Thus, some assumptions are yet needed to decide it. \citet{2019ApJ...872..149L} constructed a steady disk-outflow model in which the large-scale field forms by the advection of the external field in the disk. They showed that a moderate external field (plasma beta of several hundred at the outer disk radius) leads to a large-scale magnetic field inclined with the disk surface and supports the disk outflow. They found that the outflow is non-relativistic overall and the terminal velocity of the outflow emitted from the disk's inner radius is up to $\sim 0.3~c$, and the mass loss rate can be higher than the mass accretion rate for a certain range of the plasma beta. As seen in Figures~\ref{Mdot_rad_erad} and \ref{Mdot_rad_az}, our time-dependent model also demonstrates that the mass outflow rate is higher than the mass accretion rate, although their time evolution can change by $\epsilon_{\rm rad}$, $\bar{\alpha}_{r\phi}$ and $\bar{\alpha}_{z\phi}$. For exploring the plasma-beta dependence on this result, we need to decide on the magnetic field evolution. We will extend our time-dependent model by solving the induction equation in the future.

There are some X-ray TDEs, such as ASASSN-15oi \citep{2017ApJ...851L..47G}, OGLE16aaa \citep{2020A&A...639A.100K}, and AT 2019azh \citep{2022ApJ...925...67L}, which show the late-time X-ray brightening with the optical and UV flares at early times. OGLE16aaa was not detected in the soft X-ray waveband at discovery, while ASASSN-15oi and AT2019azh were also detected at early times. This early-time X-ray emission is very weak at $10^{41}$ erg/s. In contrast, the late-time X-ray luminosity is one or two orders of magnitude higher to be $10^{42-43}$ erg/s, but still significantly lower than the early-time optical and UV luminosities. \cite{2021ApJ...921...20H} proposed the simple analytical model that the late-time X-ray flare is caused by the viscous accretion from the disk circularization, while the optical and UV flares occurs due to the stream-stream collision at early times. The time in our model is normalized by a viscous timescale $\tau_0=14.3\,{\rm yr}$. The X-ray brightens at $t/\tau_0\lesssim1$ for Models I, II, and III, while it brightens at $t/\tau_0<1$ in Model IV because magnetic braking (non-zero value of $\bar{\alpha}_{z\phi}$) causes faster accretion due to the shorter viscous timescale. In addition, Models I and IV show that the X-ray luminosity has a peak of $\sim10^{41}$ erg/s from panel (b) of Figure~\ref{lum_band}. These results indicate that Model IV can explain the observed X-ray rebrightening. The X-ray luminosity is still low compared to the observed one, but, e.g., higher values of $\epsilon_{\rm rad}$ and $\bar{\alpha}_{z\phi}$ would allow for higher luminosity. The detailed comparison between our model and the observed X-ray rebrightening is our future task.

Several TDEs such as PS1-10jh \citep{2012Natur.485..217G}, ASASSN-14ae \citep{2014MNRAS.445.3263H}, OGLE16aaa \citep{2017MNRAS.465L.114W}, iPTF16axa \citep{2017ApJ...842...29H}, iPTF16fnl \citep{2017ApJ...844...46B}, AT2017eqx \citep{2019MNRAS.488.1878N}, AT2019qiz \citep{2020MNRAS.499..482N}, and ASASSN-18pg \citep{2020ApJ...898..161H} has been shown to have significantly brighter optical and UV emissions compared to its observationally insignificant X-ray emissions. The strong dominance of optical and UV radiation in the observational spectrum is interpreted by the disk-wind model, in which an optically thick outflow obscures the disk emission, and the emitted X-ray photons are reprocessed into lower-energy optical or UV photons \citep{2020SSRv..216..114R}.
%ASASSN-18pg \citep{2020ApJ...898..161H} have shown that their optical/UV emissions are significantly brighter than their observationally insignificant X-ray emissions. The strong dominance of optical/UV radiation in the observational spectrum is interpreted by the disk-wind model, where an optically thick outflow obscures the disk emission, and the emitted X-ray photons are reprocessed into lower-energy optical/UV photons \citep{2020SSRv..216..114R}. 
A spherical outflow model from a super-Eddington accretion flow has been constructed for TDEs by \cite{2009MNRAS.400.2070S,2020ApJ...894....2P,2023arXiv231217417M}. Following these studies, assuming a spherically symmetric wind launched from a radius $r_{\rm l}$ with a mass loss rate $\dot{M}_{\rm w}$, wind velocity $v_{\rm w}$, and  wind density $\rho(r) = \rho_{\rm l} (r/r_{\rm l})^{-2}$, where $\rho_{\rm l} = \dot{M}_{\rm w} / 4\pi r_{\rm l}^2 v_{\rm w}$, we can estimate the optical depth of the wind as
\begin{eqnarray}
\tau(r) 
&&
= \int_{r}^{r_{\rm out}} \rho \kappa_{\rm es} \, {\rm d}r 
= \frac{\kappa_{\rm es}\dot{M}_{\rm w}}{4\pi r_{\rm l} v_{\rm w}}
\left( 
\frac{r}{r_{\rm l}}
\right)^{-1}
\nonumber 
\\
&&
\sim 
50 
\left(\frac{\dot{M}_{\rm w}}{\dot{M}_{\rm Edd}}\right) 
\left(\frac{v_{\rm w}}{10^{4}~{\rm km~s^{-1}}}\right)^{-1}
\left(\frac{r_{\rm l}}{r_{\rm ISCO}}\right)^{-1} 
\left(\frac{r}{r_l}\right)^{-1},
\nonumber
\end{eqnarray}
where we also assumed that $r_{\rm out} \gg r_{\rm l}$. If $\dot{M}_{\rm w} \ll \dot{M}_{\rm Edd}$, the wind is optically thin, and thus reprocessing is inefficient. This corresponds to cases where the initial disk mass is small or in the later times of our models. In contrast, if $\dot{M}_{\rm w} \gtrsim \dot{M}_{\rm Edd}$, the wind is optically thick so that X-ray photons emitted from the disk would be absorbed and re-emitted in the lower energy band. However, we cannot quantitatively evaluate the reprocessed spectrum without solving the radiative transport equation. 
%The simultaneous application of the disk-wind and photoionization radiative transfer models to the observed continuum flux and emission lines, respectively, is an effective method to elucidate the physical properties of TDEs. 
This is a future task that we will explore.
\section{Conclusions}
\label{sec:conclusion}
%%%%%%%%%%%%%%

%We have constructed a one-dimensional model of a time-dependent, geometrically thin accretion disk with a magnetically driven non-relativistic outflow. The MHD turbulent viscosity and magnetic braking are characterized by the two parameters: $\bar{\alpha}_{r\phi}$ and $\bar{\alpha}_{z\phi}$, respectively. These are derived by applying the extended $\alpha$ parameter prescription to the MHD momentum equations. Moreover, we have introduced two other parameters: one is $C_{\rm w,0}$ to control the vertical mass flux, and another is $\epsilon_{\rm rad}$ to be the ratio of the thermal cooling to the disk heating fluxes. We have found numerical disk-wind solutions for the basic equations with these four parameters. \red{The angular momentum of the disk is conserved in the absence of wind ($\epsilon_{\rm rad} = 1$) due to the zero viscous torque at the inner boundary. However, the angular momentum decreases with time due to mass loss by the wind. In particular, the $\bar{\alpha}_{z\phi}$ parameter causes the  angular momentum of the disk to decrease more efficiently than the $\epsilon_{\rm rad}$ parameter. We also confirm that $C_{\rm w,0}$ is smaller than $C_{\rm w, sim}$ over the entire disk region within a reasonable time; the mass flux of the disk wind is determined by the energetic constraint.} Our primary conclusions are summarized as follows:

We have constructed a one-dimensional model of a time-dependent, geometrically thin accretion disk with a magnetically driven non-relativistic wind. The MHD turbulent viscosity and magnetic braking are characterized by two parameters: $\bar{\alpha}_{r\phi}$ and $\bar{\alpha}_{z\phi}$, respectively. These are derived by applying an extended $\alpha$ parameter prescription to the MHD momentum equations. In addition, we have introduced two other parameters: $C_{\rm w,0}$, which controls the vertical mass flux, and $\epsilon_{\rm rad}$, which is the ratio of the thermal cooling flux to the disk heating flux. We have found numerical disk-wind solutions for the basic equations with these four parameters. In the absence of the wind, the angular momentum of the disk is conserved due to the zero viscous torque at the inner boundary. However, the angular momentum decreases with time due to mass loss by the wind. In particular, when magnetic braking is present (i.e., $\bar{\alpha}_{z\phi}\neq0$), the disk loses angular momentum more efficiently than in the absence of magnetic braking (i.e., $\bar{\alpha}_{z\phi}=0$). We also confirm that $C_{\rm w,0}$ remains smaller than $C_{\rm w, sim}$ over the entire disk region within a reasonable time, indicating that the mass flux of the disk wind is determined by the energetic constraint. Our primary conclusions are summarized as follows:

%We have constructed a one-dimensional model of a time-dependent, geometrically thin accretion disk with a magnetically driven non-relativistic outflow. The MHD turbulent viscosity and magnetic braking are characterized by the two parameters: $\bar{\alpha}_{r\phi}$ and $\bar{\alpha}_{z\phi}$, respectively. These are derived by adopting extended the $\alpha$ parameter prescription to the MHD momentum equations. Moreover, we introduce two other parameters: one is $C_{\rm w,0}$ to control the vertical mass flux, and another is $\epsilon_{\rm rad}$ to be the ratio of the thermal cooling to disk heating fluxes. We have found numerical disk-wind solutions for the basic equations with these four parameters. \blue{The angular momentum of the disk is conserved when the wind is absent ($\epsilon_{\rm rad} = 1$) due to the zero viscous torque at the inner boundary. However, the angular momentum decreases with time due to mass loss by the wind. Notably, the $\bar{\alpha}_{z\phi}$ parameter makes the disk's angular momentum decrease more efficiently than the $\epsilon_{\rm rad}$ parameter. We confirm that $C_{\rm w,0}$ is smaller than $C_{\rm w, sim}$ over the disk's entire region within a reasonable time; the mass flux of the disk wind is determined by the energetics constraint.} Our primary conclusions are summarized as follows:

\begin{enumerate}

\item
The mass accretion rate follows the power law of time $t^{-19/16}$ at late times if the wind is absent ($\epsilon_{\rm rad} = 1$). This result corresponds to the classical solution of \citet{1990ApJ...351...38C}. 

\item
In the case that the wind is present ($0<\epsilon_{\rm rad}<1$) without magnetic braking ($\bar\alpha_{z\phi}=0$), the mass accretion rate follows the power law of time at late times and is steeper than the classical solution: $t^{-19/16}$. In addition, the mass accretion rate becomes steeper as $\epsilon_{\rm rad}$ approaches 0. When magnetic braking is on (i.e., $\bar\alpha_{z\phi}=0.001$), the mass accretion rate decays rapidly with time and the power-law index evolves with time: for $\dot{M}\propto t^{n}$, $|n|$ becomes larger with time and $n<-2$ at late times. 

\item
The bolometric luminosity is not proportional to the mass accretion rate due to mass loss by the disk wind. In fact, we find that the bolometric light curve is flatter than the mass accretion rate but steeper than the mass loss rate in the absence of magnetic braking: the bolometric decay index is distributed over $-5/3<{n_{\rm l}}\lesssim-19/16$ for $L\propto{t}^{n_{\rm l}}$. We also confirm that the bolometric decay index asymptotes to $-19/16$ in the absence of wind at late times.

%We find a new regime of the late-time bolometric light curve in a TDE that it is steeper than the classical $t^{-19/16}$ solution due to the magnetically driven winds: for $L\propto{t^{n_{\rm l}}}$, $-5/3<{n_{\rm l}}\lesssim-19/16$ in the case without magnetic braking ($\bar\alpha_{z\phi}=0$) or $n_{\rm l}<-2$ in the case with magnetic braking ($\bar\alpha_{z\phi}\gtrsim0.001$).
\item
We identify a new regime in the late-time bolometric light curve of a TDE, where it is steeper than the classical $t^{-19/16}$ solution due to magnetically driven winds. For $L\propto{t^{n_{\rm l}}}$, we find $-5/3<{n_{\rm l}}\lesssim-19/16$ in the absence of magnetic braking ($\bar\alpha_{z\phi}=0$), and $n_{\rm l}<-2$ when magnetic braking is present ($\bar\alpha_{z\phi}\gtrsim0.001$).

\item 
In the disk emission, the UV luminosity is the highest among the optical, UV, and X-ray luminosities. The X-ray emission shows a significant rebrightening at late times. In particular, for $\bar{\alpha}_{z\phi} \neq 0$, the X-ray emission reaches the highest peak and exhibits the shortest time interval between the optical or UV and X-ray peaks. These results highlight the critical role of $\bar{\alpha}_{z\phi}$ in modulating the timing and intensity of disk emission, which is of observational importance for understanding the underlying physical mechanisms.

\item Our model predicts that late-time bolometric light curves steeper than $t^{-19/16}$ in UV-bright TDEs are potential evidence for magnetically driven winds.

\end{enumerate}

\begin{acknowledgments}
We thank the referee for the constructive suggestions that have improved the paper. M.T. and K.H. have been supported by the Basic Science Research Program through the National Research Foundation of Korea (NRF) funded by the Ministry of Education (2016R1A5A1013277 to K.H. and 2020R1A2C1007219 to K.H. and M.T.). This work was financially supported by the Research Year of Chungbuk National University in 2021. This work is also supported by Grant-inAid for Scientific Research from the MEXT/JSPS of Japan, 22H01263 to T.K.S. This research was supported in part by grant no. NSF PHY-2309135 to the Kavli Institute for Theoretical Physics (KITP).
\end{acknowledgments}

%% To help institutions obtain information on the effectiveness of their 
%% telescopes the AAS Journals has created a group of keywords for telescope 
%% facilities.
%
%% Following the acknowledgments section, use the following syntax and the
%% \facility{} or \facilities{} macros to list the keywords of facilities used 
%% in the research for the paper.  Each keyword is check against the master 
%% list during copy editing.  Individual instruments can be provided in 
%% parentheses, after the keyword, but they are not verified.

%\vspace{5mm}
%\facilities{HST(STIS), Swift(XRT and UVOT), AAVSO, CTIO:1.3m,CTIO:1.5m,CXO}

%% Similar to \facility{}, there is the optional \software command to allow 
%% authors a place to specify which programs were used during the creation of 
%% the manuscript. Authors should list each code and include either a
%% citation or url to the code inside ()s when available.

%\software{astropy \citep{2013A&A...558A..33A,2018AJ....156..123A},  
%          Cloudy \citep{2013RMxAA..49..137F}, 
%          Source Extractor \citep{1996A&AS..117..393B}
%          }

%% Appendix material should be preceded with a single \appendix command.
%% There should be a \section command for each appendix. Mark appendix
%% subsections with the same markup you use in the main body of the paper.

%% Each Appendix (indicated with \section) will be lettered A, B, C, etc.
%% The equation counter will reset when it encounters the \appendix
%% command and will number appendix equations (A1), (A2), etc. The
%% Figure and Table counter will not reset.

\appendix

%
%%%%%%%%%%%%%%%%%%%%%%%%
\section{{Derivation of Basic Equations for MHD Disk and Wind Evolution}}
%\label{app_sdeq}
%%%%%%%%%%%%%%%%%%%%%%%%
%

{
In Appendix, we introduce the detailed derivation of the evolutionary equations for the mass and energy of a disk with the wind by adding in the formalism of \citet{2016A&A...596A..74S}, a modification so as to fit our problem for TDE disks.
}

%
%%%%%%%%%%%%%%%%%%%%%%%%
\subsection{{Surface Density Evolution}}
\label{app_sdeq}
%%%%%%%%%%%%%%%%%%%%%%%%
%

The mass conservation and momentum conservation equations of the general magneto-hydrodynamics (MHD) 
are given by \citep{1998RvMP...70....1B} as
\begin{equation}
\frac{\partial \rho}{\partial t} + \vec{\nabla}\cdot (\rho \vec{v}) = 0
\label{mcons}
\end{equation}
and
\begin{equation}
\rho \frac{\partial \vec{v}}{\partial t} + (\rho \vec{v}\cdot \vec{\nabla}) \vec{v} = - \vec{\nabla}\left(p + \frac{B^2}{8 \pi}\right) - \rho \vec{\nabla}\Phi + \left(\frac{\vec{B}}{4\pi} \cdot \vec{\nabla}\right) \vec{B} +  \eta_v \left(\nabla^2 \vec{v} + \frac{1}{3} \vec{\nabla}(\vec{\nabla}\cdot \vec{v})\right),
\label{momcons}
\end{equation}
respectively, where $\vec{v}$ is the fluid velocity, $p$ is the pressure, 
\begin{eqnarray}
\Phi =-\frac{GM}{\sqrt{r^2+z^2}}
%\approx -\frac{GM}{r}
\label{eq:gp}
\end{eqnarray}
is the gravitational potential, $B$ is the magnetic-field vector, and $\eta_v$ is the microscopic kinematic shear viscosity. 
\citet{1998RvMP...70....1B} assumed that the bulk viscosity due to the microscopic kinematic shear viscosity $\eta_v$ vanishes. 

%We write the MHD equations
Assuming that the disk and wind are axisymmetric, we rewrite equations~\ref{mcons} and \ref{momcons} with cylindrical coordinates as
\begin{equation}
\frac{\partial \rho}{\partial t} + \frac{1}{r} \frac{\partial }{\partial r}(r \rho v_r) + \frac{\partial}{\partial z} (\rho v_z) = 0,
\label{cylmcons}
\end{equation}
and
\begin{equation}
\frac{\partial}{\partial t}(r \rho v_{\phi}) + \frac{1}{r} \frac{\partial}{\partial r}\left[r^2 \left\{\rho v_r v_{\phi} - \frac{B_r B_{\phi}}{4\pi}\right\}\right] + \frac{\partial}{\partial z}\left[r\left\{\rho v_z v_{\phi} - \frac{B_z B_{\phi}}{4\pi}\right\}\right] = 0,
\label{cylang1}
\end{equation}
{respectively.}

{
For the purpose of adopting the $\alpha$ prescription \citep{1973A&A....24..337S} for our model, we decompose the azimuthal velocity, $v_{\phi}$, into the mean Keplerian flow and perturbation components as 
\begin{eqnarray}
v_{\phi} = r \Omega + \delta v_{\phi},
\label{eq:vphi}
\end{eqnarray} 
where} 
$\delta v_{\phi} \ll r \Omega$ {and $\Omega$ is the Keplerian frequency:
\begin{eqnarray}
\Omega=\sqrt{\frac{GM}{r^3}}.
\label{eq:okep}
\end{eqnarray} 
Note that $\Phi=-r^2\Omega^2$ at the disk mid-plane ($z=0$) from equation~(\ref{eq:gp}).
In addition, we vertically integrate equations (\ref{cylmcons}) and (\ref{cylang1})
} to get 
\begin{equation}
\frac{\partial \Sigma}{\partial t} + \frac{1}{r}\frac{\partial}{\partial r}(r \Sigma v_r) + \dot{\Sigma}_{\rm w}  = 0
\label{cylmcons1}
\end{equation}

and
\begin{equation}
\frac{\partial}{\partial t}(r^2\Omega \Sigma) + \frac{1}{r} \frac{\partial}{\partial r} \left[r^2 \Sigma\left\{v_r r \Omega + \bar{\alpha}_{r\phi} c_s^2 \right\}\right] + r \left[\dot{\Sigma}_{\rm w}  r \Omega + \bar{\alpha}_{z\phi} \rho c_s^2\right] = 0,
\label{cylang4}
\end{equation}
where 
\begin{eqnarray}
\dot{\Sigma}_{\rm w} = 2\rho v_{z,H}
\label{eq:vmflux}
\end{eqnarray}
is the vertical mass flux, $v_{z,H} \equiv v_{z}(r,H)$ is the vertical velocity evaluated at the disk scale-height $H$, and $\bar{\alpha}_{r\phi}$ 
and $\bar{\alpha}_{z\phi}$ are defined by
\begin{eqnarray}
\bar\alpha_{r\phi}
&\equiv&
\frac{1}{c_{\rm s}^2}
\int_{-H}^{H} \rho \left[v_r \delta v_{\phi} - \frac{B_r B_{\phi}}{4\pi \rho}\right] 
\, dz 
\biggr/
\int_{-H}^{H} \rho\, dz, \nonumber \\
\label{eq:alrphi0}
\\
\bar\alpha_{z\phi}
&\equiv&
\frac{1}{c_{\rm s}^2}
\biggr[
v_z \delta v_{\phi} - \frac{B_z B_{\phi}}{4\pi \rho}
\biggr]_{z=H}, 
\label{eq:alzphi0} 
\end{eqnarray}
respectively.

Equation (\ref{cylang4}) is reduced to
\begin{equation}
r \Sigma v_r = -\frac{2}{r\Omega} \left[\frac{\partial}{\partial r} 
\left(
\bar{\alpha}_{r\phi}
r^2 \Sigma c_s^2 
\right) 
+ 
\bar{\alpha}_{z\phi} r^2 \rho c_s^2
\right],
\label{rsvreqn}
\end{equation} 
where equations~(\ref{eq:okep}) and (\ref{cylmcons1}) are used for the derivation. Equation~(\ref{rsvreqn}) makes it possible to evaluate the mass accretion rate of the disk: $\dot{M} = -2 \pi r \Sigma v_r$. Substituting equation (\ref{rsvreqn}) into equation (\ref{cylmcons1}), we obtain the evolutionary equation of the surface density as
\begin{equation}
\frac{\partial \Sigma}{\partial t} 
- 
\frac{2}{r}\frac{\partial}{\partial r}\left[\frac{1}{r\Omega}
\left\{
\frac{\partial}{\partial r} 
\left(
\bar{\alpha}_{r\phi}
r^2 \Sigma c_s^2 
\right) 
+ 
\bar{\alpha}_{z\phi} r^2 \rho c_s^2 
\right\}
\right] 
+ \dot{\Sigma}_{\rm w} = 0,
\label{sigteqn}
\end{equation}

%
%%%%%%%%%%%%%%%%%%%%%%%%
\subsection{{Energy Equation}}
\label{app_eneq}
%%%%%%%%%%%%%%%%%%%%%%%%
%

The energy conservation equation is given by \citep{1998RvMP...70....1B}

\begin{equation}
\frac{\partial}{\partial t}\left[\frac{1}{2} \rho v^2 + \rho \Phi + \frac{p}{\gamma -1} + \frac{B^2}{8\pi}\right] + \vec{\nabla} \cdot \left[\vec{v} \left(\frac{1}{2} \rho v^2 + \rho \Phi + \frac{\gamma}{\gamma -1}p \right)  + \frac{\vec{B}}{4\pi} \times \left(\vec{v} \times \vec{B}\right) + \vec{Q} \right] = 0,
\nonumber
\end{equation}

\noindent where $\gamma$ is a ratio of specific heats and {$\vec{Q}=(Q_r,Q_\phi,Q_z)$} is other contributions to energy flux in addition to the MHD energy, such as thermal conduction and radiative heating or cooling. 
The above equation is rewritten in cylindrical coordinates with the axisymmetric assumption as
\begingroup
\allowdisplaybreaks
\begin{multline}
\frac{\partial}{\partial t}\left[\frac{1}{2} \rho v^2 + \rho \Phi + \frac{p}{\gamma -1} + \frac{B^2}{8\pi}\right] + \frac{1}{r} \frac{\partial}{\partial r}\left[r \left\{v_r \left(\frac{1}{2} \rho v^2 + \rho \Phi + \frac{\gamma}{\gamma -1}p + \frac{B_{\phi}^2 + B_z^2}{4\pi}\right) - \frac{B_r}{4\pi} \left(v_{\phi}B_{\phi} + v_z B_z\right) + Q_{r}\right\}\right]  \\ + \frac{\partial}{\partial z}\left[v_z \left(\frac{1}{2} \rho v^2 + \rho \Phi + \frac{\gamma}{\gamma -1}p + \frac{B_{\phi}^2 + B_r^2}{4\pi}\right) - \frac{B_z}{4\pi} \left(v_{\phi}B_{\phi} + v_r B_r\right) + Q_{z}\right] = 0.
\label{eneqn}
\end{multline}
\endgroup
Assuming $r \Omega \gg v_r,~\delta v_{\phi},~v_z,~c_s,~B/\sqrt{4\pi \rho}$, the second and third terms of equation (\ref{eneqn}) are reduced to
\begingroup
\allowdisplaybreaks
\begin{multline}
\frac{\partial}{\partial r}\left[r \left\{v_r \left(\frac{1}{2} \rho v^2 + \rho \Phi + \frac{\gamma}{\gamma -1}p + \frac{B_{\phi}^2 + B_z^2}{4\pi}\right) - \frac{B_r}{4\pi} \left(v_{\phi}B_{\phi} +  v_z B_z\right)\right\}\right] =\frac{\partial }{\partial r}\left[r \left\{- \frac{1}{2} \rho r^2 \Omega^2 v_r + \right. \right. \\ \left. \left. \rho r \Omega \left(v_r \delta v_{\phi} - \frac{B_r B_{\phi}}{4\pi \rho}\right) \right\}\right]
\label{mnt3}
\end{multline}
\endgroup
and
\begin{equation}
\frac{\partial}{\partial z}\left[v_z \left(\frac{1}{2} \rho v^2 + \rho \Phi + \frac{\gamma}{\gamma -1}p + \frac{B_{\phi}^2 + B_r^2}{4\pi}\right) - \frac{B_z}{4\pi} \left(v_{\phi}B_{\phi} + v_r B_r\right) \right] = \frac{\partial}{\partial z}(\rho v_z E_{\rm w}),
\label{mnt4}
\end{equation}
respectively, where equations~(\ref{eq:gp}), (\ref{eq:vphi}), and (\ref{eq:okep}) are adopted for these modifications, 
and $E_{\rm w}$ is given as the wind energy by
\begin{equation}
E_{\rm w} = \frac{1}{2} v^2 + \Phi + \frac{\gamma c_s^2}{\gamma -1} + \frac{B_{\phi}^2 + B_r^2}{4\pi \rho} - \frac{B_z}{4\pi\rho v_z} \left(v_{\phi}B_{\phi} + v_r B_r\right).
\label{eq:ew0}
\end{equation}

{Substituting} equations (\ref{mnt3}) and (\ref{mnt4}) {into} equation (\ref{eneqn}), we get
\begin{equation}
\frac{\partial}{\partial t}\left[-\frac{1}{2} \rho r^2 \Omega^2\right] + \frac{1}{r}\frac{\partial }{\partial r}\left[r \left\{- \frac{1}{2} \rho r^2 \Omega^2 v_r + \rho r \Omega 
\left(
v_r \delta v_{\phi} - \frac{B_r B_{\phi}}{4\pi \rho}
\right) 
\right\} \right] + \frac{\partial}{\partial z}(\rho v_z E_{\rm w} + Q_{z}) = 0,
\label{eq:ene}
\end{equation}
where $Q_r=0$ is adopted because there is no energy dissipation in the radial direction. Integrating equation~(\ref{eq:ene}) vertically leads to
\begin{equation}
\frac{\partial}{\partial t}\left[-\frac{1}{2} \Sigma r^2 \Omega^2\right] 
+ 
\frac{1}{r}\frac{\partial }{\partial r}\left[r \left\{- \frac{1}{2} \Sigma r^2 \Omega^2 v_r 
+ 
\bar{\alpha}_{r\phi}
\Sigma r \Omega c_s^2 \right\}\right]  
+ \dot{\Sigma}_{\rm w} E_{\rm w} + {Q}_{\rm rad} = 0,
\label{zeneqn}
\end{equation}
where $Q_{\rm rad} = \int_{-H}^{H}\,Q_{z}\,dz$ is the radiative cooling rate.

Combining equation (\ref{zeneqn}) with equation~(\ref{sigteqn}) results in 
\begin{equation}
\dot{\Sigma}_{\rm w} \left[E_{\rm w} + \frac{r^2 \Omega^2}{2} \right] 
+ {Q}_{\rm rad} 
= 
\frac{3}{2}
\bar{\alpha}_{r\phi}
\Omega \Sigma c_s^2 
+ 
\bar{\alpha}_{z\phi} r \Omega \rho c_s^2,
\end{equation}
where equation~(\ref{eq:okep}) is adopted for the derivation.

%%%%%%%%%%%%%%%%%%%%%%%%%%%%%%%%%%%%%%%%%%%%%%%%%%

%% For this sample we use BibTeX plus aasjournals.bst to generate the
%% the bibliography. The sample631.bib file was populated from ADS. To
%% get the citations to show in the compiled file do the following:
%%
%% pdflatex sample631.tex
%% bibtext sample631
%% pdflatex sample631.tex
%% pdflatex sample631.tex

\bibliography{reference}{}

\begin{thebibliography}{}
\expandafter\ifx\csname natexlab\endcsname\relax\def\natexlab#1{#1}\fi
\providecommand{\url}[1]{\href{#1}{#1}}
\providecommand{\dodoi}[1]{doi:~\href{http://doi.org/#1}{\nolinkurl{#1}}}
\providecommand{\doeprint}[1]{\href{http://ascl.net/#1}{\nolinkurl{http://ascl.net/#1}}}
\providecommand{\doarXiv}[1]{\href{https://arxiv.org/abs/#1}{\nolinkurl{https://arxiv.org/abs/#1}}}

\bibitem[{{Abramowicz} {et~al.}(1988){Abramowicz}, {Czerny}, {Lasota}, \&
  {Szuszkiewicz}}]{1988ApJ...332..646A}
{Abramowicz}, M.~A., {Czerny}, B., {Lasota}, J.~P., \& {Szuszkiewicz}, E. 1988,
  \apj, 332, 646, \dodoi{10.1086/166683}

\bibitem[{{Auchettl} {et~al.}(2017){Auchettl}, {Guillochon}, \&
  {Ramirez-Ruiz}}]{2017ApJ...838..149A}
{Auchettl}, K., {Guillochon}, J., \& {Ramirez-Ruiz}, E. 2017, \apj, 838, 149,
  \dodoi{10.3847/1538-4357/aa633b}

\bibitem[{{Bagnulo} {et~al.}(1999){Bagnulo}, {Landolfi}, \& {Landi
  Degl'Innocenti}}]{1999A&A...343..865B}
{Bagnulo}, S., {Landolfi}, M., \& {Landi Degl'Innocenti}, M. 1999, \aap, 343,
  865

\bibitem[{{Balbus} \& {Hawley}(1991)}]{1991ApJ...376..214B}
{Balbus}, S.~A., \& {Hawley}, J.~F. 1991, \apj, 376, 214,
  \dodoi{10.1086/170270}

\bibitem[{{Balbus} \& {Hawley}(1998)}]{1998RvMP...70....1B}
---. 1998, Reviews of Modern Physics, 70, 1, \dodoi{10.1103/RevModPhys.70.1}

\bibitem[{{Blagorodnova} {et~al.}(2017){Blagorodnova}, {Gezari}, {Hung},
  {Kulkarni}, {Cenko}, {Pasham}, {Yan}, {Arcavi}, {Ben-Ami}, {Bue}, {Cantwell},
  {Cao}, {Castro-Tirado}, {Fender}, {Fremling}, {Gal-Yam}, {Ho}, {Horesh},
  {Hosseinzadeh}, {Kasliwal}, {Kong}, {Laher}, {Leloudas}, {Lunnan}, {Masci},
  {Mooley}, {Neill}, {Nugent}, {Powell}, {Valeev}, {Vreeswijk}, {Walters}, \&
  {Wozniak}}]{2017ApJ...844...46B}
{Blagorodnova}, N., {Gezari}, S., {Hung}, T., {et~al.} 2017, \apj, 844, 46,
  \dodoi{10.3847/1538-4357/aa7579}

\bibitem[{{Blandford} \& {Payne}(1982)}]{1982MNRAS.199..883B}
{Blandford}, R.~D., \& {Payne}, D.~G. 1982, \mnras, 199, 883,
  \dodoi{10.1093/mnras/199.4.883}

\bibitem[{{Bonnerot} {et~al.}(2017{\natexlab{a}}){Bonnerot}, {Price}, {Lodato},
  \& {Rossi}}]{2017MNRAS.469.4879B}
{Bonnerot}, C., {Price}, D.~J., {Lodato}, G., \& {Rossi}, E.~M.
  2017{\natexlab{a}}, \mnras, 469, 4879, \dodoi{10.1093/mnras/stx1210}

\bibitem[{{Bonnerot} {et~al.}(2017{\natexlab{b}}){Bonnerot}, {Rossi}, \&
  {Lodato}}]{2017MNRAS.464.2816B}
{Bonnerot}, C., {Rossi}, E.~M., \& {Lodato}, G. 2017{\natexlab{b}}, \mnras,
  464, 2816, \dodoi{10.1093/mnras/stw2547}

\bibitem[{{Bonnerot} {et~al.}(2016){Bonnerot}, {Rossi}, {Lodato}, \&
  {Price}}]{2016MNRAS.455.2253B}
{Bonnerot}, C., {Rossi}, E.~M., {Lodato}, G., \& {Price}, D.~J. 2016, \mnras,
  455, 2253, \dodoi{10.1093/mnras/stv2411}

\bibitem[{{Brandenburg} {et~al.}(1995){Brandenburg}, {Nordlund}, {Stein}, \&
  {Torkelsson}}]{1995ApJ...446..741B}
{Brandenburg}, A., {Nordlund}, A., {Stein}, R.~F., \& {Torkelsson}, U. 1995,
  \apj, 446, 741, \dodoi{10.1086/175831}

\bibitem[{{Cannizzo} {et~al.}(1990){Cannizzo}, {Lee}, \&
  {Goodman}}]{1990ApJ...351...38C}
{Cannizzo}, J.~K., {Lee}, H.~M., \& {Goodman}, J. 1990, \apj, 351, 38,
  \dodoi{10.1086/168442}

\bibitem[{{Cao} \& {Gu}(2015)}]{2015MNRAS.448.3514C}
{Cao}, X., \& {Gu}, W.-M. 2015, \mnras, 448, 3514, \dodoi{10.1093/mnras/stv269}

\bibitem[{{Chandrasekhar}(1961)}]{1961hhs..book.....C}
{Chandrasekhar}, S. 1961, {Hydrodynamic and hydromagnetic stability} (Oxford
  University Press)

\bibitem[{{Cufari} {et~al.}(2022){Cufari}, {Coughlin}, \&
  {Nixon}}]{2022ApJ...924...34C}
{Cufari}, M., {Coughlin}, E.~R., \& {Nixon}, C.~J. 2022, \apj, 924, 34,
  \dodoi{10.3847/1538-4357/ac32be}

\bibitem[{{Curd} \& {Narayan}(2019)}]{2019MNRAS.483..565C}
{Curd}, B., \& {Narayan}, R. 2019, \mnras, 483, 565,
  \dodoi{10.1093/mnras/sty3134}

\bibitem[{{Curd} \& {Narayan}(2023)}]{2023MNRAS.518.3441C}
---. 2023, \mnras, 518, 3441, \dodoi{10.1093/mnras/stac3330}

\bibitem[{{Dai} {et~al.}(2018){Dai}, {McKinney}, {Roth}, {Ramirez-Ruiz}, \&
  {Miller}}]{2018ApJ...859L..20D}
{Dai}, L., {McKinney}, J.~C., {Roth}, N., {Ramirez-Ruiz}, E., \& {Miller},
  M.~C. 2018, \apjl, 859, L20, \dodoi{10.3847/2041-8213/aab429}

\bibitem[{{Donati} {et~al.}(2002){Donati}, {Babel}, {Harries}, {Howarth},
  {Petit}, \& {Semel}}]{2002MNRAS.333...55D}
{Donati}, J.~F., {Babel}, J., {Harries}, T.~J., {et~al.} 2002, \mnras, 333, 55,
  \dodoi{10.1046/j.1365-8711.2002.05379.x}

\bibitem[{{Feng} {et~al.}(2019){Feng}, {Cao}, {Gu}, \&
  {Ma}}]{2019ApJ...885...93F}
{Feng}, J., {Cao}, X., {Gu}, W.-M., \& {Ma}, R.-Y. 2019, \apj, 885, 93,
  \dodoi{10.3847/1538-4357/ab4592}

\bibitem[{{Folsom} {et~al.}(2016){Folsom}, {Petit}, {Bouvier}, {L{\`e}bre},
  {Amard}, {Palacios}, {Morin}, {Donati}, {Jeffers}, {Marsden}, \&
  {Vidotto}}]{2016MNRAS.457..580F}
{Folsom}, C.~P., {Petit}, P., {Bouvier}, J., {et~al.} 2016, \mnras, 457, 580,
  \dodoi{10.1093/mnras/stv2924}

\bibitem[{{Frank} {et~al.}(2002){Frank}, {King}, \&
  {Raine}}]{2002apa..book.....F}
{Frank}, J., {King}, A., \& {Raine}, D.~J. 2002, {Accretion Power in
  Astrophysics: Third Edition} (Cambridge University Press)

\bibitem[{{Gezari} {et~al.}(2017){Gezari}, {Cenko}, \&
  {Arcavi}}]{2017ApJ...851L..47G}
{Gezari}, S., {Cenko}, S.~B., \& {Arcavi}, I. 2017, \apjl, 851, L47,
  \dodoi{10.3847/2041-8213/aaa0c2}

\bibitem[{{Gezari} {et~al.}(2012){Gezari}, {Chornock}, {Rest}, {Huber},
  {Forster}, {Berger}, {Challis}, {Neill}, {Martin}, {Heckman}, {Lawrence},
  {Norman}, {Narayan}, {Foley}, {Marion}, {Scolnic}, {Chomiuk}, {Soderberg},
  {Smith}, {Kirshner}, {Riess}, {Smartt}, {Stubbs}, {Tonry}, {Wood-Vasey},
  {Burgett}, {Chambers}, {Grav}, {Heasley}, {Kaiser}, {Kudritzki}, {Magnier},
  {Morgan}, \& {Price}}]{2012Natur.485..217G}
{Gezari}, S., {Chornock}, R., {Rest}, A., {et~al.} 2012, \nat, 485, 217,
  \dodoi{10.1038/nature10990}

\bibitem[{{Gogichaishvili} {et~al.}(2018){Gogichaishvili}, {Mamatsashvili},
  {Horton}, \& {Chagelishvili}}]{2018ApJ...866..134G}
{Gogichaishvili}, D., {Mamatsashvili}, G., {Horton}, W., \& {Chagelishvili}, G.
  2018, \apj, 866, 134, \dodoi{10.3847/1538-4357/aadbad}

\bibitem[{{Golightly} {et~al.}(2019){Golightly}, {Coughlin}, \&
  {Nixon}}]{2019ApJ...872..163G}
{Golightly}, E. C.~A., {Coughlin}, E.~R., \& {Nixon}, C.~J. 2019, \apj, 872,
  163, \dodoi{10.3847/1538-4357/aafd2f}

\bibitem[{{Gu} \& {Lu}(2007)}]{2007ApJ...660..541G}
{Gu}, W.-M., \& {Lu}, J.-F. 2007, \apj, 660, 541, \dodoi{10.1086/512967}

\bibitem[{{Guillochon} \& {McCourt}(2017)}]{2017ApJ...834L..19G}
{Guillochon}, J., \& {McCourt}, M. 2017, \apjl, 834, L19,
  \dodoi{10.3847/2041-8213/834/2/L19}

\bibitem[{{Hawley} {et~al.}(1995){Hawley}, {Gammie}, \&
  {Balbus}}]{1995ApJ...440..742H}
{Hawley}, J.~F., {Gammie}, C.~F., \& {Balbus}, S.~A. 1995, \apj, 440, 742,
  \dodoi{10.1086/175311}

\bibitem[{{Hayasaki} \& {Jonker}(2021)}]{2021ApJ...921...20H}
{Hayasaki}, K., \& {Jonker}, P.~G. 2021, \apj, 921, 20,
  \dodoi{10.3847/1538-4357/ac18c2}

\bibitem[{{Hayasaki} {et~al.}(2013){Hayasaki}, {Stone}, \&
  {Loeb}}]{2013MNRAS.434..909H}
{Hayasaki}, K., {Stone}, N., \& {Loeb}, A. 2013, \mnras, 434, 909,
  \dodoi{10.1093/mnras/stt871}

\bibitem[{{Hayasaki} {et~al.}(2016){Hayasaki}, {Stone}, \&
  {Loeb}}]{2016MNRAS.461.3760H}
---. 2016, \mnras, 461, 3760, \dodoi{10.1093/mnras/stw1387}

\bibitem[{{Hayasaki} {et~al.}(2018){Hayasaki}, {Zhong}, {Li}, {Berczik}, \&
  {Spurzem}}]{2018ApJ...855..129H}
{Hayasaki}, K., {Zhong}, S., {Li}, S., {Berczik}, P., \& {Spurzem}, R. 2018,
  \apj, 855, 129, \dodoi{10.3847/1538-4357/aab0a5}

\bibitem[{{Hills}(1975)}]{1975Natur.254..295H}
{Hills}, J.~G. 1975, \nat, 254, 295, \dodoi{10.1038/254295a0}

\bibitem[{{Holoien} {et~al.}(2014){Holoien}, {Prieto}, {Bersier}, {Kochanek},
  {Stanek}, {Shappee}, {Grupe}, {Basu}, {Beacom}, {Brimacombe}, {Brown},
  {Davis}, {Jencson}, {Pojmanski}, \& {Szczygie{\l}}}]{2014MNRAS.445.3263H}
{Holoien}, T.~W.~S., {Prieto}, J.~L., {Bersier}, D., {et~al.} 2014, \mnras,
  445, 3263, \dodoi{10.1093/mnras/stu1922}

\bibitem[{{Holoien} {et~al.}(2016){Holoien}, {Kochanek}, {Prieto}, {Grupe},
  {Chen}, {Godoy-Rivera}, {Stanek}, {Shappee}, {Dong}, {Brown}, {Basu},
  {Beacom}, {Bersier}, {Brimacombe}, {Carlson}, {Falco}, {Johnston}, {Madore},
  {Pojmanski}, \& {Seibert}}]{2016MNRAS.463.3813H}
{Holoien}, T.~W.~S., {Kochanek}, C.~S., {Prieto}, J.~L., {et~al.} 2016, \mnras,
  463, 3813, \dodoi{10.1093/mnras/stw2272}

\bibitem[{{Holoien} {et~al.}(2020){Holoien}, {Auchettl}, {Tucker}, {Shappee},
  {Patel}, {Miller-Jones}, {Mockler}, {Groenewald}, {Hinkle}, {Brown},
  {Kochanek}, {Stanek}, {Chen}, {Dong}, {Prieto}, {Thompson}, {Beaton},
  {Connor}, {Cowperthwaite}, {Dahmen}, {French}, {Morrell}, {Buckley},
  {Gromadzki}, {Roy}, {Coulter}, {Dimitriadis}, {Foley}, {Kilpatrick}, {Piro},
  {Rojas-Bravo}, {Siebert}, \& {van Velzen}}]{2020ApJ...898..161H}
{Holoien}, T. W.~S., {Auchettl}, K., {Tucker}, M.~A., {et~al.} 2020, \apj, 898,
  161, \dodoi{10.3847/1538-4357/ab9f3d}

\bibitem[{{Hung} {et~al.}(2017){Hung}, {Gezari}, {Blagorodnova}, {Roth},
  {Cenko}, {Kulkarni}, {Horesh}, {Arcavi}, {McCully}, {Yan}, {Lunnan},
  {Fremling}, {Cao}, {Nugent}, \& {Wozniak}}]{2017ApJ...842...29H}
{Hung}, T., {Gezari}, S., {Blagorodnova}, N., {et~al.} 2017, \apj, 842, 29,
  \dodoi{10.3847/1538-4357/aa7337}

\bibitem[{{Jafari} \& {Vishniac}(2018)}]{2018ApJ...854....2J}
{Jafari}, A., \& {Vishniac}, E.~T. 2018, \apj, 854, 2,
  \dodoi{10.3847/1538-4357/aaa75b}

\bibitem[{{Kajava} {et~al.}(2020){Kajava}, {Giustini}, {Saxton}, \&
  {Miniutti}}]{2020A&A...639A.100K}
{Kajava}, J. J.~E., {Giustini}, M., {Saxton}, R.~D., \& {Miniutti}, G. 2020,
  \aap, 639, A100, \dodoi{10.1051/0004-6361/202038165}

\bibitem[{Kato {et~al.}(2008)Kato, Fukue, \& Mineshige}]{kato_black-hole_2008}
Kato, S., Fukue, J., \& Mineshige, S. 2008, Black-{Hole} {Accretion} {Disks}
  --- {Towards} a {New} {Paradigm} ---.
\newblock \url{https://ui.adsabs.harvard.edu/abs/2008bhad.book.....K}

\bibitem[{{Li} \& {Cao}(2019)}]{2019ApJ...872..149L}
{Li}, J., \& {Cao}, X. 2019, \apj, 872, 149, \dodoi{10.3847/1538-4357/ab0207}

\bibitem[{{Lightman} \& {Eardley}(1974)}]{1974ApJ...187L...1L}
{Lightman}, A.~P., \& {Eardley}, D.~M. 1974, \apjl, 187, L1,
  \dodoi{10.1086/181377}

\bibitem[{{Liu} {et~al.}(2022){Liu}, {Dou}, {Chen}, \&
  {Shen}}]{2022ApJ...925...67L}
{Liu}, X.-L., {Dou}, L.-M., {Chen}, J.-H., \& {Shen}, R.-F. 2022, \apj, 925,
  67, \dodoi{10.3847/1538-4357/ac33a9}

\bibitem[{{Lodato} {et~al.}(2009){Lodato}, {King}, \&
  {Pringle}}]{2009MNRAS.392..332L}
{Lodato}, G., {King}, A.~R., \& {Pringle}, J.~E. 2009, \mnras, 392, 332,
  \dodoi{10.1111/j.1365-2966.2008.14049.x}

\bibitem[{{Mageshwaran} {et~al.}(2023){Mageshwaran}, {Shaw}, {Bhattacharyya},
  \& {Hayasaki}}]{2023arXiv231217417M}
{Mageshwaran}, T., {Shaw}, G., {Bhattacharyya}, S., \& {Hayasaki}, K. 2023,
  arXiv e-prints, arXiv:2312.17417, \dodoi{10.48550/arXiv.2312.17417}

\bibitem[{{Matsumoto} \& {Tajima}(1995)}]{1995ApJ...445..767M}
{Matsumoto}, R., \& {Tajima}, T. 1995, \apj, 445, 767, \dodoi{10.1086/175739}

\bibitem[{{Mummery} \& {Balbus}(2019)}]{2019MNRAS.489..132M}
{Mummery}, A., \& {Balbus}, S.~A. 2019, \mnras, 489, 132,
  \dodoi{10.1093/mnras/stz2141}

\bibitem[{{Nicholl} {et~al.}(2019){Nicholl}, {Blanchard}, {Berger}, {Gomez},
  {Margutti}, {Alexander}, {Guillochon}, {Leja}, {Chornock}, {Snios},
  {Auchettl}, {Bruce}, {Challis}, {D'Orazio}, {Drout}, {Eftekhari}, {Foley},
  {Graur}, {Kilpatrick}, {Lawrence}, {Piro}, {Rojas-Bravo}, {Ross}, {Short},
  {Smartt}, {Smith}, \& {Stalder}}]{2019MNRAS.488.1878N}
{Nicholl}, M., {Blanchard}, P.~K., {Berger}, E., {et~al.} 2019, \mnras, 488,
  1878, \dodoi{10.1093/mnras/stz1837}

\bibitem[{{Nicholl} {et~al.}(2020){Nicholl}, {Wevers}, {Oates}, {Alexander},
  {Leloudas}, {Onori}, {Jerkstrand}, {Gomez}, {Campana}, {Arcavi},
  {Charalampopoulos}, {Gromadzki}, {Ihanec}, {Jonker}, {Lawrence}, {Mandel},
  {Schulze}, {Short}, {Burke}, {McCully}, {Hiramatsu}, {Howell}, {Pellegrino},
  {Abbot}, {Anderson}, {Berger}, {Blanchard}, {Cannizzaro}, {Chen},
  {Dennefeld}, {Galbany}, {Gonz{\'a}lez-Gait{\'a}n}, {Hosseinzadeh}, {Inserra},
  {Irani}, {Kuin}, {M{\"u}ller-Bravo}, {Pineda}, {Ross}, {Roy}, {Smartt},
  {Smith}, {Tucker}, {Wyrzykowski}, \& {Young}}]{2020MNRAS.499..482N}
{Nicholl}, M., {Wevers}, T., {Oates}, S.~R., {et~al.} 2020, \mnras, 499, 482,
  \dodoi{10.1093/mnras/staa2824}

\bibitem[{{Park} \& {Hayasaki}(2020)}]{2020ApJ...900....3P}
{Park}, G., \& {Hayasaki}, K. 2020, \apj, 900, 3,
  \dodoi{10.3847/1538-4357/ab9ebb}

\bibitem[{{Pascucci} {et~al.}(2023){Pascucci}, {Cabrit}, {Edwards}, {Gorti},
  {Gressel}, \& {Suzuki}}]{2023ASPC..534..567P}
{Pascucci}, I., {Cabrit}, S., {Edwards}, S., {et~al.} 2023, in Astronomical
  Society of the Pacific Conference Series, Vol. 534, Protostars and Planets
  VII, ed. S.~{Inutsuka}, Y.~{Aikawa}, T.~{Muto}, K.~{Tomida}, \& M.~{Tamura},
  567, \dodoi{10.48550/arXiv.2203.10068}

\bibitem[{{Petit} {et~al.}(2008){Petit}, {Wade}, {Drissen}, {Montmerle}, \&
  {Alecian}}]{2008MNRAS.387L..23P}
{Petit}, V., {Wade}, G.~A., {Drissen}, L., {Montmerle}, T., \& {Alecian}, E.
  2008, \mnras, 387, L23, \dodoi{10.1111/j.1745-3933.2008.00474.x}

\bibitem[{{Piro} \& {Lu}(2020)}]{2020ApJ...894....2P}
{Piro}, A.~L., \& {Lu}, W. 2020, \apj, 894, 2, \dodoi{10.3847/1538-4357/ab83f6}

\bibitem[{{Rees}(1988)}]{1988Natur.333..523R}
{Rees}, M.~J. 1988, \nat, 333, 523, \dodoi{10.1038/333523a0}

\bibitem[{{Roth} {et~al.}(2020){Roth}, {Rossi}, {Krolik}, {Piran}, {Mockler},
  \& {Kasen}}]{2020SSRv..216..114R}
{Roth}, N., {Rossi}, E.~M., {Krolik}, J., {et~al.} 2020, \ssr, 216, 114,
  \dodoi{10.1007/s11214-020-00735-1}

\bibitem[{{Schmidt} {et~al.}(2003){Schmidt}, {Harris}, {Liebert}, {Eisenstein},
  {Anderson}, {Brinkmann}, {Hall}, {Harvanek}, {Hawley}, {Kleinman}, {Knapp},
  {Krzesinski}, {Lamb}, {Long}, {Munn}, {Neilsen}, {Newman}, {Nitta},
  {Schlegel}, {Schneider}, {Silvestri}, {Smith}, {Snedden}, {Szkody}, \&
  {Vanden Berk}}]{2003ApJ...595.1101S}
{Schmidt}, G.~D., {Harris}, H.~C., {Liebert}, J., {et~al.} 2003, \apj, 595,
  1101, \dodoi{10.1086/377476}

\bibitem[{{Shakura} \& {Sunyaev}(1973)}]{1973A&A....24..337S}
{Shakura}, N.~I., \& {Sunyaev}, R.~A. 1973, \aap, 24, 337

\bibitem[{{Starling} {et~al.}(2004){Starling}, {Siemiginowska}, {Uttley}, \&
  {Soria}}]{2004MNRAS.347...67S}
{Starling}, R. L.~C., {Siemiginowska}, A., {Uttley}, P., \& {Soria}, R. 2004,
  \mnras, 347, 67, \dodoi{10.1111/j.1365-2966.2004.07167.x}

\bibitem[{{Strubbe} \& {Quataert}(2009)}]{2009MNRAS.400.2070S}
{Strubbe}, L.~E., \& {Quataert}, E. 2009, \mnras, 400, 2070,
  \dodoi{10.1111/j.1365-2966.2009.15599.x}

\bibitem[{{Suzuki} \& {Inutsuka}(2009)}]{2009ApJ...691L..49S}
{Suzuki}, T.~K., \& {Inutsuka}, S.-i. 2009, \apjl, 691, L49,
  \dodoi{10.1088/0004-637X/691/1/L49}

\bibitem[{{Suzuki} \& {Inutsuka}(2014)}]{2014ApJ...784..121S}
---. 2014, \apj, 784, 121, \dodoi{10.1088/0004-637X/784/2/121}

\bibitem[{{Suzuki} {et~al.}(2010){Suzuki}, {Muto}, \&
  {Inutsuka}}]{2010ApJ...718.1289S}
{Suzuki}, T.~K., {Muto}, T., \& {Inutsuka}, S.-i. 2010, \apj, 718, 1289,
  \dodoi{10.1088/0004-637X/718/2/1289}

\bibitem[{{Suzuki} {et~al.}(2016){Suzuki}, {Ogihara}, {Morbidelli}, {Crida}, \&
  {Guillot}}]{2016A&A...596A..74S}
{Suzuki}, T.~K., {Ogihara}, M., {Morbidelli}, A., {Crida}, A., \& {Guillot}, T.
  2016, \aap, 596, A74, \dodoi{10.1051/0004-6361/201628955}

\bibitem[{{Tchekhovskoy} {et~al.}(2014){Tchekhovskoy}, {Metzger}, {Giannios},
  \& {Kelley}}]{2014MNRAS.437.2744T}
{Tchekhovskoy}, A., {Metzger}, B.~D., {Giannios}, D., \& {Kelley}, L.~Z. 2014,
  \mnras, 437, 2744, \dodoi{10.1093/mnras/stt2085}

\bibitem[{{Velikhov}(1959)}]{Velikhov1959}
{Velikhov}, E.~P. 1959, Zh. Eksp. Teor. Fiz., 36, 1398

\bibitem[{{Wyrzykowski} {et~al.}(2017){Wyrzykowski}, {Zieli{\'n}ski},
  {Kostrzewa-Rutkowska}, {Hamanowicz}, {Jonker}, {Arcavi}, {Guillochon},
  {Brown}, {Koz{\l}owski}, {Udalski}, {Szyma{\'n}ski}, {Soszy{\'n}ski},
  {Poleski}, {Pietrukowicz}, {Skowron}, {Mr{\'o}z}, {Ulaczyk}, {Pawlak},
  {Rybicki}, {Greiner}, {Kr{\"u}hler}, {Bolmer}, {Smartt}, {Maguire}, \&
  {Smith}}]{2017MNRAS.465L.114W}
{Wyrzykowski}, {\L}., {Zieli{\'n}ski}, M., {Kostrzewa-Rutkowska}, Z., {et~al.}
  2017, \mnras, 465, L114, \dodoi{10.1093/mnrasl/slw213}

\bibitem[{{Xue} {et~al.}(2011){Xue}, {Sadowski}, {Abramowicz}, \&
  {Lu}}]{2011ApJS..195....7X}
{Xue}, L., {Sadowski}, A., {Abramowicz}, M.~A., \& {Lu}, J.-F. 2011, \apjs,
  195, 7, \dodoi{10.1088/0067-0049/195/1/7}

\bibitem[{{Zhong} {et~al.}(2023){Zhong}, {Hayasaki}, {Li}, {Berczik}, \&
  {Spurzem}}]{2023ApJ...959...19Z}
{Zhong}, S., {Hayasaki}, K., {Li}, S., {Berczik}, P., \& {Spurzem}, R. 2023,
  \apj, 959, 19, \dodoi{10.3847/1538-4357/ad0122}

\end{thebibliography}
\bibliographystyle{aasjournal}

%% This command is needed to show the entire author+affiliation list when
%% the collaboration and author truncation commands are used.  It has to
%% go at the end of the manuscript.
%\allauthors

%% Include this line if you are using the \added, \replaced, \deleted
%% commands to see a summary list of all changes at the end of the article.
%\listofchanges

\end{document}